\documentclass{optica-article}

\journal{opticajournal} 

\articletype{Research Article}

\usepackage{lineno}

\usepackage{xcolor}
\usepackage[caption=false]{subfig}
\usepackage{siunitx}
\usepackage{braket}

\usepackage{orcidlink}

\usepackage[final]{changes}

\begin{document}

\title{Number-State Reconstruction with a Single Single-Photon Avalanche Detector}

\author{Patrick R. Banner\,\orcidlink{0009-0006-9957-4996},\authormark{1,* } Deniz Kurdak\,\orcidlink{0000-0003-4076-3013},\authormark{1} Yaxin Li\,\orcidlink{0000-0001-8734-0136},\authormark{1} Alan Migdall\,\orcidlink{0000-0002-8444-8288},\authormark{1} J. V. Porto\,\orcidlink{0000-0002-6290-7535},\authormark{1} and S. L. Rolston\,\orcidlink{0000-0003-1671-4190}\authormark{1}}

\address{\authormark{1}Joint Quantum Institute, NIST/University of Maryland, College Park, MD 20742, USA}
\email{\authormark{*}pbanner@umd.edu}


\begin{abstract*}
\added{Single-photon avalanche detectors (SPADs) are crucial sensors of light for many fields and applications. However, they are not able to resolve photon number, so typically more complex and more expensive experimental setups or devices must be used to measure the number of photons in a pulse. Here }\deleted{W}\added{w}e present a methodology for performing photon number-state \deleted{tomography}\added{reconstruction} with only one \added{SPAD}\deleted{single-photon avalanche detector (SPAD)}. The methodology, which is cost-effective and easy to implement, \added{uses maximum-likelihood techniques with a detector model whose parameters are measurable.}\deleted{uses a detector model with measurable parameters together with maximum-likelihood analysis.} We achieve excellent agreement between known input pulses and their reconstructions for coherent states with up to $\approx 10$ photons and peak input photon rates up to several Mcounts/s. W\added{hen detector imperfections are small, w}e maintain good agreement for coherent pulses with peak input photon rates of over 40 Mcounts/s, greater than one photon per detector dead time. For anti-bunched light, the reconstructed and independently measured \added{pulse-averaged} values of $g^{(2)}(0)$ are also consistent\added{ with one another}. Our algorithm is applicable to \deleted{coherent pulses and non-classical light with $g^{(2)}(0) < 1$,}\added{light pulses} whose \deleted{total} pulse width\added{ and correlation timescales} \deleted{is}\added{are both} at least a few detector dead times. These results, achieved with single commercially available SPADs, provide an inexpensive number-state \deleted{tomography}\added{reconstruction} method and expand the capabilities of single-photon detectors.
\end{abstract*}

\section{Introduction}

Single-photon avalanche detectors (SPADs) are critical for many areas of research in atomic physics, quantum optics, and quantum communication as well as a variety of other fields, including biology at the single-photon level~\cite{Atoms1, Atoms2, Atoms3, OWL, Entanglement1, Entanglement2, QRNG, QKD1, QKD2, QKD3, QComm1, HofbauerBook, LIDAR1, VisLightComm, Bio1, Bio2, Bio3, Imaging1, Space1, AddlIntro1, AddlIntro2}. The operating principles of SPADs have been described in detail~\cite{MigdallSPDs, QuenchingCircuits}. Briefly, SPADs are photodiodes reverse biased well above the\added{ir} breakdown voltage. In this state, when a single photon impinges on the photodiode and excites an electron to the conduction band, the large bias accelerates the electron to high enough energies to cause impact ionization, resulting in an ``avalanche'' of current. This avalanche is sensed by a quenching circuit~\cite{QuenchingCircuits}, which lowers the bias until the avalanche ends, and then increases the bias back to its normally operating value. Importantly, during the avalanche, an additional photon may arrive and excite a new electron, but the noise on the avalanche is so large that this \deleted{single }additional excitation is unresolvable. Thus, SPADs are inherently unable to resolve photon number, unlike other detectors that have at least some photon-number-resolving capability~\cite{TES1, SNSPD1, SNSPD2}. This complicates what might be one of the most obvious uses of devices that can detect single photons: number-state \deleted{tomography}\added{reconstruction}. To obtain photon-number information, a common approach is to assume a SPAD produces either zero or one click per experimental attempt, and this "click/no-click" information is the basis of several techniques that use multiple detectors or multiplex a single detector~\cite{TDM1, TDM2, TDM3, SPADArray2, SPADArray1, Multi1, Multi2, OH1, ManyEtas, Landweber}. These techniques by their nature increase costs or complexity of implementation.

In this work, we perform photon number-state \deleted{tomography}\added{reconstruction} using a single SPAD with no\added{ inherent} photon-number-resolving capability. In our scheme, light pulses and correlation timescales longer than the detector dead time allow the SPAD to click more than once in a measurement period. This simple scheme, when combined with maximum-likelihood analysis, allows for remarkable performance with minimum cost and complexity, while the requirement that light pulses are longer than the detector dead time is \deleted{applicable to}\added{compatible with} many modern atomic physics and quantum optics experiments. Our algorithm also accounts for various SPAD imperfections, \deleted{including}\added{namely} detector efficiency less than unity, dark counts~\cite{DC1}, afterpulsing~\cite{AP1, AP_Modeling, CountingStats}, and dead times~\cite{MigdallSPDs}.

\deleted{In this paper, we use this information as input to a maximum-likelihood algorithm to perform number-state \deleted{tomography}\added{reconstruction}.} We\added{ first} develop a model for the SPAD, discussed in Section~\ref{sec:model} (and detailed in Section A of Supplement 1). Our model requires detector characterization, which we detail in Sections B and C of Supplement 1. We use this data to perform the reconstruction via the expectation-maximization-entropy (EME) algorithm~\cite{EME}, reviewed in Section~\ref{sec:recon}. The implementation of this model and the EME algorithm in code have been made open access~\cite{Git}. To validate the algorithm, we use carefully calibrated coherent pulses, whose number distribution we know \emph{a priori}. The results are discussed in Section~\ref{sec:results}. For coherent-state pulses with peak count rates of up to a few Mcounts/s, the total variation distance we achieve is below $10^{-2}$, both for simple square pulses and pulses with nontrivial time profiles. We also explore the algorithm and its limitations at high count rates, approaching one photon per dead time. Even at these high count rates, reasonably good agreement is achieved\added{, depending on detector parameters}. We also show that the reconstruction of non-classical light achieves good agreement with an independently measured value of the second-order pulse-averaged autocorrelation function $g^{(2)}(0)$. Thus we show that a single inexpensive SPAD is capable of performing number-state \deleted{tomography}\added{reconstruction} for a wide range of input light statistics.

\section{Description and Model of SPADs\label{sec:model}}

Given that the clicks produced by a SPAD do not correspond one-to-one to photons incident on the detector, we adopt the following formalism. Consider a pulse of light with a number distribution $\mathbf{P}$ (equivalent to the diagonal of the Fock-basis density matrix of the light). The ensemble measurement of this light produces a corresponding click number distribution $\mathbf{C}$. We model the detector as a matrix $\mathbb{D}$ relating these two vectors:
\begin{equation}
    \mathbf{C} = \mathbb{D}\mathbf{P}.
\end{equation}
(Note that we perform a truncation of the number basis to $n \in [0,n_{\text{max}}]$, so that all of the objects here are finite in size.) Each element of $\mathbb{D}$ is interpreted as a probability that a number of incident photons (given by the element's column) produces a number of clicks (given by the element's row). Since any given number of photons must produce some number of clicks (including zero), each column of $\mathbb{D}$ must sum to unity. 

The matrix $\mathbb{D}$ in our model is the product of four individual matrices, each of which contains information about a particular imperfection of the SPAD. (1) The effects of the non-unit detector efficiency are captured by a matrix $\mathbb{L}(\eta_0)$, where we define $\eta_0$ as the detector efficiency when the detector is operating normally, i.e. recovered from all previous clicks. (2) The effects of background counts are accounted for by a matrix $\mathbb{B}(p_b)$, where $p_b$ is the probability of a background count in the data collection window. Here background events result from both dark counts~\cite{DC1} and ambient, non-signal photons. (3) Recovery time effects~\cite{MigdallSPDs, TwilightDelays, CountingStats} are due to the quenching circuit turning the bias off and back on after a click, which takes a total time called the ``recovery time.'' There are two such effects, both accounted for in the matrix $\mathbb{R}$. One is losses due to the detector dead time, when the detector bias is below breakdown so that avalanching does not occur. The other effect is twilight counts, when photons produce clicks while the detector bias is turning back on. We treat the system efficiency as time-dependent after each click (see Fig. S1 of Supplement 1), and account for the time delay typical of twilight counts~\cite{TwilightDelays}. Rather than measuring both the time-dependence of the detector efficiency and the delay profile of the twilight counts~\cite{TwilightDelays}, we assume a model, and measure the effective parameters of that model (see Supplement 1 for details). Thus we write this matrix as $\mathbb{R}[\gamma(t), D(\tau)]$ to signify that it depends on both the photon profile $\gamma(t)$ and a model for the time-dependent detector efficiency $D(\tau)$. (4) Finally, afterpulsing effects~\cite{AP1, AP_Modeling} are modeled by the matrix $\mathbb{A}(p_a)$, where $p_a$ is the probability of one afterpulse click in the data collection window given the input photon profile. This probability is constructed from the detector's temporal afterpulsing profile as well as its total probability of afterpulsing $\tilde{p}_a$. The measured values of the detector parameters relevant for all four matrices are summarized in Table~\ref{tab:DetValues}; the methodology for these measurements is described in Section B of Supplement 1.

\begin{table*}[t]
\centering
\caption{\label{tab:DetValues}%
Measured parameters for two SPADs under test. The properties are detector efficiency (DE, $\eta_0$), background count rate $r_b$, total probability for the detector to afterpulse after a click ($\tilde{p}_a$), and the dead, reset, and recovery times $t_{\text{dead}}$, $t_{\text{reset}}$, and $t_{\text{rec}}$ respectively. The given uncertainties are one standard deviation. See Sections B and C of Supplement 1 for measurement details and the uncertainty calculations, respectively.}
\sisetup{
table-alignment-mode = format,
table-number-alignment = center,
}
\begin{tabular}{@{}
l
S[table-format = 1.3(1)]
S[table-format = 3.0(1)]
S[table-format = 1.5(1)]
S[table-format = 2.2(1)]
S[table-format = 1.2(1)]
S[table-format = 2.2(1)]
@{}}
\hline\hline
Detector & {DE, $\eta_0$} & {$r_b$ (counts/s)} & {$\tilde{p}_a$} & {$t_{\text{dead}}$ (ns)} & {$t_{\text{reset}}$ (ns)} & {$t_{\text{rec}}$ (ns)} \\
\hline
SPAD1 & 0.633(3) & 137(1) & 0.00602(2) & 14.05(8) & 8.67(2) & 22.72(8) \\
SPAD2 & 0.660(3) & 205(1) & 0.02482(3) & 13.47(8) & 8.26(2) & 21.73(8) \\
\hline\hline
\end{tabular}
\end{table*}

These matrices are combined into $\mathbb{D}$ by multiplying in a physically motivated order. First, incoming photons are subjected to the loss of $\mathbb{L}$. Then background counts are added using $\mathbb{B}$. Both background clicks and clicks produced by signal photons are subject to recovery time effects encoded in $\mathbb{R}$. Finally, the clicks that remain may afterpulse, modeled by $\mathbb{A}$. Thus we compute
\begin{equation}
\mathbb{D} = \mathbb{A}(p_a) \mathbb{R}[\gamma(t), D(\tau)] \mathbb{B}(p_B) \mathbb{L}(\eta_0).
\end{equation}
The details of the construction of the four matrices are found in Supplement 1, and our implementation in code is publicly available in Code 1~\cite{Git}.

\section{Reconstruction Algorithm\label{sec:recon}}

Given the relation $\mathbf{C} = \mathbb{D}\mathbf{P}$, and given the measured click data $\mathbf{C}$ and the matrix $\mathbb{D}$, it seems natural simply to invert $\mathbb{D}$ to find $\mathbf{P}$. However, this direct matrix inversion may not produce a physically reasonable result, i.e. one which sums to unity and has no negative elements. If nothing else, physically unreasonable results may arise due to sampling error. This problem is exacerbated when, for instance, the detection efficiency is low.

Instead, we use the maximum-likelihood principle. The likelihood can be constrained via Lagrange multipliers and can be maximized over the relevant number of variables using any maximization method. Here we use an iterative procedure, a variant of the expectation-maximization (EM) algorithm known as the expectation-maximization-entropy (EME) algorithm~\cite{EME}. We initialize the algorithm with a uniform distribution $\mathbf{P}^{(0)}_n = 1/(n_{\text{max}}+1)$. Then for all subsequent iterations, the distribution is
\begin{equation}
    \mathbf{P}^{(k+1)}_n = \sum_{m=0}^{n_{\text{max}}} \frac{\mathbf{C}_m}{\sum_j \mathbb{D}_{mj} \mathbf{P}_j^{(k)}} \mathbb{D}_{mn} \mathbf{P}_n^{(k)} - \alpha \left( \ln \mathbf{P}_n^{(k)} - S^{(k)} \right),
\end{equation}
where $S^{(k)} = \sum_{n=0}^{n_{\text{max}}} \mathbf{P}_n^{(k)} \ln \mathbf{P}_n^{(k)}$, and $\mathbf{P}_n^{(k)}$ is the probability of $n$ photons predicted by the $k^{\text{th}}$ iteration. The parameter $\alpha$ controls the strength of the entropy term; this term smoothes the reconstruction and, for classical and bunched light, \deleted{increases}\added{improves} the fidelity of the reconstruction~\cite{MaxEntBook1, EME}. We use \deleted{$\alpha = 5 \times 10^{-3}$}\added{$\alpha = 10^{-3}$} for all of our reconstructions\deleted{ of coherent states, and $\alpha = 0$ for our reconstruction of anti-bunched light}, following~\cite{EME}\deleted{ (also see discussion below)}. The algorithm is iterated until a stop condition is met, namely
\begin{equation}
    \sqrt{\sum_n \left( \mathbf{P}_n^{(k+1)} - \mathbf{P}_n^{(k)} \right)^2} < \epsilon
\end{equation}
where, in this work, $\epsilon=10^{-12}$ is used, as in~\cite{EME}. Convergence typically takes no more than $\approx 10^4$ iterations, which happens in a fraction of a second using an ordinary laptop. Our implementation can be found in Code 1~\cite{Git}.

\section{Results and Discussion\label{sec:results}}

\begin{figure*}[t!]
\includegraphics[width=\textwidth]{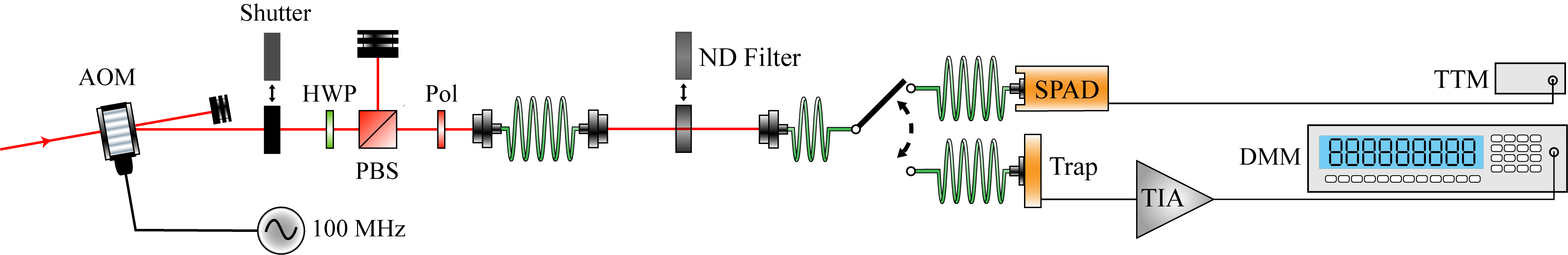}
\caption{\label{fig:ExperimentalSetup} The experimental setup for \deleted{tomography}\added{reconstruction} measurements. 780 nm light is sent through an acousto-optic modulator (AOM) and a shutter, then a system for power control consisting of a half-waveplate (HWP), a polarizing beam splitter (PBS), and a polarizer (Pol) on a motorized rotation mount. The light is then coupled into a fiber, whose output is passed free-space through a calibrated ND filter on a motorized flip mount, then coupled into another fiber, which may be connected to a trap detector~\cite{Traps} or to a SPAD. The trap detector is attached to a high-gain precision transimpedance amplifier (TIA) whose output voltage is read by an externally triggered digital multimeter (DMM), while the SPAD signals are sent to a time tagger module (TTM).}
\end{figure*}

To validate our algorithm, we sent calibrated coherent light pulses into our detectors using the setup shown in Fig.~\ref{fig:ExperimentalSetup}. The light ($\lambda=780$ nm) is sent through a set of polarization optics for power control, as well as an acousto-optic modulator (AOM) to shape the pulses and a slow shutter for protecting the SPADs. The light is coupled into a fiber, whose output is passed through a calibrated neutral density (ND) filter with transmittance $T_{\text{ND}} = 0.001412(1)$. The light is then coupled into another fiber, which is inserted into either a SPAD or into a trap detector~\cite{Traps} used for calibration. The trap consists of multiple photodiodes arranged to allow very high efficiency and uniformity, making it well-suited as a calibration transfer standard. The trap is read through a high-gain precision transimpedance amplifier (TIA) and an 8-1/2-digit digital multimeter (DMM). SPAD signals are sent to a time tagger module (TTM). \added{Our TTM has a resolution of about 164 ps, and we use bins of about 1 ns width (6 TTM bins) for all of our reconstructions.}

To calibrate the average photon number per pulse, we connect the fiber to the trap detector, with no ND filter in the path, and record the voltage. The power incident on the trap is $P_0 = V_{\text{meas}}/(\eta_{\text{trap}}RG)$, where $\eta_{\text{trap}}$ is the efficiency of the trap, $R$ is the detector responsivity, $G$ is the TIA gain, and $V_{\text{meas}}$ is the average voltage reading. The calibrated ND filter is then inserted to reduce the power at the SPAD to $P_{\text{in}} = T_{\text{ND}}P_0$. Finally, while the measurement of $P_0$ is done with continuous-wave (cw) light, the final \deleted{tomography}\added{reconstruction} measurements are done on pulsed light, so we convert the cw power to a total average energy in a pulse via a factor $f_s$, which accounts for the pulse shape relative to the cw measurement and depends on the time window for the data collection (see Supplement 1). Thus the expected average number of photons in each coherent pulse is
\begin{equation}
    \bar{n}_{\text{exp}} = \frac{P_{\text{in}}f_s}{hc/\lambda} = \frac{V_{\text{meas}}T_{\text{ND}}f_s}{\eta_{\text{trap}}RG} \frac{1}{hc/\lambda}.
    \label{eqn:n_exp}
\end{equation}

To differentiate calibration errors from reconstruction errors, we report two metrics for coherent state reconstruction. First, we fit the reconstructed result to a Poissonian distribution and extract an average photon number $\bar{n}_{\text{fit}}$. We then report the fractional difference between $\bar{n}_{\text{fit}}$ and $\bar{n}_{\text{exp}}$, calculated as $\delta\bar{n} = (\bar{n}_{\text{exp}} - \bar{n}_{\text{fit}})/\bar{n}_{\text{exp}}$, which gives an estimate of the calibration error. Second, to quantify how well the reconstruction matches the fitted Poissonian, thus estimating the reconstruction error, we report a total variation distance from the fitted distribution, defined by
\begin{equation}
    \Delta \equiv \frac{1}{2} \sum_n \left| \mathbf{P}_n^{\text{(1)}} - \mathbf{P}_n^{\text{(2)}} \right|,
\end{equation}
where $\mathbf{P}_n^{\text{(1)}}$ and $\mathbf{P}_n^{\text{(2)}}$ are two distributions. The factor of $1/2$ normalizes $\Delta$ to the range $0$ to $1$, where $0$ indicates a perfect match, and $1$ means the two distributions have no overlap. For the coherent pulses considered here, the two distributions are the reconstructed distribution and the fitted Poissonian.

In addition, for all the reconstructions performed in this work, we report a reconstructed value of the pulse-averaged second-order autocorrelation function $g^{(2)}(\tau = 0)$, estimated as~\cite{MigdallSPDs}
\begin{equation}
    g^{(2)}_{\text{recon}} = \frac{\braket{n(n-1)}}{\braket{n}^2},
    \label{eqn:g2recon}
\end{equation}
with the expectation values computed using the reconstructed distribution.

\begin{figure*}
\includegraphics[width=\textwidth]{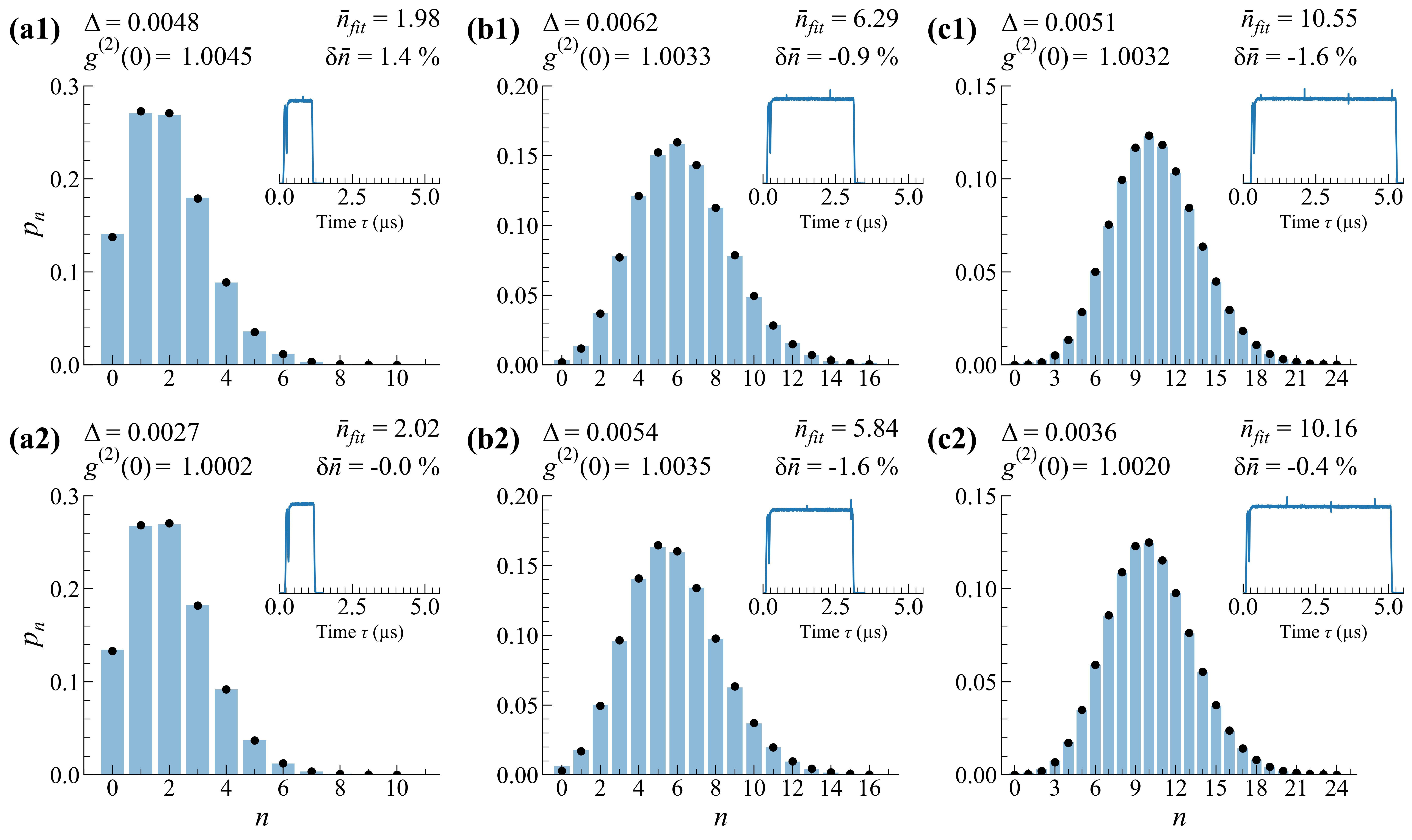}
\caption{\label{fig:SquarePulseRecons} \deleted{Results of the reconstruction}\added{Reconstructed distributions (blue bars) and fitted Poissonians (black dots)} for square coherent pulses for SPAD1 (top row) and SPAD2 (bottom row). A peak input photon rate of $\approx 2$ Mcounts/s is used for all pulses. For both SPADs and all average photon numbers, the distances $\Delta$ are less than $10^{-2}$ and all calibration differences $\delta\bar{n}$ are less than 2~\% in magnitude. The recovery time corrections are taken to second order for all results in this figure. The insets show the pulse as measured by the SPADs, with 1 ns bins, which is what we used in the reconstructions. The dip in power at about 100 ns after the start of the pulse is likely an artifact of the AOM drive turning on.}
\end{figure*}

\begin{figure}
\includegraphics[width=\textwidth]{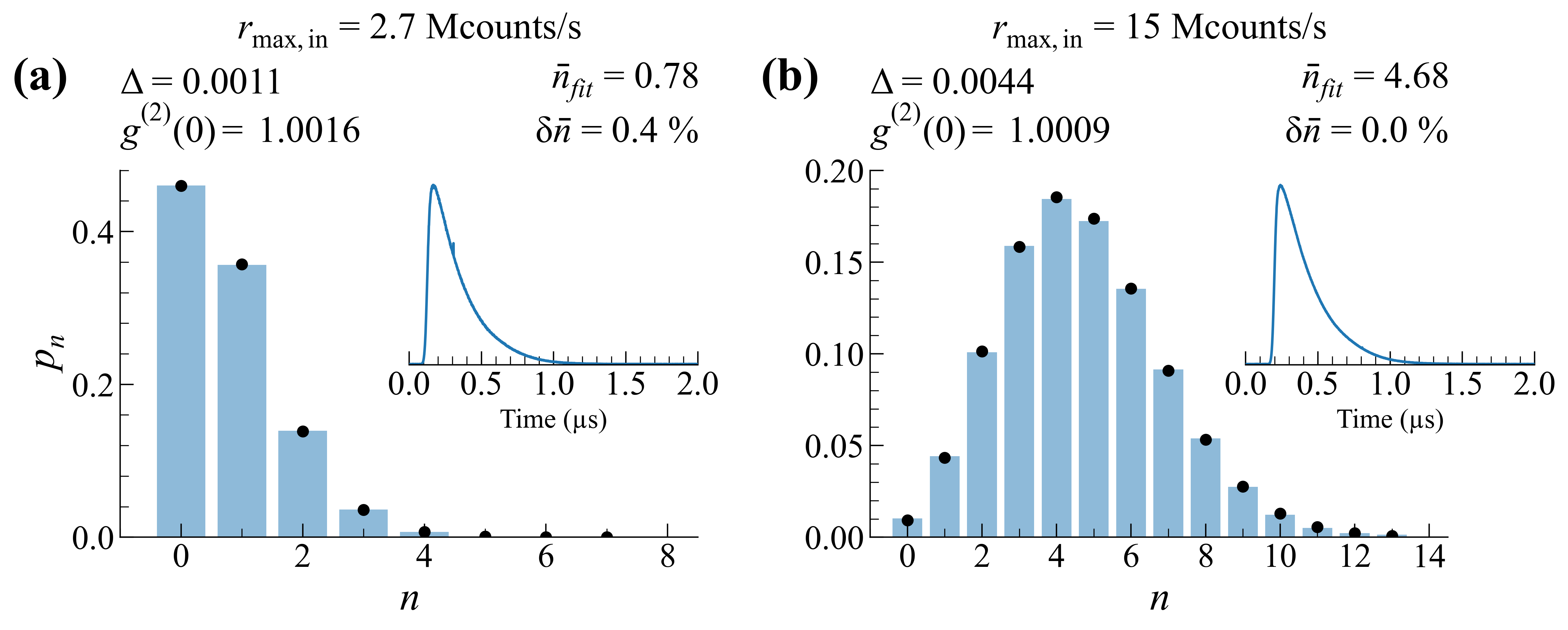}
\caption{\label{fig:ShapedPulseRecons} \deleted{Results of the reconstruction}\added{Reconstructed distributions (blue bars) and fitted Poissonians (black dots)} of shaped pulses with SPAD1 at two different input photon rates. The peak input rates $r_{\text{max,in}}$ are as indicated. The insets show the pulse as seen by the SPADs, with 1 ns bins. The recovery time corrections are taken to (a) third and (b) sixth order.}
\end{figure}

In Fig.~\ref{fig:SquarePulseRecons} we present the results of our reconstruction algorithm for square coherent pulses, all with roughly the same peak input count rate of about 2 Mcounts/s. Results are shown for both SPAD1 and SPAD2. We change the length of the pulse in order to vary the average photon number, and show the results for three average photon numbers. For each of these pulses, and all coherent pulses used here, we collected $3 \times 10^7$ cycles of data. The pulse shapes as measured by the SPADs are shown in the insets. We used 1 ns binning for the reconstruction process (both here and for all of the reconstructions done in this work) and for the insets of the figure. For all pulses measured here, we find $\Delta~<~10^{-2}$, and the calibration differences $\delta\bar{n}$ are all within 2~\% of the independently measured values\added{, showing the accuracy of the algorithm}.

\begin{figure*}
\includegraphics[width=\textwidth]{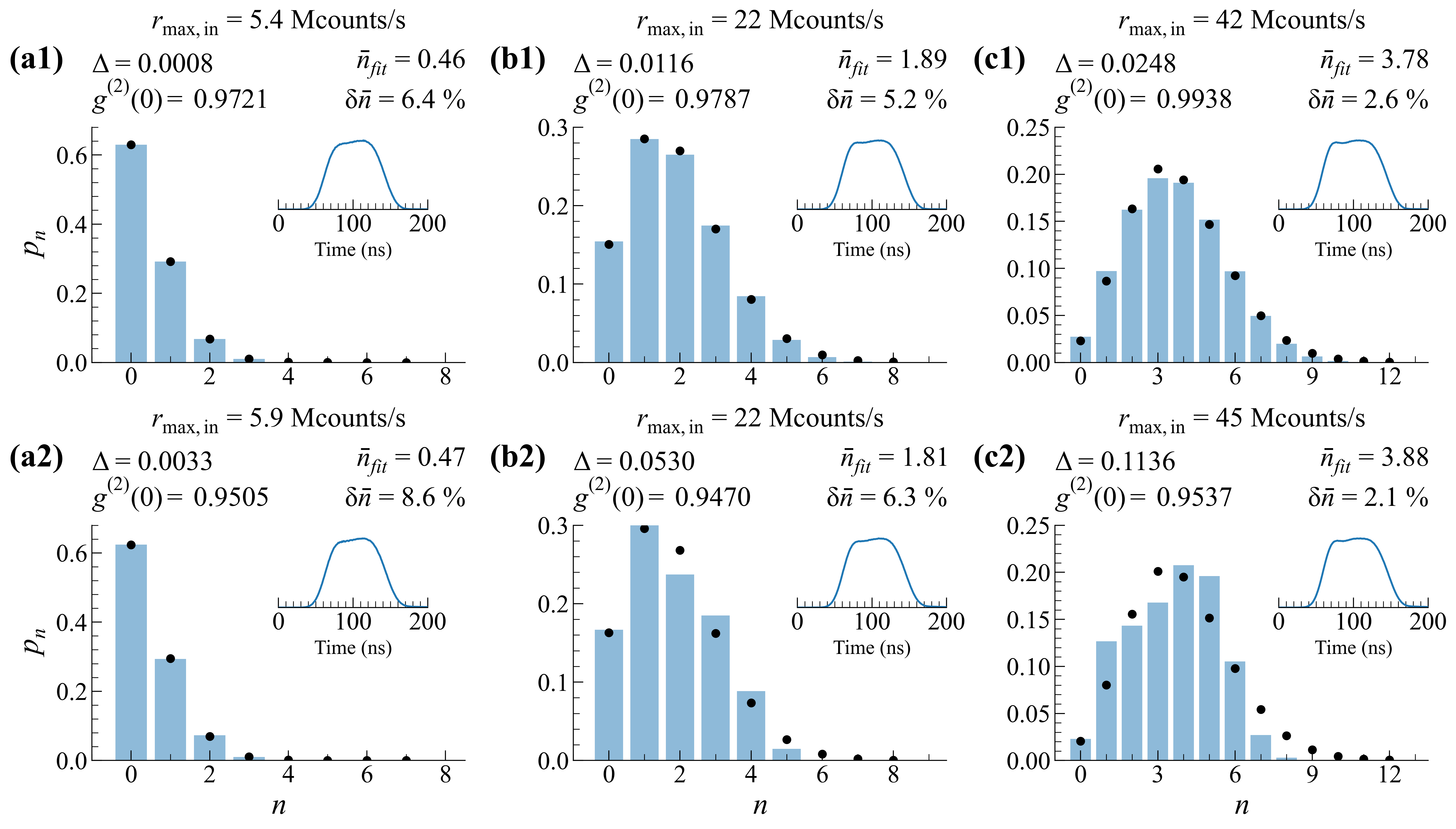}
\caption{\label{fig:ShortPulseRecons} \deleted{Results of the reconstruction}\added{Reconstructed distributions (blue bars) and fitted Poissonians (black dots)} for coherent pulses of FWHM $\approx 85$ ns, using SPAD1 (top) and SPAD2 (bottom). The insets show the pulse as seen by the SPADs, with 1 ns bins. The recovery time corrections are taken to (a) seventh and (b, c) eighth order. The peak input count rates $r_{\text{max,in}}$ are as indicated in the top center of each figure, corresponding to roughly (a) one photon per 7 recovery times, (b) one photon per 2 recovery times, and (c) one photon per recovery time. The distances $\Delta$ increase with input count rate. We believe this increase is due to interactions between twilight counts and afterpulsing, as discussed in the main text.}
\end{figure*}

In addition, in Fig.~\ref{fig:ShapedPulseRecons} we validate the time-dependent portions of the formalism by feeding non-square pulses to the SPADs. The pulses are shaped by using an arbitrary waveform generator to control the amplitude of the AOM drive. The shape was chosen to mimic single photons produced by cold-atom ensembles. To control the total energy delivered, we change the power through the first fiber in the setup using the polarizer. In Fig.~\ref{fig:ShapedPulseRecons} we show the reconstructions for $\bar{n} \approx 1, 5$ for SPAD1. Again we find $\Delta < 10^{-2}$, and calibration differences within 0.\deleted{2}5~\% in magnitude, validating the time-dependent formalism.

One of the contributions to the distances $\Delta$ of the reconstructions in Figs.~\ref{fig:SquarePulseRecons} and \ref{fig:ShapedPulseRecons} is drifts of up to a few percent in the \added{optical }power over the course of the hour-long measurement. The average of a drifting coherent state \deleted{does not produce a Poissonian distribution}\added{produces a distribution slightly wider than a Poissonian of the same average number}, contributing to the observed deviations\added{ and the slight broadening of the reconstructed distributions, as indicated by $g^{(2)}(0)$ values slightly greater than 1}. Based on Monte Carlo simulations, we believe these drifts contribute to \deleted{deviations}\added{total variation distances} at the level of $1$-$3 \times 10^{-3}$, \added{representing }a significant portion of the observed $\Delta$ values. 

Further, we test the limits of the algorithm at high input photon rates with $\approx85$ ns full-width half-max (FWHM) pulses. The results are shown in Fig.~\ref{fig:ShortPulseRecons} for both SPAD1 and SPAD2 for three input rates. The reconstruction algorithm still performs reasonably well\added{ for SPAD1 (top row of Fig.~\ref{fig:ShortPulseRecons}), whose afterpulsing probability is small}\deleted{, with $\Delta < 6 \times 10^{-2}$ for all the data in Fig.~\ref{fig:ShortPulseRecons}}. Note that the highest photon rate here, 4 photons in 85 ns, is greater than 1 input photon per recovery time---i.e. the detector is dead for most of the pulse. This makes the near-agreements\added{ of the SPAD1 data} with the expected distributions in Fig.~\ref{fig:ShortPulseRecons} particularly noteworthy. \added{For SPAD2 (bottom row of Fig.~\ref{fig:ShortPulseRecons}), which has a higher afterpulsing probability, the distributions are less faithful to the expected Poissonians. For both datasets, we observe increasing $\Delta$ between the reconstructed and expected distributions as the input count rate increases; we discuss possible reasons for this in the next section. In addition, we note the counterintuitive result that the largest values of $\delta\bar{n}$ in Fig.~\ref{fig:ShortPulseRecons} occur for the lowest count rates, where corrections for afterpulsing, background counts, and recovery time effects are small. We believe these large $\delta\bar{n}$ values signify an unidentified systematic calibration error of $\sim 5$-$10$~\% in $\bar{n}_{\text{exp}}$ which affects all short pulses for both SPADs. This positive systematic error, combined with a decrease of $\bar{n}_{\text{fit}}$ away from $\bar{n}_{\text{exp}}$ (leading to a negative contribution to $\delta\bar{n}$), results in the appearance of improving $\delta\bar{n}$ with increasing count rate.}

\deleted{Nevertheless, the reconstructed distributions increasingly deviate from the expected Poissonian distribution as the input count rate increases. We observe (Fig.~\ref{fig:ShortPulseRecons}) that the algorithm performs better for SPAD1 than SPAD2. Table~\ref{tab:DetValues} shows that the only significant difference between the detectors is that SPAD1 afterpulses much less than SPAD2. Given this, our hypothesis for the limit reached in the reconstructions of Fig.~\ref{fig:ShortPulseRecons} is the following: when a click is followed immediately by another click, the second click can be either an afterpulse or a twilight count. In principle, both of these can occur for a given click, but only one click will register. Thus, although twilight counts and afterpulsing have independent origins, the detector does not treat them independently, as it cannot register simultaneous twilight and afterpulsing events. However, our model assumes independence of these effects. This logic implies that the reconstruction algorithm overcompensates, and this effect increases with input count rate. Indeed, as expected, we observe $\bar{n}_{\text{fit}}$ increasing faster than $\bar{n}_{\text{exp}}$ as the count rate increases. The overall magnitude of this effect is limited by the afterpulsing probability, which is smaller for SPAD1 than for SPAD2. This effect is the primary limitation of the reconstruction algorithm in the high-count-rate limit.}

\begin{figure*}
\centering
\includegraphics[width=0.95\textwidth]{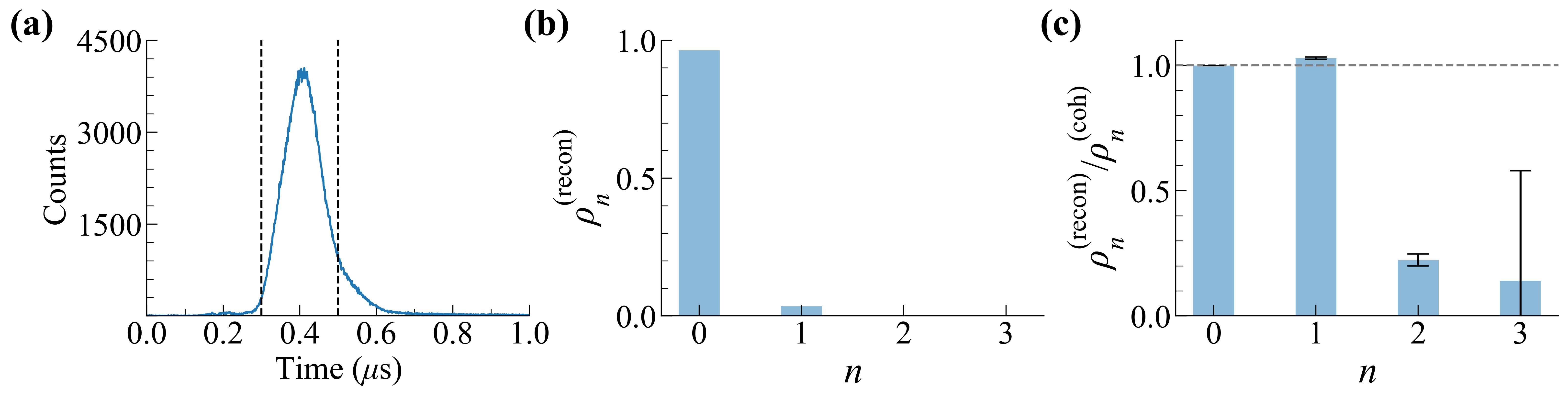}
\caption{\label{fig:SinglePhotonResults} Reconstruction results for anti-bunched light using SPAD2. (a) The light pulse as seen by the SPAD, with 1 ns bins. The vertical black dashed lines indicate the 200-ns window used for reconstruction. (b) The reconstructed distribution $\rho_n^{\text{(recon)}}$. The recovery time corrections are taken to second order. (c) The ratio of the reconstructed state to a coherent state of the same average photon number, $\bar{n} = 0.036$, whose number distribution we denote $\rho_n^{\text{(coh)}}$. The $n=2$ and $n=3$ components of the reconstructed distribution are much smaller than those expected of a coherent state (gray dashed line), indicating anti-bunching. The one-standard-deviation uncertainties shown are computed through Monte Carlo sampling, as detailed in Supplemental 1.}
\end{figure*}

Finally, we investigated our ability to reconstruct anti-bunched light. Since we do not know the incident photon number distribution \emph{a priori}, the metrics used above to evaluate our reconstruction of coherent states do not apply. As an alternative, we measure the \added{pulse-averaged }value of $g^{(2)}(\tau = 0)$ using a two-detector Hanbury Brown-Twiss (HBT) setup~\cite{MigdallSPDs, RbRySinglePhotons}. We compare the independently measured $g^{(2)}_{\text{HBT}}$ to the value estimated from the reconstructed distribution using Eqn.~\ref{eqn:g2recon}.

To produce anti-bunched light, we used a Rydberg atomic ensemble described previously~\cite{RbRySinglePhotons}, with size $\approx 20~\mu\text{m} \times 8~\mu\text{m} \times 8~\mu\text{m}$ and Rydberg state $n=112$, and we performed about $2 \times 10^7$ data collection cycles. 
\deleted{The results of this single-photon reconstruction are shown in Fig.~\ref{fig:SinglePhotonResults}.} The light we produce is measured with the HBT setup to have a raw $g^{(2)}_{\text{HBT, raw}}(0) = 0.25(1)$ and $g^{(2)}_{\text{HBT}}(0) = 0.21(1)$ after background subtraction~\cite{RbRySinglePhotons}. \added{The light pulse shape is shown in Fig.~\ref{fig:SinglePhotonResults}(a).}

\added{The reconstructed distribution is shown in Fig.~\ref{fig:SinglePhotonResults}(b). To perform a reconstruction whose value of the correlation function $g^{(2)}_{\text{recon}}$ can be compared to that measured by the HBT setup, we reduce the reconstruction window to the one shown in vertical dashed lines in Fig.~\ref{fig:SinglePhotonResults}(a), which contains most of the photon pulse but avoids the tail. This tail is primarily coherent leakage into the SPADs, which is removed by the background subtraction in the calculation of $g^{(2)}_{\text{HBT}}$ but not by the background correction of our reconstruction algorithm. The resulting reconstructed distribution is compared to a coherent state of the same average photon number in Fig.~\ref{fig:SinglePhotonResults}(c). In the latter, the $n=2$ and $n=3$ components are well below what is expected of a coherent state, indicating the anti-bunched nature of the light.} The reconstructed distribution is calculated to have $g^{(2)}_{\text{recon}}(0) = 0.2\deleted{5(3)}\added{2(2)}$\deleted{.}\added{, which is in excellent agreement with the independently measured value of 0.21(1). Thus we demonstrate our algorithm's ability to accurately reconstruct anti-bunched light.} \deleted{Our methodology for reaching this number involved two important modifications from the coherent state reconstruction procedure. First, as shown in Fig.~\ref{fig:SinglePhotonResults}(a), we reduced the window over which we reconstruct to contain most of the pulse but to avoid the long tail. The long tail is primarily coherent leakage into the SPADs, which is removed by the background subtraction of the $g^{(2)}_{\text{HBT}}$ computation, but not by the background correction of our reconstruction algorithm. Second, we set the entropy regularization $\alpha$ to zero. This is consistent with Ref. [40], which found that lower entropy regularization improves the reconstruction of non-classical light; the regularization acts to smooth the number distribution in a way suitable for coherent and thermal state reconstruction, but not for Fock states, which are not smooth.}

\deleted{We note that both our work and the work of Ref. [40] suggest that the choice of $\alpha$ is based only on whether the light is anti-bunched. If it is known prior to reconstruction that the light to be measured is anti-bunched, we recommend setting $\alpha = 0$. If it is not known in advance whether the light is anti-bunched, we recommend performing a reconstruction with the value we used for coherent states, $\alpha = 5 \times 10^{-3}$, and if the resulting reconstructed $g^{(2)}$ is less than unity, redoing the reconstruction with $\alpha = 0$. }

\deleted{The resulting reconstructed distribution is shown in Fig.~\ref{fig:SinglePhotonResults}(b), and it is compared to a coherent state of the same average photon number in Fig.~\ref{fig:SinglePhotonResults}(c). In the latter, the $n=2$ and $n=3$ components are below what is expected of a coherent state, indicating the anti-bunched nature of the light. The final value of $g^{(2)}_{\text{recon}}(0) = 0.2\deleted{5(3)}\added{2(2)}$ is in reasonable agreement with the independently measured value of 0.21(1). We believe some of the discrepancy may be due to the nature of the recovery effects corrections, a subject of further investigation.}

\section{Limits of the Algorithm}

\added{While the results presented here indicate a successful reconstruction process, the algorithm involves a number of assumptions and limitations. We describe the most important of these here, with a detailed discussion in Section A.V. of Supplement 1.}

\added{First, our algorithm is effective when both the light pulse and its correlations have timescales greater than a few detector recovery times. We have demonstrated the algorithm's effectiveness for coherent pulses of lengths of at least a few detector recovery times, and for a pulse of non-classical light with pulse-averaged $g^{(2)}(0) < 1$. If the pulse or its correlations have timescales less than several detector recovery times, we expect the algorithm to fail. A clear example is that of broadband thermal light, whose bunching timescale is typically much less than 1~ns; in this case, photons ``bunched'' with previous photons will not register additional clicks, as the detector will be dead, and this will prevent us from correctly reconstructing the light statistics. Some light sources produce bunching with longer timescales~\cite{LongBunching}, which we believe will be able to be reconstructed by our algorithm.}

\added{Second, our model of the detector weakens at high count rates. We believe the most important example of this breakdown, at least for our datasets, is that the model of independence between afterpulsing and twilight counts loses validity at high count rates. Although these two effects have independent origins, they  ``interact'' by occurring on a single detector. This means that, if the detector is dead, afterpulses from previous clicks may be blocked; it also means that, in the few nanoseconds immediately after the end of a recovery time, the detector can only register one click even though both an afterpulse and a twilight count are theoretically possible. Both of these effects lead to an overcompensation of afterpulsing by our model, which should lead both to a decreasing $\delta\bar{n}$ with increasing count rate and a narrower reconstructed distribution. Both of these effects are observed in Fig.~\ref{fig:ShortPulseRecons}. This and other effects, such as non-Markovianity in afterpulsing, are discussed in detail in subsection A.V. of Supplement 1. Thus, our algorithm, while still working well even approaching one photon per dead time, performs best for count rates that are small compared to the inverse of the detector dead time.}

\section{Conclusion}

In conclusion, we have presented a method for performing number-state \deleted{tomography}\added{reconstruction} with a single SPAD. The method works when the pulses of interest span at least a few multiples of the detector dead time, a regime that fits many of today's quantum optics experiments. Further, the algorithm performs reasonably well even with input rates up to about one photon per recovery time. \deleted{We believe a primary limitation of our algorithm is its assumption of the independence of afterpulsing and recovery time effects}\added{At these very high count rates, we have explored some of the ways in which the algorithm loses accuracy}, which may be remedied in future work. This method could be extended to two or more detectors, yielding more information than standard multi-detector techniques and allowing reconstruction over a broader array of input light sources, such as sources with correlations shorter than the detector dead time.

\begin{backmatter}
\bmsection{Funding}
Content in the funding section will be generated entirely from details submitted to Prism. We acknowledge the support of the Maryland-ARL Quantum Partnership (MAQP). This material is also based upon work supported by the National Science Foundation Graduate Research Fellowship Program under Grant No. DGE 1840340.

\bmsection{Acknowledgments}
The authors thank Alessandro Restelli for helpful discussions. \added{We are also very grateful for this work's peer reviewers, who contributed substantial improvements and additional rigor to the work. }Any opinions, findings, and conclusions or recommendations expressed in this material are those of the author(s) and do not necessarily reflect the views of the National Science Foundation.

\bmsection{Disclosures}
The authors declare no conflicts of interest.

\bmsection{Data availability}
Data underlying the results presented in this paper are not publicly available at this time but may be obtained from the authors upon reasonable request.

\bmsection{Supplemental document}
See Supplemental 1 for supporting content.

\end{backmatter}

\raggedbottom

\bibliography{refs}

\begin{thebibliography}{10}
\bibitem{NISTDisc}
{Disclaimer: Certain commercial equipment, instruments, or materials are
  identified in this paper to foster understanding. Such identification does
  not imply recommendation or endorsement by the National Institute of
  Standards and Technology, nor does it imply that the materials or equipment
  identified are necessarily the best available for the purpose.}
\end{thebibliography}

\end{document}


\maketitle

\section{Construction of the Detector Matrix}

As discussed in the main text, a pulse of light with photon-number distribution $\mathbf{P}$ under repeated measurements produces a click number distribution $\mathbf{C}$. All of the effects of the detector are contained in a matrix relating the two:
\begin{equation}
    \mathbf{C} = \mathbb{D}\mathbf{P}.
\end{equation}
In turn, in our model $\mathbb{D}$ is composed of four matrices:
\begin{equation}
\mathbb{D} = \mathbb{A}(p_a) \mathbb{R}[\gamma(t), D(\tau)] \mathbb{B}(p_B) \mathbb{L}(\eta_0),
\end{equation}
where each of these matrices encodes the effects of a particular SPAD imperfection. In this section we detail the methods we use to construct these matrices. We have made the code that implements everything in this section open-access, as well as provided some sample data and example reconstruction implementations, available in Code 1 \cite{Git}.

\subsection{Detection Efficiency}

The simplest, and typically largest, correction to be applied to the click distribution to reach the incident photon number distribution is the non-unit detector efficiency (DE). If the detector is fully armed when a photon arrives, then the probability of generating a click is a Bernoulli process; thus if all of the photons of a pulse arrive when the detector is fully armed, the probability that a given number of clicks is produced is
\begin{equation}
P(\text{$m$ clicks given $n$ photons}) = {n \choose m} \eta_0^m (1-\eta_0)^{n-m},
\end{equation}
where $\eta_0$ is the DE when the detector is fully armed. Note that the binomial coefficient is zero when $m > n$. From the above equation, we define the matrix $\mathbb{L}(\eta_0)$ through its elements $\mathbb{L}_{mn}(\eta_0)$, which are the probabilities that, of $n$ incident photons, $m$ produce clicks and $n-m$ are lost. Note that $\mathbb{L}(\eta_0)$ is upper triangular.

\subsection{Background Counts}

We model background counts, including dark counts, as a Poisson process \cite{DC1} with a total rate $r_b$, which we take to be constant over the data collection time scale. Thus if a data collection window has duration $T$, then the mean number of background events is $p_b = r_bT$. The probability of $k$ background events is then
\begin{equation}
P(\text{$k$ background counts}) = \frac{p_b^k e^{-p_b}}{k!}.
\end{equation}
We use this to construct the background count matrix $\mathbb{B}(p_b)$, whose elements $\mathbb{B}_{mn}$ are the probabilities that there will be $m-n$ background counts in the data collection window, so that $n$ photon-produced clicks yield $m$ total (photon-produced and background) clicks. Since background counts add to the total counts, this matrix is lower triangular. Note also that since background counts add clicks, the possible total number of clicks is unbounded. Therefore truncating the basis produces probabilities whose sum differs from one. \deleted{In practice, we set the diagonals of the matrix manually to meet the condition that each column should sum to unity.}\added{We manually set $\mathbb{B}_{n_{\text{max}}n}$, the last row of the matrix, so that each column sums to unity.}

\subsection{\label{sec:RT_app}Recovery Time Effects}

The construction of the recovery time effects matrix $\mathbb{R}$ requires detailed consideration, because recovery time effects are correlated with previous clicks, and this memory has to be taken into account.

\begin{figure}
\centering
\includegraphics[width=0.7\textwidth]{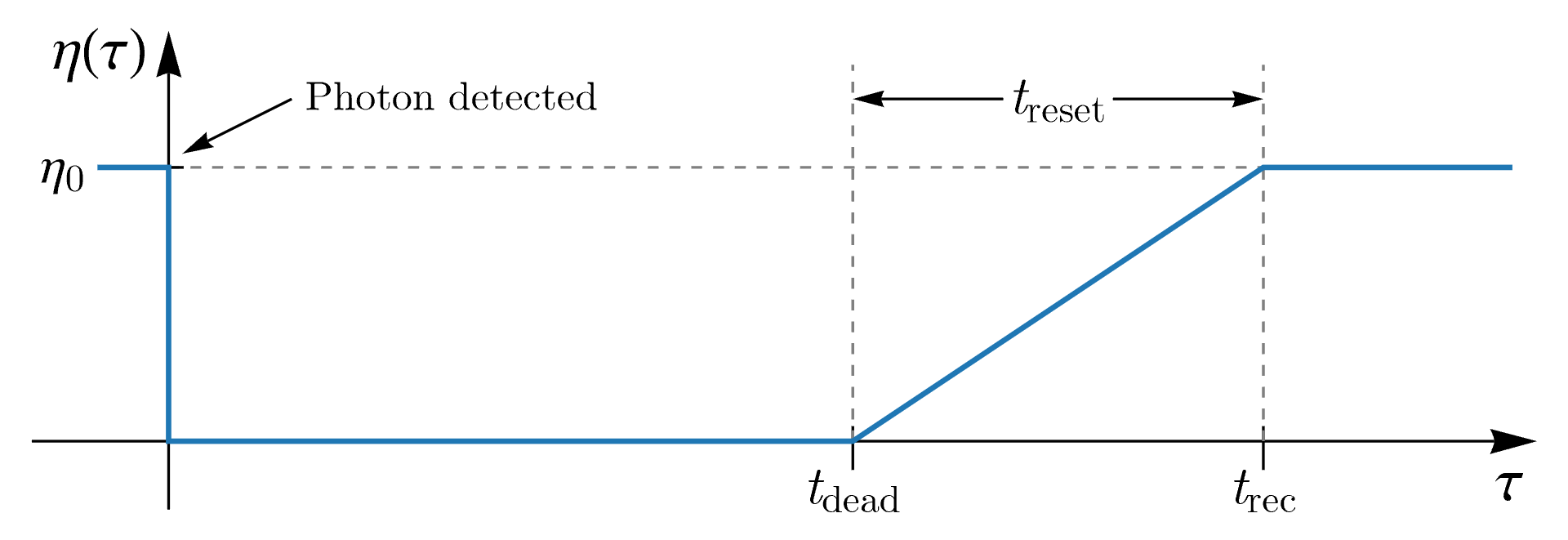}
\caption{\label{fig:Dtau} Illustration of our model of SPAD recovery effects \cite{MigdallSPDs}. The function $\eta(\tau)$ is the efficiency of the detector immediately following a photon detection, with $\eta_0$ being the fully armed efficiency. After a photon arrives and causes an avalanche ($\tau=0$), the detector becomes dead for a time $t_{\text{dead}}$ set by the quenching circuit. The circuit then resets the bias over a time $t_{\text{reset}}$ and the detector starts to be able to sense photons. The total time to return to full bias is the recovery time $t_{\text{rec}} = t_{\text{dead}} + t_{\text{reset}}$. We define $D(\tau) = 1-\eta(\tau)/\eta_0$ as the loss profile.}
\end{figure}

When a fully armed detector clicks, the system efficiency changes due to the action of the quenching circuit. We model this behavior with an efficiency function $\eta(\tau)$ shown in Fig.~\ref{fig:Dtau}. (Although it is possible to measure $\eta(\tau)$ \cite{TwilightDelays}, we simply assume the model here.) For this piece of the formalism, we want to correct for recovery time effects only, so we use a scaled version of $\eta(\tau)$, namely $D(\tau) = 1-\eta(\tau)/\eta_0$, a function that is zero if the detector loses no clicks to recovery time effects and 1 if it has no ability to click due to recovery time effects. Using this function treats all losses as caused by the dead time.

Additionally, let $\gamma(t)$ be the input photon profile. We measure \deleted{this}\added{$\gamma$} by histogramming the arrival times of the clicks in only the runs where exactly one click occurred\deleted{; this profile is immune to time-dependent afterpulsing and recovery time effects.}\added{, and normalizing the histogram. As discussed below, this is an approximation and suffers from distortions due to recovery time and afterpulsing effects, but these distortions are generally small, and often a separate measurement can be made to reduce their effects.} \deleted{(We will normalize out the scalings due to a constant detector efficiency and a constant afterpulsing probability.)}This measured profile also includes background counts, but this is \deleted{what we want}\added{desirable}, since the combination of background and photon counts is what we model to be affected by the dead time.

With this paradigm, imagine an experimental run in which two photons are incident on the detector. According to the above discussion, we assume the first photon generates a click. Then there are three possibilities for the second photon, enumerated in Table~\ref{tab:RTSymbols}:
\begin{enumerate}
   \item It arrives during the recovery time triggered by the previous click, and gets lost (i.e. does not produce a click);
   \item It arrives during the recovery time of the previous click, and produces a click, which is delayed until the end of the recovery time (note that these are twilight counts \cite{TwilightDelays});
   \item It arrives after the recovery time of the first photon, and therefore produces a click, since the detector has recovered and is fully armed.
\end{enumerate}
The probabilities of each of these events can be calculated using $\gamma(t)$ and $D(\tau)$ as:
\begingroup
\allowdisplaybreaks
\begin{align}
    p_1 &= \frac{1}{N_2}\int_0^T \gamma(t_1) \int_{t_{1}}^{t_1+t_{\text{rec}}} \gamma(t_2) D(t_2-t_1) \, dt_2 \, dt_1, \\
    p_2 &= \frac{1}{N_2}\int_0^T \gamma(t_1) \int_{t_{1}}^{t_1+t_{\text{rec}}} \gamma(t_2) [1-D(t_2-t_1)] \, dt_2 \, dt_1, \\
    p_3 &= \frac{1}{N_2}\int_0^T \gamma(t_1) \int_{t_1+t_{\text{rec}}}^T \gamma(t_2) \, dt_2 \, dt_1.
\end{align}%
\endgroup
Here $T$ is the data collection time duration and $N$ is a normalization factor,

\begin{equation}
    N_2 = \int_0^T \gamma(t_1) \int_{t_1}^T \gamma(t_2) \,dt_2 \, dt_1.
\end{equation}
The second integral in each equation is a factor that determines the probability of the relevant event \emph{given that} a photon arrived at time $t_1$, and $\gamma(t_1)$ is the probability that the first photon arrives at time $t_1$. We then integrate over the arrival time of the first click. Note that $\gamma(t)$ does not have to be a true probability density in the sense of integrating to unity; the normalization factor $N_2$ handles this. By inspection, $p_1+p_2+p_3 = N_2/N_2 = 1$. Note that, when $\gamma(t)$ does integrate to 1 over the window from $t=0$ to $t=T$, we get $N_2 = 1/2$, consistent with an identical particles formulation of the problem.

For higher numbers of incident photons, the possible combinations of the three types of events grows quickly, and at high count rates all of these possibilities have to be considered. Thus, to treat large Fock bases, we need an algorithm that determines all of the possibilities and the necessary integrals. In the following, we detail (a) an encoding method for photon events, (b) an algorithm for generating all possible events, (c) an algorithm for constructing the needed integral for each event, and finally (d) the method for including the result of the integral in the recovery effects matrix $\mathbb{R}$.

\begin{table*}
\caption{\label{tab:RTSymbols}%
Summary of photon events for calculating recovery effects. The detector may be fully armed or recovering, and if the latter the photon may or may not produce a click, resulting in three different event types for every photon that arrives at the detector. This table also gives the symbol for encoding each possible event and a justification for the symbol. Note that in the accompanying Code 1 \cite{Git}, these symbols are replaced with (from top to bottom) 1, 2, and 3.}
\begin{tabular}{p{0.40\textwidth} p{0.07\textwidth} p{0.45\textwidth}}
\hline\hline
Description of event & \Centering{Symbol} & Symbol explanation \\
\hline
\RaggedRight{Photon arrives during the recovery time triggered by the previous click, and gets lost (i.e. does not produce a click)} & \Centering{$\circ$} & \RaggedRight{A circle is used for photons arriving while the detector is recovering; the hollow circle indicates the loss of the photon.} \\[0.75em]
\RaggedRight{Photon arrives during the recovery time triggered by the previous click, and itself produces a click, which is delayed until the end of the recovery time (i.e. a twilight count)} & \Centering{$\bullet$} & \RaggedRight{A circle is used for photons arriving while the detector is recovering; the filled circle indicates the photon produces a click.} \\[1.75em]
\RaggedRight{Photon arrives when the detector is fully armed (i.e. after the recovery times of all previous clicks), and therefore produces a click} & \Centering{$\bigstar$} & \RaggedRight{The star indicates this photon is different from other events in that it arrives when the detector is fully armed.} \\
\hline\hline
\end{tabular}
\end{table*}

\paragraph{Encoding.} We begin by coding all possible combinations of the three types of events. Every experimental run contains a set of photons, each of which arrived in one of the three conditions described in Table~\ref{tab:RTSymbols}. Thus we can describe the photons of any run by a string of symbols encoding the arrival conditions of each photon, in order of arrival. We use the symbols $\circ$, $\bullet$, and $\bigstar$, as mapped and justified in Table~\ref{tab:RTSymbols}.

In writing down all possible events, we assume the first photon of a run arrives when the detector is in normal operating mode, i.e. the first photon is always a $\bigstar$. This is a good assumption when there is no signal prior to the start of the detection window.

For example, the string $\bigstar\bullet$ describes a run in which the first photon arrives when the detector is fully armed and produces a click, and then a second photon arrives within the first click's recovery time and produces a delayed twilight count. And $\bigstar\circ\bullet$ is an experimental run in which the first photon produces a click because it arrives when the detector is fully armed, and then the second photon arrives during that first click's recovery time and is lost, and then a third photon arrives, also during the first click's recovery time, and produces a delayed twilight count.

There is one more factor to determining possible events, which we can see by considering the string $\bigstar\bullet\bullet$. When a $\bullet$ appears in the string, a photon arrived during the recovery time of the previous click, and generated a new, delayed click. But this formulation means the string $\bigstar\bullet\bullet$ has an ambiguity of interpretation: either (a) both of the $\bullet$ photons arrived during the recovery time of the same click (the first of the cycle), or (b) the second photon arrived during the recovery time of the first click, and produced a new click, and then the third photon arrived during the recovery time of that second click. In the first case, note that the three photons produce only two clicks total, but in the second case all three photons produce a click. There are many strings with similar ambiguities.

To disambiguate these possibilities, we add square brackets $\left[ ~ \right]$ to our strings. These brackets indicate photons that come in the same recovery periods, and we refer to the photons grouped by these brackets simply as ``groups.'' For example, scenario (a) in the previous paragraph will be written $\left[\bigstar\bullet\bullet\right]$ (one group), while (b) will be written $\left[ \bigstar\bullet \right] \left[\bullet\right]$ (two groups). We  use the term ``event'' to refer to a specific, unambiguous set of photon events, and the term ``event string'' to refer to the encoding of the event, while we use the term ``string'' to refer simply to a possible combination of $\circ$'s, $\bullet$'s, and $\bigstar$'s with no grouping. For more than two incident photons, typically one string (like $\bigstar\bullet\bullet$) corresponds to more than one event, like the two written above.

Note that $\bigstar$ photons require no disambiguation, so long as we enforce the definition: a $\bigstar$ happens only after all previous recovery times have finished. This means $\bigstar$ photons are always at the beginning of their group (though not all groups begin with a $\bigstar$). For instance possible events include $\left[\bigstar\bullet \right] \left[ \bigstar \right]$ and $\left[\bigstar\bullet \right] \left[ \bigstar \circ \right]$. In addition, $\circ$ photons require no disambiguation, as they produce no click, and therefore do not affect the detector dynamics.

\begin{figure}
\centering
\includegraphics[width=0.7\textwidth]{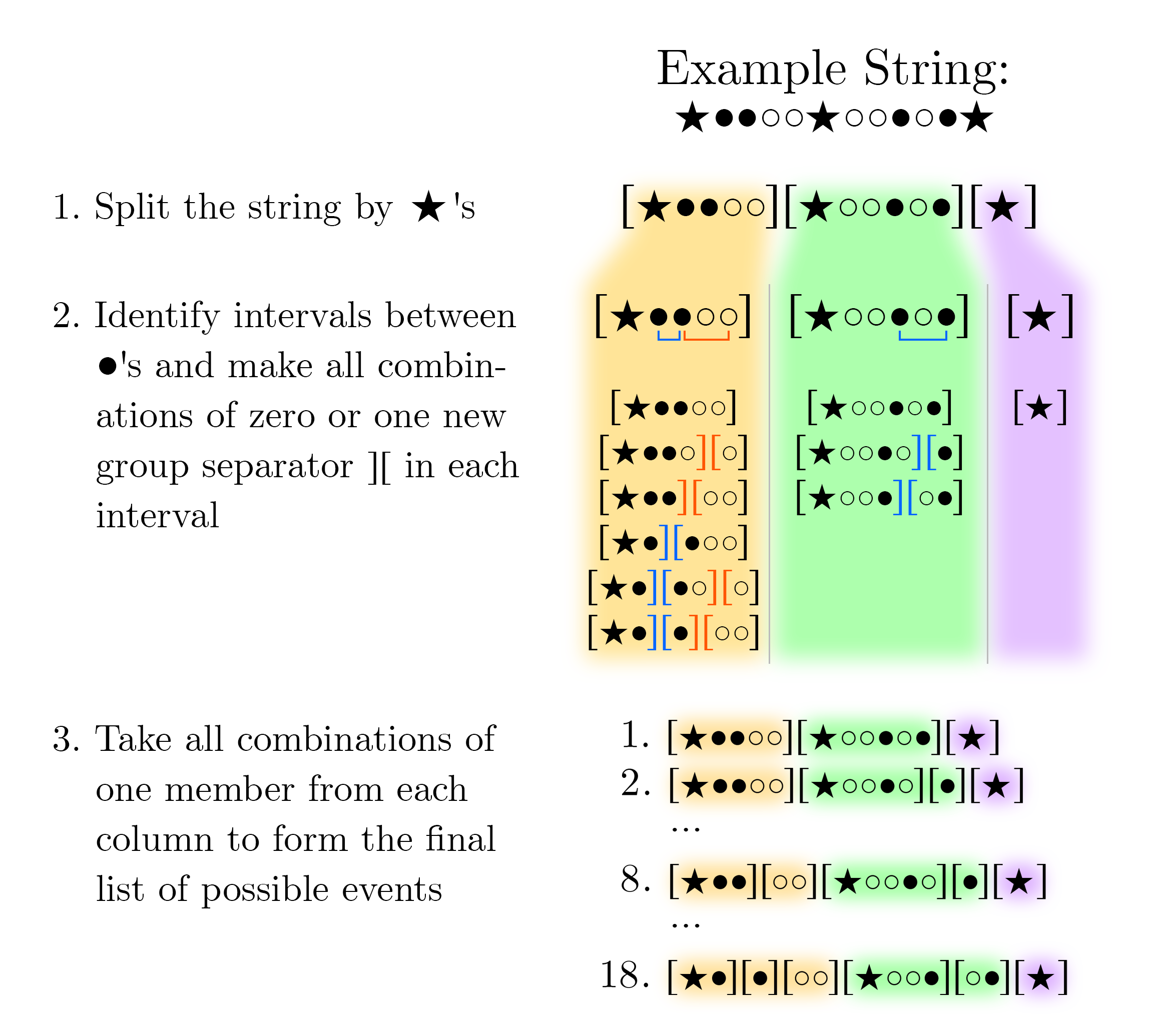}
\caption{\label{fig:RT_Algorithm_B} An illustration of the algorithm for generating all possible events corresponding to a given string, using an example string to show the process.}
\end{figure}

\paragraph{Generating Possible Events.} Using this encoding, we can discuss how to generate possible events. A first step is to form all possible combinations of up to $n_{\text{max}}$-symbol strings of $\bigstar$, $\bullet$, and $\circ$ which begin with a $\bigstar$. (Here $n_{\text{max}}$ is the largest $n$ considered in the truncated basis.) For each of these strings, we generate the set $S$ of possible events. The general algorithm for doing this is illustrated for an example string in Fig.~\ref{fig:RT_Algorithm_B}. Starting with a given string and the empty set $S$, there are three basic steps:
\begin{enumerate}
    \item Split the string by $\bigstar$'s, so that each substring begins with exactly one $\bigstar$. This splitting corresponds to the definition: when a $\bigstar$ occurs, it means all previous recovery times have finished.
    \item Account for the ambiguities caused by $\bullet$'s: for each substring, find $\bullet$'s. Generate new events using the rule that, between every $\bullet$ in the substring and the next $\bullet$ in that substring (or the end of the substring), there may be no additional splittings, or there may be exactly one additional split $][$ between any two photons. This generates a set of possible events $S_i$ for the photons of each group from step 1, where $i$ indexes the substrings from Step 1.
    \item The set $S$ of possible events for the entire string is generated by forming all possible combinations of exactly one member of each of the sets $S_i$ and concatenating the choices, in order.
\end{enumerate}
Performing this algorithm on all combinations of $\circ$'s, $\bullet$'s, and $\bigstar$'s up to length $n$ results in a set of all possible events of up to $n$ photons.

\paragraph{Constructing the Integral for an Event.} Each of the events from the algorithm above is a possible experimental realization with an occurrence probability, which we compute by performing integrals over $\gamma(t)$.

As each photon arrives at a time $t_i$, and these times are integrated over, $t_i$ become integration variables. Since each photon must come after the previous photon, the integrals are nested such that an inner integral in general depends on the previous $t_i$. There are three things to be determined for each of the integrals: the integral bounds, the integrand, and, if the integrand includes a $D$ function, the argument of that $D$ function.

For the integrand, the rules are: for a $\bigstar$, integrate over $\gamma$ only; for a $\circ$, integrate over $\gamma D$; and for a $\bullet$, integrate over $\gamma (1-D)$. The argument of $\gamma$ is always the arrival time of the photon (the integration variable). The argument of $D$ will be the integration variable minus some ``reference'' time, which is the time of the previous click. Algorithmically we can determine the reference time as follows:
\begin{itemize}
    \item If the photon is preceded by another $\circ$ or $\bullet$ in the same group, use the same reference as the $D$ function in the previous integral, because this photon arrived during the recovery time of the same click as the last photon. 
    \item If the photon is at the beginning of a group or is preceded by a $\bigstar$, the reference should be the start of the integral, because this photon is the first one to arrive in this recovery time.
\end{itemize}

The integral bounds are set as follows. For the start of the integral,
\begin{enumerate}
    \item The very first photon's integral starts at $t=0$.
    \item If the photon is a $\circ$ or $\bullet$ that is not at the beginning of its group, use the previous integral's integration variable as the start time.
    \item If the photon is a $\circ$ or $\bullet$ at the beginning of its group, use the time of the previous $\bigstar$ plus $nt_{\text{rec}}$, where $n$ is the number of group separators $][$ between the last $\bigstar$ and the current photon.
    \item If the photon is a $\bigstar$, look to the most recent photon that is not a $\circ$. If that photon is a $\bigstar$, start the integral at the time of that $\bigstar$ plus $t_{\text{rec}}$. If that photon is a $\bullet$, start the integral at the time of the time reference of the $D$ function of the $\bullet$'s integral plus $2t_{\text{rec}}$, leaving an extra recovery time for the click produced by the $\bullet$.
\end{enumerate}
Finally, for the end of the integral,
\begin{enumerate}
    \item If the photon is a $\circ$ or $\bullet$ that is not at the beginning of its group but is not immediately preceded by a $\bigstar$, use the previous integral's end time.
    \item If the photon is a $\circ$ or $\bullet$ that is at the start of its group, or is not at the start of its group but is immediately preceded by a $\bigstar$, end the integral $t_{\text{rec}}$ after the integral's start time.
    \item If the photon is a $\bigstar$, end the integral at $T$.
\end{enumerate}

We note here that these integral bounds imply a simplifying assumption, namely that all twilight counts come at exactly the end of the recovery time of the previous click, with no spread in time. Although it is possible to measure the spread of delays of twilight counts \cite{TwilightDelays} and account for it, we do not do so here.

After the integral is constructed according to the above rules, it needs to be divided by the normalization factor
\begin{equation}
    N_{m} = \int_0^T \gamma(t_1) \int_{t_1}^T \gamma(t_2) \cdots \int_{t_{m-1}}^T \gamma(t_m) \,dt_m \cdots \, dt_1.
\end{equation}
where $m$ is the number of photons in the event.

\noindent \begin{minipage}{\textwidth}
We give a few examples of higher-order integrals, which together demonstrate all of the rules given in this section:
\begin{subequations}

\begin{eqnarray}
    \left[ \bigstar \bullet \circ \right] &:& p = \frac{1}{N_3}\int_0^T \gamma(t_1) \int_{t_1}^{t_1+t_{\text{rec}}} \gamma(t_2)[1-D(t_2-t_1)] \int_{t_2}^{t_1+t_{\text{rec}}} \gamma(t_3) D(t_3-t_1) \,dt_3\,dt_2\,dt_1 \nonumber \\
    \left[ \bigstar \bullet \right] \left[ \circ \right] &:& p =  \frac{1}{N_3}\int_0^T \gamma(t_1) \int_{t_1}^{t_1+t_{\text{rec}}} \gamma(t_2)[1-D(t_2-t_1)] \int_{t_1+t_{\text{rec}}}^{t_1+2t_{\text{rec}}} \gamma(t_3) D(t_3-(t_1+t_{\text{rec}})) \,dt_3\,dt_2\,dt_1 \nonumber \\
    \left[ \bigstar \circ \right] \left[ \bigstar \right] &:& p = \frac{1}{N_3}\int_0^T \gamma(t_1) \int_{t_1}^{t_1+t_{\text{rec}}} \gamma(t_2)D(t_2-t_1) \int_{t_1+t_{\text{rec}}}^{T} \gamma(t_3) \,dt_3\,dt_2\,dt_1 \nonumber \\
    \left[ \bigstar \bullet \right] \left[ \bigstar \circ \right] &:& p = \frac{1}{N_4}\int_0^T \gamma(t_1) \int_{t_1}^{t_1+t_{\text{rec}}} \gamma(t_2)[1-D(t_2-t_1)] \int_{t_1+2t_{\text{rec}}}^{T} \gamma(t_3) \int_{t_3}^{t_3+t_{\text{rec}}} \gamma(t_4) D(t_4-t_3) \,dt_4\,dt_3\,dt_2\,dt_1 \nonumber
\end{eqnarray}

\end{subequations}
\end{minipage}
\vspace{0.5em}

\paragraph{Contributing to the Matrix.} Finally, the integral for each event is interpreted as a probability that a given number of photons results in a particular number of clicks. Therefore each integral is a contribution to some element of the matrix $\mathbb{R}$. For each event, the number of photons, and therefore the column of the element to add the integral result to, is simply the total number of $\circ$'s, $\bullet$'s, and $\bigstar$'s in the event's string. The number of clicks, which determines the row of the matrix element to add the integral to, is the number of groups in the event string, with one additional caveat to correctly include twilight counts:
\begin{enumerate}
    \item For each $\bigstar$, look for the closest previous photon that is not a $\circ$. If it is a $\bullet$, and that $\bullet$ is in the group immediately preceding the $\bigstar$, then add one to the total number of clicks.
    \item If there is a $\bullet$ after the last $\bigstar$ of the entire event string, and this $\bullet$ is in the last group of the string, add one to the total number of clicks.
\end{enumerate}
For instance, the string $\bigstar \bullet \circ \bullet \bullet$ codes to six events, all of which have five photons. The event $\left[ \bigstar \bullet \circ \bullet \bullet \right]$ has two clicks, the events $\left[ \bigstar \bullet \circ \bullet \right] \left[ \bullet \right]$, $\left[ \bigstar \bullet \circ \right] \left[ \bullet \bullet \right]$, and $\left[ \bigstar \bullet \right] \left[ \circ \bullet \bullet \right]$ have three clicks, and $\left[ \bigstar \bullet \circ \right] \left[ \bullet \right] \left[ \bullet \right]$ and $\left[ \bigstar \bullet \right] \left[ \circ \bullet \right] \left[ \bullet \right]$ have four clicks, the maximum possible number for this string (the $\circ$ is guaranteed to be lost).

By combining all of this---producing strings that code possible events, using those strings to construct the integrals for the probabilities of those events, and then evaluating those integrals and adding the results to the appropriate elements of the matrix---we produce the recovery effects matrix $\mathbb{R}$.

\begin{figure}
\centering
\subfloat{\includegraphics[width=0.4\textwidth,valign=c]{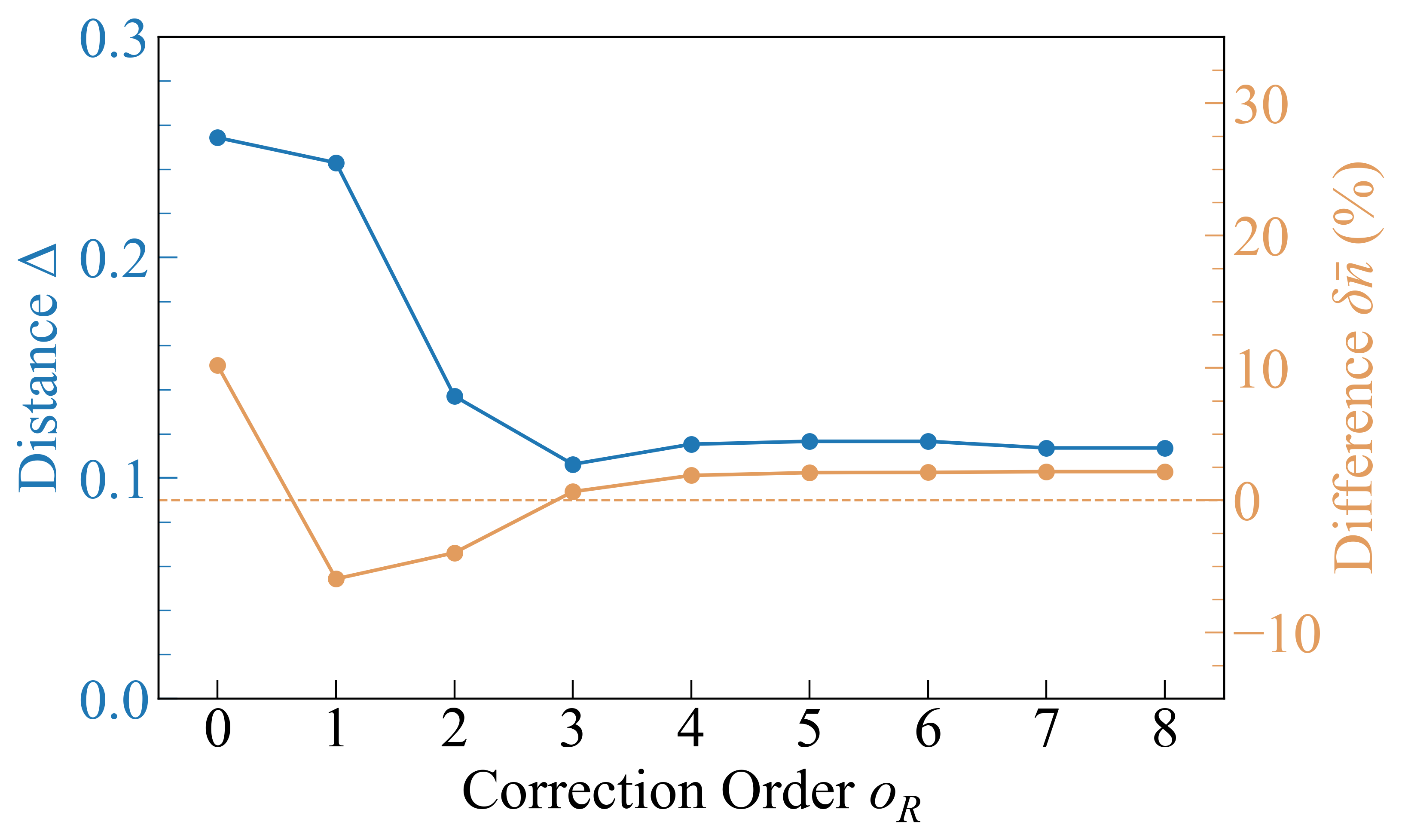}}
\subfloat{\includegraphics[width=0.58\textwidth,valign=c]{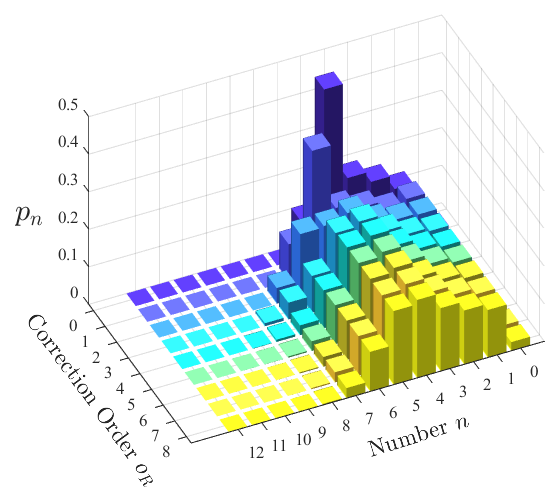}}
\caption{\label{fig:o_R_effects} A demonstration of the reconstruction of the short (85 ns full-width half-max) pulse of Fig.~4(c2) of the main text, which had an input average photon number $\bar{n}_{\text{fit}} = 3.97$, or about one photon per recovery time. In this limit the recovery time effects are large, and the calculation must be taken to high orders. The 3D bar plot shows the reconstructions as a function of the order parameter $o_R$. For low $o_R$, the fraction of photons put into low numbers varies \deleted{significantly}\added{somewhat} with $o_R$\deleted{, as can be seen in the ``back'' of the figure, for example by the ``bald spot'' of no population around $n=2$ for $o_R = 1,2$}; meanwhile there is no population at high numbers, because recovery time effects mean we almost never observe events with more than a few clicks, and $o_R$ is not large enough that the matrix $\mathbb{R}$ can ``add back'' those clicks. But as $o_R$ is increased, the matrix gains more power to add back clicks, and also results in a distribution that is stable to changes in $o_R$. The inset\added{ on the left} shows the convergence of both the total variation distance $\Delta$ and the difference $\delta\bar{n}$ as we increase $o_R$.}
\end{figure}

We note that, if all possible events for a basis of a given size are computed, the columns of the matrix should sum to unity, as the integrals for $p_1$, $p_2$, and $p_3$ did. However, in practice one may want to reduce the time taken to compute the matrix when it is known that the recovery time effects are not too large, e.g. for low count rates. In this case one can save time by implementing an ``order'' parameter $o_R$, which is defined such that at most $o_R$ photons can arrive during the recovery times of all clicks. (It is so named because it indicates what is typically meant by the ``order'' of the calculation.) This is equivalent to reducing the initial set of all $\circ$'s, $\bullet$'s, and $\bigstar$'s to just the set containing at most $o_R$ $\circ$'s and $\bullet$'s (and any number of $\bigstar$'s). When $o_R$ is less than the size of the matrix, we partially populate the first $o_R$ diagonals of $\mathbb{R}$ above the true diagonal, but we do not compute all possible events, so the columns do not add to one. We fix this manually, \deleted{by setting the diagonal elements so that the columns sum to unity.}\added{in the following way: for each column $n$, if $n-o_R \leq 1$, all events in the column have been calculated and the column should already be normalized, but if $n-o_R > 1$, we set the element in row $n-o_R-1$ of the column so that the column sums to one. (Here rows and columns are zero-indexed.) This corresponds physically to approximating that all the uncalculated events fall into the first uncalculated order of corrections.} The effects of $o_R$ on the reconstruction of an input pulse whose recovery time effects are significant are shown in Fig.~\ref{fig:o_R_effects}.

\subsection{\label{sec:AP_app}Afterpulsing Effects}

\begin{figure}
\centering
\includegraphics[width=0.7\textwidth]{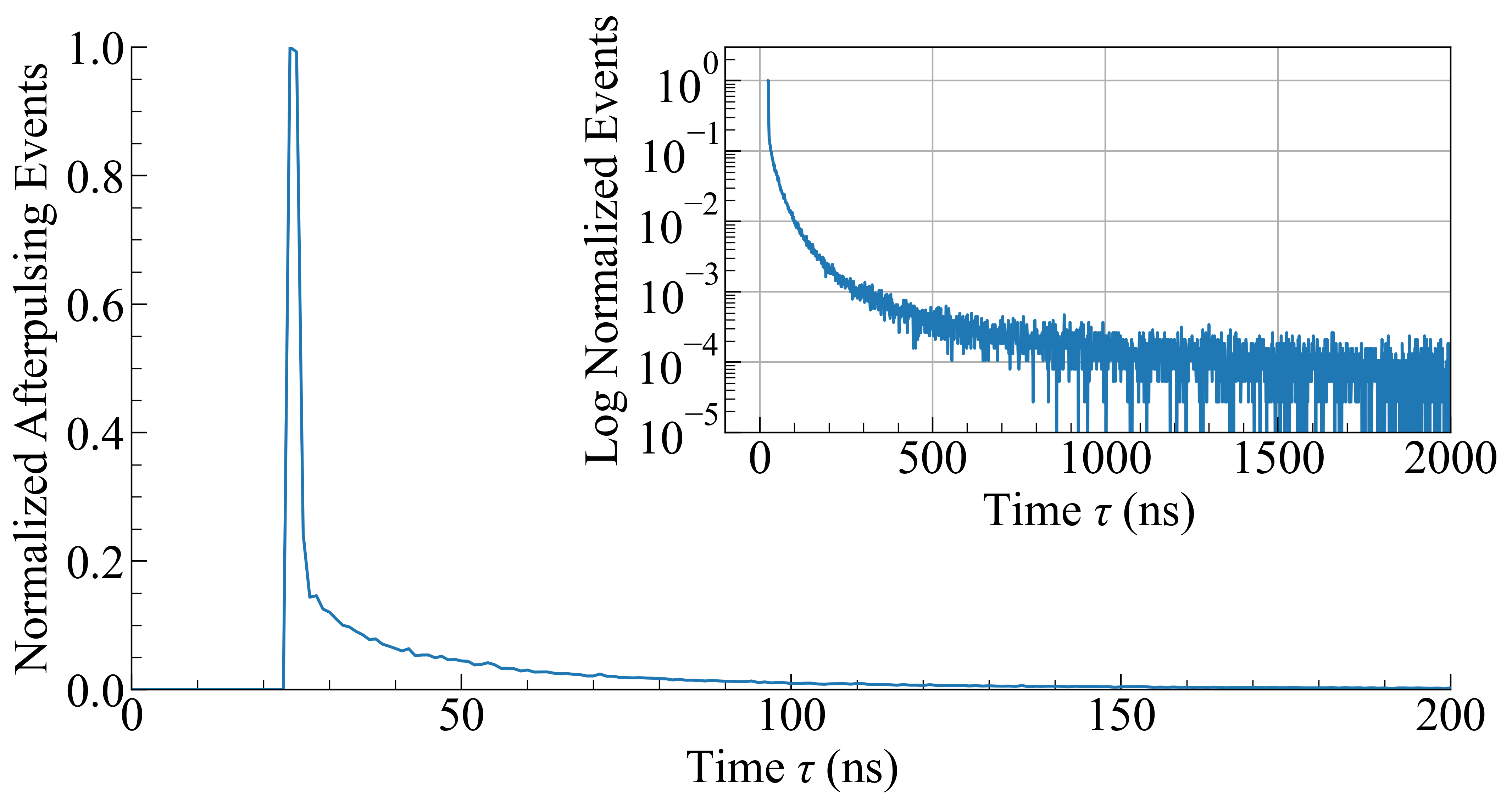}
\caption{\label{fig:AP_prof} A measured afterpulsing profile. The main figure shows the details of the first 200 ns of afterpulsing, which contains 87.3~\% of the afterpulsing feature. The inset shows, on a log scale, the first 2 $\mu$s of the afterpulsing, showing the common hyperexponential-like behavior \cite{AP_Modeling}. The first 2 $\mu$s contain 94.8~\% of the afterpulsing feature for this detector, SPAD2. The counts have been normalized here for simplicity, but this detector is measured to have about 2.4~\% total afterpulsing probability. Note the initial period of zero events, which is the due to the recovery time of the detector. Our methodology for retrieving the afterpulsing profile and probabilities is described in Section~\ref{sec:Calibration}.}
\end{figure}

We correct for afterpulsing by finding the probability $p_a$ of one afterpulse occurring in our data collection window, and use this to generate all afterpulsing corrections. To find $p_a$, we convolve the time-dependent afterpulsing probability with the photon profile. As an example of why this is necessary, consider performing \deleted{tomography}\added{reconstruction} on a light pulse that fits inside a 2 $\mu$s data collection window, using a SPAD that has the afterpulsing profile of Fig.~\ref{fig:AP_prof}. If a click occurs at the beginning of the window, it has a 2.3~\% chance of afterpulsing within the window; if the click occurs 1.8 $\mu$s into the window, it drops to 2.1~\% . To account for this, we convolve the time profile $a(\tau)$ with the photon profile. This gives the probability \deleted{of one afterpulse}\added{of an afterpulse click in the data collection window as}:
\begin{equation}
    p_a = \int_0^T \gamma(t) \int_t^{T} a(t'-t) \,dt' \,dt.
    \label{eqn:AP_prof}
\end{equation}
where $T$ is the data collection window duration. We use this probability to generate the afterpulsing matrix $\mathbb{A}(p_a)$. The elements of $\mathbb{A}$ relate a number of ``real'' clicks (clicks caused by signal photons or by background events) to total (``real'' plus afterpulsing) clicks. 

Given the above, we need to calculate the probability that $m$ real counts generate $n$ afterpulses, giving $m+n$ total clicks. \deleted{If the probability of one afterpulse click in a data collection window is $p_a$, we assume the probability of one click afterpulsing $n$ times is $p_a^n$ (we are describing afterpulses of afterpulses). This is a good approximation when $p_a \ll 1$. The probability that a click does not produce an afterpulse is then $1-\sum_{i=1}^\infty p_a^i = (1-2p_a)/(1-p_a)$. If two real counts arrive in the same experimental cycle, each one can independently afterpulse once with probability $p_a$, twice with probability $p_a^2$, and so on, so that the probability of any event with $n$ afterpulses is proportional to $p_a^n$. This model assumes that the afterpulses do not interact with other events like twilight counts; as discussed in the main text, this becomes the primary limitation of the model at high count rates, but at low count rates it is a good assumption since $p_a$ is small.}\added{Any particular click may afterpulse or not with probabilities $p_a$ and $1-p_a$ respectively. Thus the probability that a click generates zero afterpulses is $1-p_a$. Then the probability that a click generates exactly one afterpulse is $p_a(1-p_a)$, because the original click afterpulses, but the afterpulse click does not afterpulse. Proceeding in this way, the probability of a click producing $n \geq 0$ afterpulses is $p_a^n(1-p_a)$.}

Using the above model and assumptions, we determine the probability that $m$ real clicks generate $n$ total afterpulses. We can think of grouping the $n$ afterpulses into $m$ groups, each of which can have inclusively between 0 and $n$ of the afterpulse clicks. For instance, two real counts may produce three afterpulses in four ways: the first click may afterpulse three times and the second one none, the first click twice and the second once, the first click once and the second twice, or the first click none and the second click three times. This example may make clear that we can formulate the problem by asking the ways that exactly $m$ nonnegative integers can sum to $n$. For the previous example, $m=2$ and $n=3$ sees the possible sums $3+0$, $2+1$, $1+2$, $0+3$. An analytical expression for the number of sums is not known to the authors, but it can be computed numerically. 

From these sums we can determine the contribution to the probability of $m$ real clicks producing $n$ afterpulses. Each sum identified above contributes a term to be summed to obtain the final expression. \deleted{For each of these terms, we multiply together factors corresponding to the integers in the sum: if the integer is 0, we get a factor $(1-2p_a)/(1-p_a)$, and if the integer is $i > 0$, we get a factor $p_a^i$. This simplifies to the product $p_a^{N_A} [(1-2p_a)/(1-p_a)]^{N_0}$ where $N_0$ is the number of real clicks that do not afterpulse and $N_A$ is the number of total afterpulse clicks. Then all of these contributions are summed to give the final result. For example, the probability that two real clicks produce three afterpulses, which is the matrix element $\mathbb{A}_{52}$ since there are five total clicks from two real clicks, is}\added{Every term has the same form: they are the product of $m$ factors of $(1-p_a)$ and $n$ factors of $p_a$. The number of sums available from the reasoning of the previous paragraph only determines the number of such factors and therefore a prefactor for each element of $\mathbb{A}$. For example, the probability that two real clicks produce three afterpulses, which is the matrix element $\mathbb{A}_{52}$ since there are five total clicks from two real clicks, is}
%
\begin{equation}
    \mathbb{A}_{52} = 4 \times (1-p_a)^2 p_a^3.
\end{equation}
\deleted{The first term corresponds to the sums $0+3$ and $3+0$, and the second term corresponds to $2+1$ and $1+2$.} In this way we can build up the entire afterpulsing matrix.

If desired, in practice one can limit the number of afterpulses accounted for by implementing an ``order'' parameter $o_a$, which we define as the maximum number of afterpulses that can be produced by all of the clicks in a single experimental run. A finite $o_a$ merely places a limit on the maximum distance from the diagonal for which we will compute matrix elements. For all the reconstructions in this work, we use $o_a = 2$.

Because afterpulsing adds clicks, the possibilities for events are infinite: a single real click can produce two total clicks, or three, or four, and so on. Since the matrices we compute are always finite, the computed probabilities will never sum to unity. This is true regardless of the chosen order parameter $o_A$. \deleted{In practice, we set the diagonals of $\mathbb{A}$ manually so the columns of the matrix, and therefore the probabilities, sum to one.}\added{To resolve this, we manually set elements as follows: for each column $n$, if $n+o_a+1 \geq n_{\text{max}}$, set the element in row $n_{max}$, and otherwise, set the element in row $n+o_a+1$ such that the column sums to one. Physically this corresponds to letting all of the uncalculated probabilities fall into the next order of the calculation.}

\subsection{Limitations of the Model}

\added{The model and calculations presented here involve a number of approximations, which we discuss in this section, in no particular order.}

\paragraph{Afterpulsing models.} \added{We note that we have not carefully considered the physical mechanisms of the afterpulsing in our SPADs, which may impact the probability distribution used for the calculation. The model we have used, in which the probability of $n$ afterpulses is geometric, is widely adopted, but the use of a Poissonian distribution is equally reasonable from arguments about the density and number of these traps \cite{CountingStats,NonMarkovianAP}. Such a model appears to produce larger total variation distances for our detectors.}

\paragraph{Calculation of the photon profile.} \added{Both the afterpulsing probability calculation and the calculation of the recovery time effects matrix $\mathbb{R}$ require a photon profile $\gamma(t)$ over which to integrate, but our measured click profiles are affected by dead times and afterpulsing. We have approximated $\gamma(t)$ by histogramming the clicks in data collection cycles where the detector registers exactly one click. This choice suffers from some distortions, e.g. if a photon impinges on the detector and produces a click, and then a second photon arrives while the detector is dead, the run is counted as producing only one click, even though there were two photons.}

\added{When the input light state is Poissonian, these effects may be analyzed and corrected by modeling the process as a Poisson point process \cite{CountingStats}. Consider an input state with a mean photon rate $\lambda(t)$, and define $\Lambda(t_1,t_2) = \int_{t_1}^{t_2} \lambda(t)\,dt$. The survival probability, the probability of no detections, from $t=0$ to later time $t$ is $e^{-\Lambda(0,t)}$. The probability of the first detection is}

\begin{equation}
    p_1(t_1) = \lambda(t_1) e^{-\Lambda(0,t_1)}.
    \label{eqn:p_1}
\end{equation}

\noindent \added{The probability of this being the only detection in the window from $t=0$ to $t=T$ is the probability of one detection at $t_1$ multiplied by the probability of no further detections, $e^{-\Lambda(t_1,T)}$, so that the probability of exactly one click in window $T$ is} 

\begin{equation}
    P_1(t_1) = \lambda(t_1) e^{-\Lambda(0,T)}.
\end{equation}

\noindent \added{If our detector were ideal, $P_1$ would be proportional to $\lambda(t_1)$ and we could use it directly to measure the photon profile. However, for instance, a dead time $t_d$ modifies the single click probability to  $P_1'(t_1) = \lambda(t_1) e^{-\Lambda(0,t_1)} e^{-\Lambda(t_1+t_d,T)}$, no longer proportional to $\lambda(t_1)$ \cite{CountingStats}.}

\added{For coherent states, an improvement to the model can be made by numerically solving Eqn.~\ref{eqn:p_1}. Analysis of our coherent-state data shows only negligible improvements, even for large count rates. For non-coherent states, the use of Eqn.~\ref{eqn:p_1} is incorrect. For instance, if the pulse is a pure two-photon Fock state, we will not recover the full photon profile by only looking at single-click experimental cycles. One solution is to take a separate measurement in which a large attenuation is placed in front of the detector; this should make the entire pulse shape accessible while leaving it unchanged.}

\paragraph{Model of the recovery time; non-Markovianity of afterpulsing and recovery time.} \added{In this work we assume a model for the quantum efficiency during the recovery time shown in Fig.~\ref{fig:Dtau}. We note that this property of the detector can actually be measured using, for instance, the methods of Ref.~\cite{TwilightDelays}, in which very short light pulses are sent to the detector with controlled delays and the resulting detector response is measured.}

\added{Our model of the detector during the recovery time loses validity at high count rates, as discussed further below and shown in Fig.~\ref{fig:ResetTimesFig}. There could be numerous reasons for this due to the non-Markovian behavior of the detector, which is not modeled in our algorithm. This behavior manifests in various ways. (1) Changes to the recovery time with increasing count rate have been observed~\cite{Autocorrelations}. In our data we observe a shift of the DEP by around 0.5~ns occurring around count rates of 10~Mcounts/s. This increase is not accounted for in our model. (2) When a twilight count occurs, the quenching circuitry of the SPAD is not activated in the typical manner, and a result of this may be that the avalanche current lasts longer than usual~\cite{Autocorrelations}. The deep traps that later result in afterpulsing may get more filled than usual, resulting in an afterpulsing probability that increases with count rate. (3) It has also been shown that each avalanche fills only a small number of the total deep traps in a diode, so that as the count rate increases, it is reasonable to imagine more of these traps getting filled, resulting again in an afterpulsing probability that increases as the count rate increases~\cite{NonMarkovianAP}.}

\paragraph{Correlations in the recovery time effects matrix.} \added{To write down the integral equations for calculating the various probabilities in the recovery time effects matrix $\mathbb{R}$, we make an assumption about the joint photon arrival time probability, namely that $\lambda(t_1,t_2,\dots,t_n) = \lambda(t_1) \lambda(t_2) \cdots \lambda(t_n)$. This is really only true for Poissonian light; in general, correlations in the light lead to correlations in the detections \cite{MandelAndWolf}.}

\added{It can be shown, either from a semiclassical or a fully quantum theory, that the probability density of a first photodetection is~\cite{MandelAndWolf}}

\begin{equation}
    P_1(t)\,dt = \alpha I(t)\,dt
\end{equation}

\noindent \added{where $\alpha$ is a constant related to detector properties. From this, the joint probability density of two photodetections is \cite{MandelAndWolf}}

\begin{equation}
    P_2(t_1,t_2)\,dt_1\,dt_2 = \alpha^2 \langle : I(t_1)I(t_2) : \rangle\,dt_1\,dt_2 = \alpha^2 P_1(t_1)P_2(t_2) g^{(2)}(t_1,t_2) \,dt_1\,dt_2
\end{equation}

\noindent \added{where we have introduced the normalized second-order autocorrelation function $g^{(2)}$. Thus when $\langle : I(t_1)I(t_2) : \rangle\ \neq \langle I(t_1) \rangle \langle I(t_2) \rangle$, or when $g^{(2)}(t_1,t_2) \neq 1$, we cannot write $P_2(t_1,t_2) = P_1(t_1)P_1(t_2)$. Arguments of this type generalize to larger numbers of photodetections as well \cite{MandelAndWolf}.}

\added{This algorithm might be improved for non-Poissonian light by constructing a more detailed mathematical theory including correlations. However,  with a single detector, the correlation information that is available is subject to all of the detector imperfections, especially afterpulsing and dead times. For instance, with just one detector with a dead time, there is simply no way to measure $g^{(2)}(t_1,t_2)$ when $t_2-t_1$ is less than the dead time. }

\added{This is one reason that the pulses being measured with this algorithm must have widths and correlations at least a few detector dead times in length; when this condition is satisfied, the error made by this approximation is small, because the recovery time effects matrix only changes the reconstructed number-state distribution for events that occur during the recovery time. In general, this approximation will push the reconstructed distribution toward a Poissonian distribution.}

\paragraph{Interactions of afterpulses and twilight counts.} \added{This model does not include ``interactions'' between afterpulses and recovery time effects as seen by the detector. Although afterpulses have an independent physical origin from recovery time effects, their occurrence on the same detector causes them to interact in important ways. Afterpulsing does not occur when the detector is dead. Thus, afterpulses due to previous clicks may be prevented by subsequent clicks, which becomes more likely when the count rate is high. Large count rates also change the temporal distribution of the afterpulsing. After a click, twilight counts and afterpulsing have temporal profiles which overlap (in the few nanoseconds immediately after the recovery time), but the detector can only register one of them. When the there is a significant probability of a twilight count, the likelihood of an afterpulse is reduced.}

\added{As discussed in the main text, we believe the fact that our algorithm does not address these interactions is the primary limitation at the highest count rates. At high count rates the DEP clicks are overcompensated, pushing the fitted average photon number per pulse below what is expected and producing reconstructed distributions that are narrower than the corresponding Poissonian.}

\section{\label{sec:Calibration}Detector Calibration Techniques}

In this section, we discuss our methodology for measuring all of the inputs to our SPAD model.

\subsection{Using Second-Order Histograms}

\begin{figure}[tb!]
\centering
\subfloat{\includegraphics[width=0.48\textwidth]{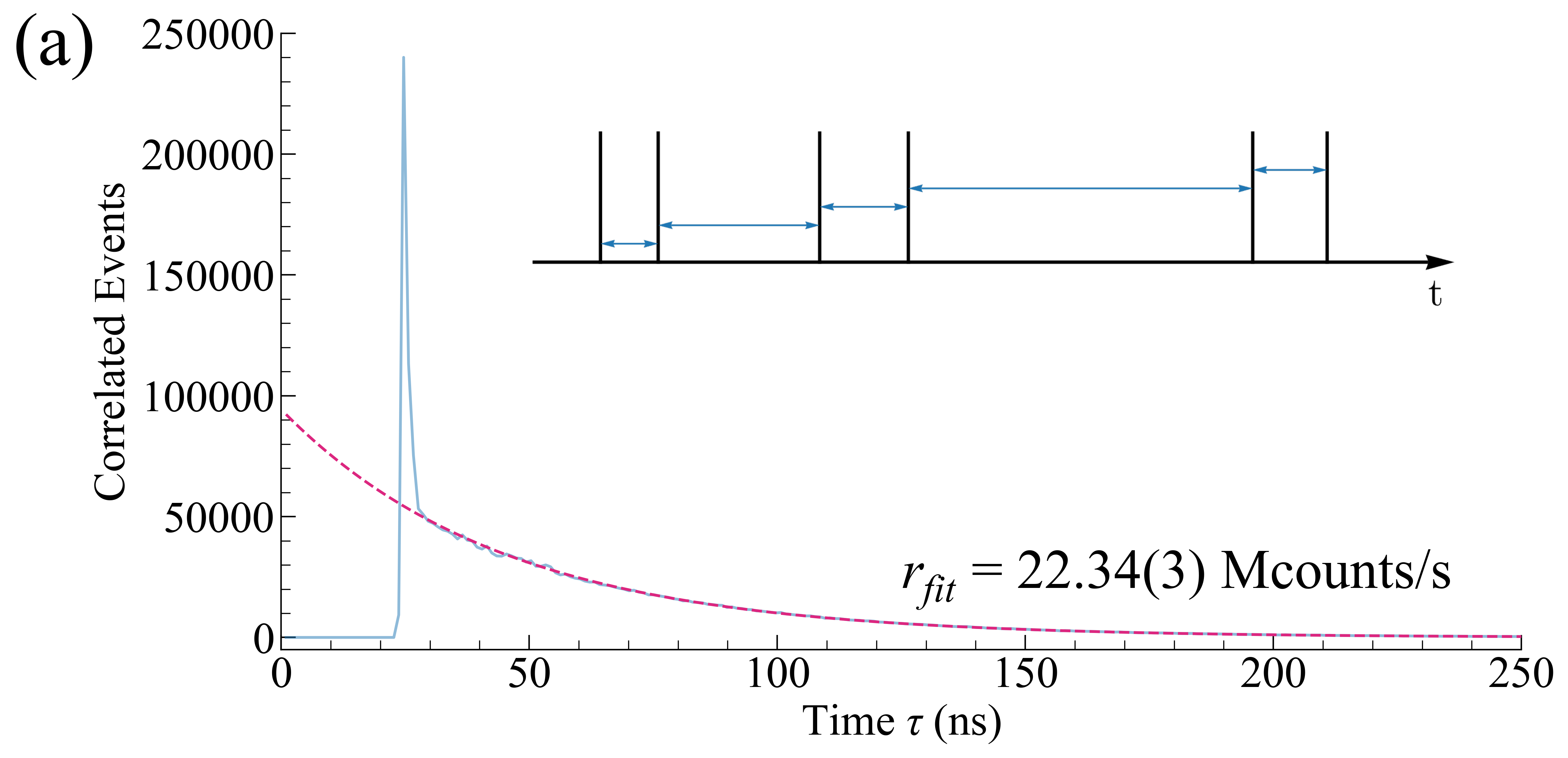} \label{fig:Hists_IAT}}
\subfloat{\includegraphics[width=0.48\textwidth]{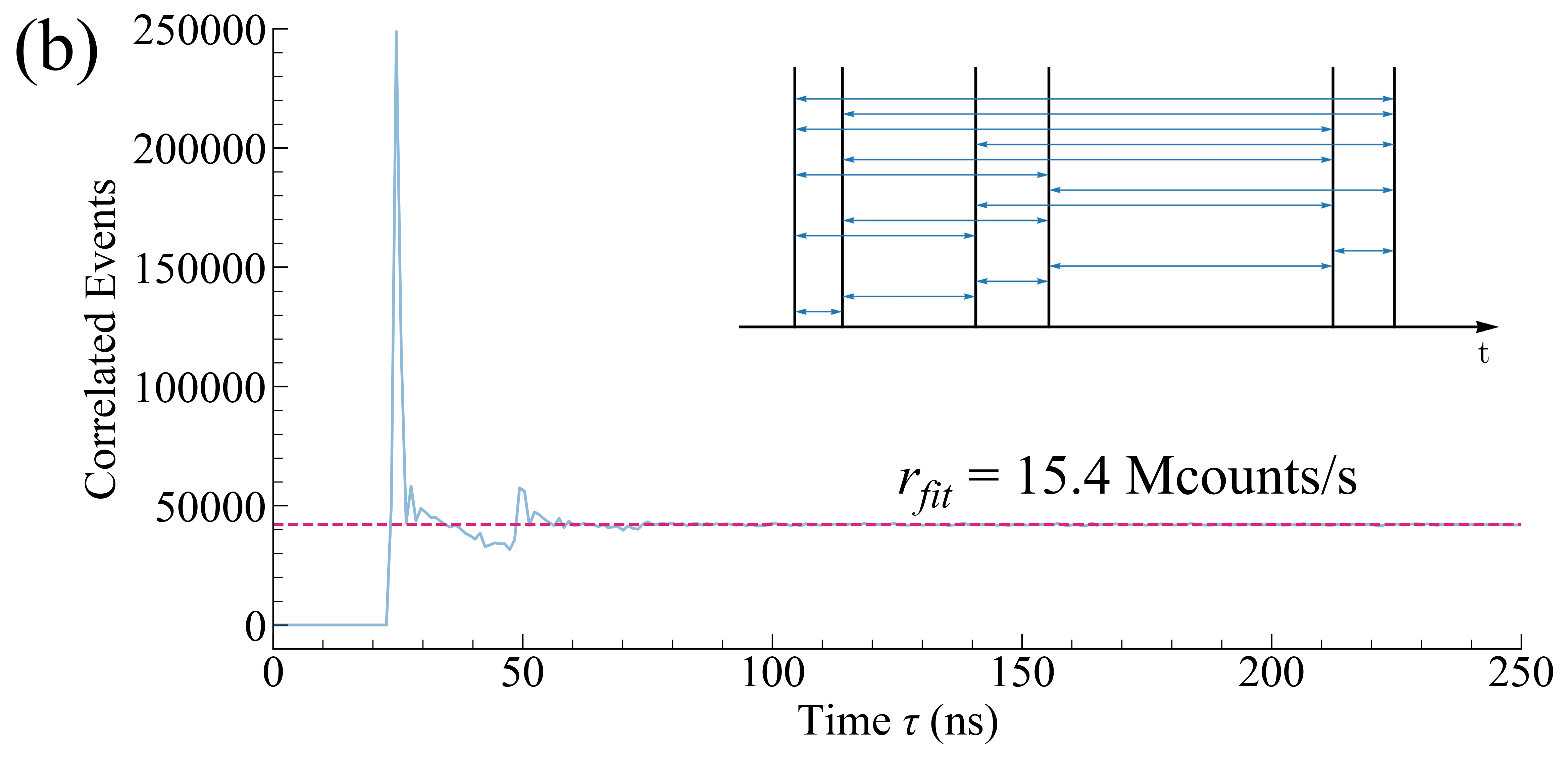} \label{fig:Hists_FC}}

\subfloat{\includegraphics[width=0.48\textwidth]{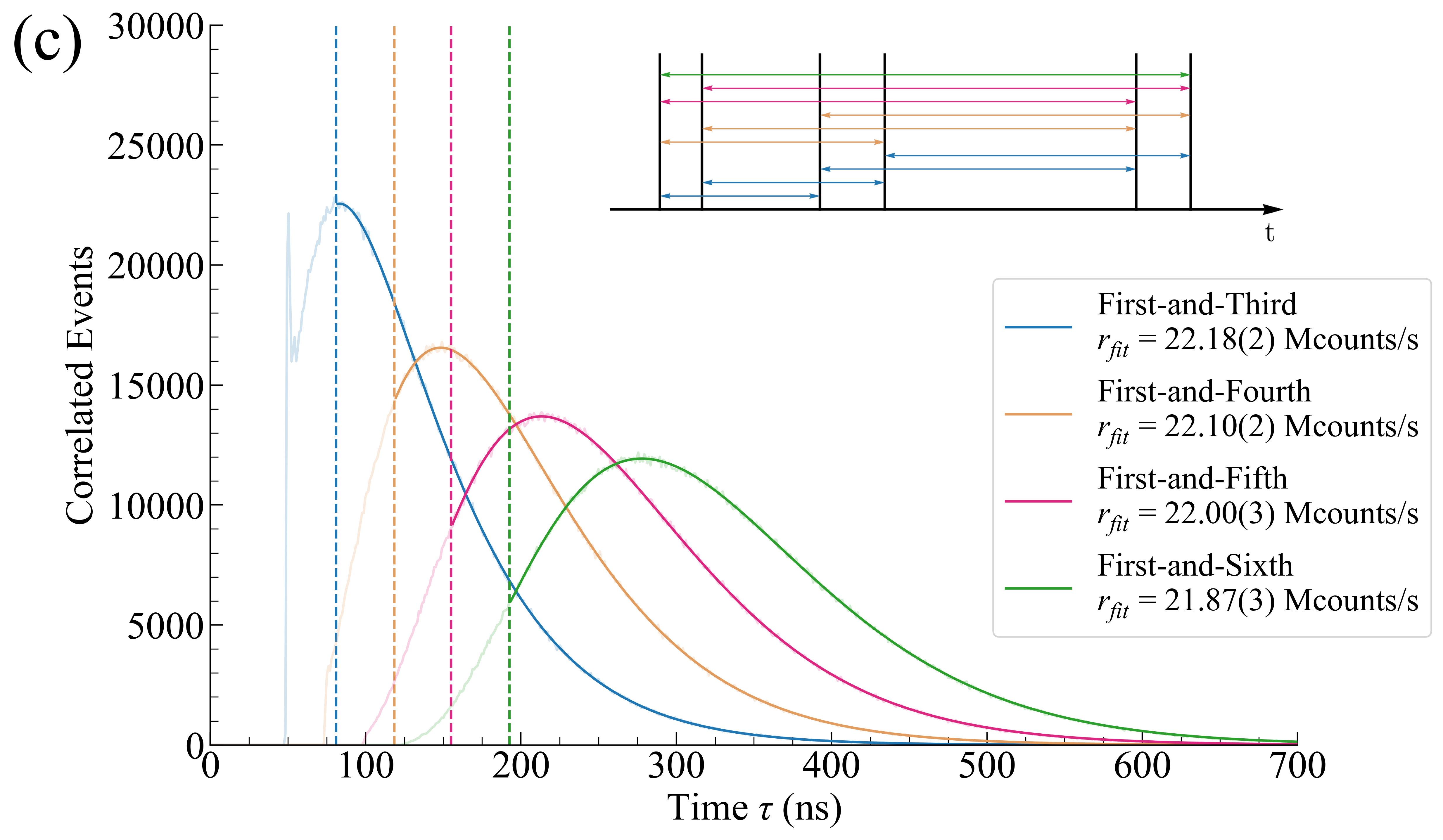} \label{fig:Hists_FN}}
\subfloat{\includegraphics[width=0.48\textwidth]{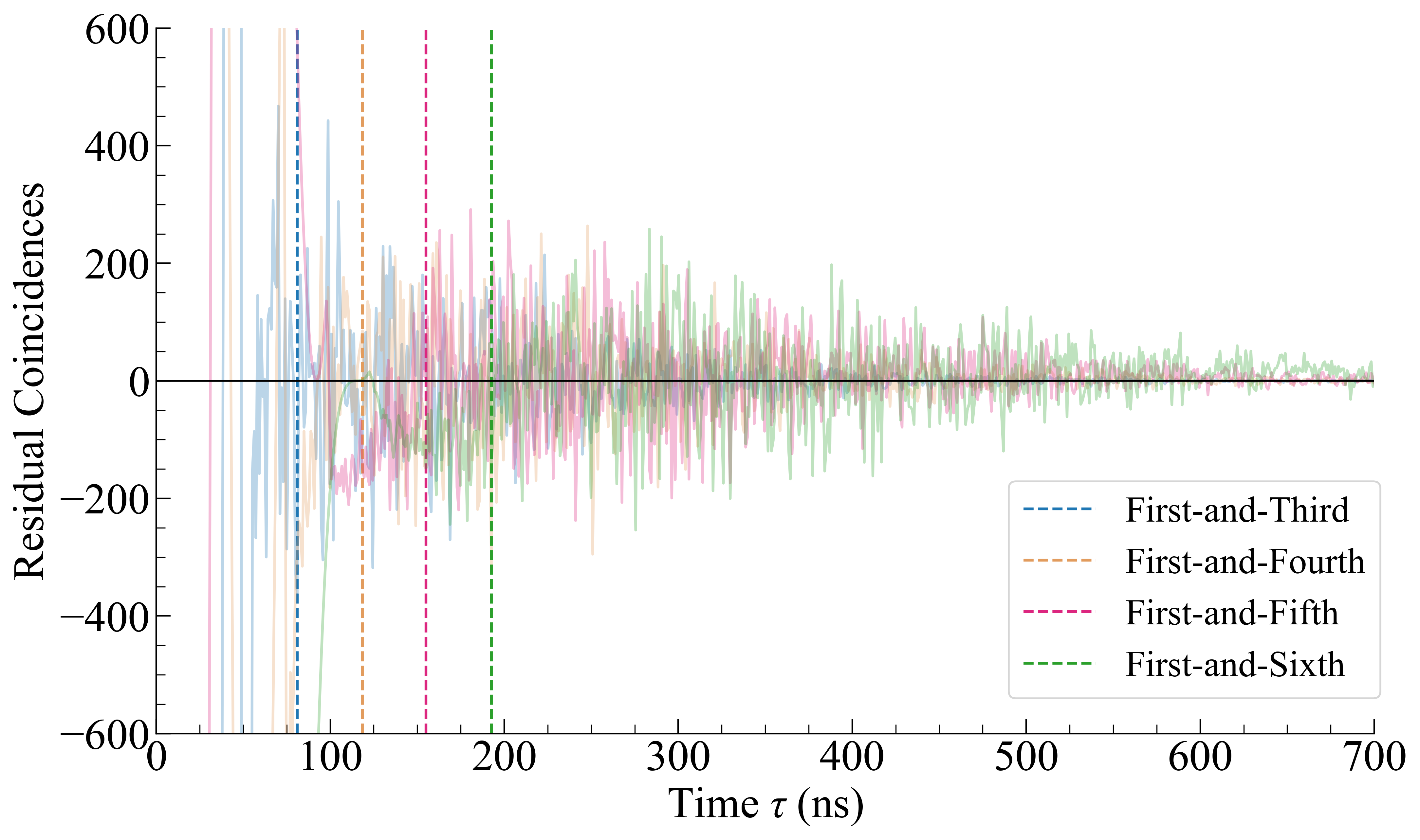} \label{fig:Res_FN}}
\caption{\label{fig:HistogramsComp} Various second-order histograms generated from the same dataset, with $\approx$~30~Mcounts/s of cw input light. \deleted{T}\added{In (a,b,c), t}he insets illustrate the click pairs used to generate the respective histograms; the horizontal axis is a time axis, and the thick vertical lines represent times when clicks occur. Pairs of clicks connected by arrows are included in the histogram, and the connecting arrows are color-coded to match the respective main figures. (a) First-and-second or ``interarrival'' data (blue) shows an exponential behavior (dashed \deleted{orange}\added{magenta}) after the detector effects peak (DEP). The fitted count rate \deleted{(starting the fit at 30 ns)}\added{using Eqn.~\ref{eqn:Prob_FirstOrder}} is \deleted{18.8}\added{22.34(3)}~Mcounts/s\deleted{, which is slightly affected by the long afterpulsing tail}. (b) Full second-order correlation data (blue) shows a linear behavior (dashed \deleted{orange}\added{magenta}) after the DEP. The fitted count rate (starting the fit at 100~$\mu$s) is \deleted{13.5}\added{15.4}~Mcounts/s, altered heavily by recovery time losses. The long-time behavior of this type of histogram is \deleted{illustrated in Fig.~\ref{fig:RT1}: it follows }a linear function whose x-intercept is the data collection time. (c) \added{Left panel:} First-and-$n$ histograms for $3 \leq n \leq 6$, with solid fit lines overlaying the data. Fitted count rates \added{using Eqn.~\ref{eqn:Prob_FirstOrder}} from $n=3$ to $n=6$ \deleted{(with fits beginning at $(n-1)\times 25$ ns) }are as indicated in the legend: \deleted{18.98, 19.01, 19.05, 19.11}\added{22.18, 22.10, 22.00, 21.87}~Mcounts/s, a difference of up to \deleted{1.6}\added{2.2}~\% from the first-and-second fit. \added{The vertical dashed lines indicate the times at which each fit is begun. Right panel: Residuals from the fits shown in the left panel, with a notable lack of structure at times greater than the fit start time, shown again with vertical dashed lines.}}
\end{figure}

Most of the calibration analysis involves the generation of various second-order histograms of times between pairs of clicks, with the only difference being which pairs of clicks to include. \deleted{Fig.~\ref{fig:HistogramsComp} shows several types of such histograms, which we discuss in detail in this section. }These histograms are used for various functions in the calibration analysis, but one of the most important is the extraction of detected count rates. Count rates are extracted from second-order histograms by fitting the histogram data to a theoretical lineshape, which is not exactly what is observed due to the time-correlated recovery time and afterpulsing effects. \added{In this subsection we discuss several types of second-order histograms, as shown in Fig.~\ref{fig:HistogramsComp}, along with the theoretical lineshapes and effects of detector imperfections on each type, and how we use each type in our detector characterization.}\deleted{The effect of dead time losses---which pair together clicks which would be separated in a zero-dead-time detector---is simply to scale all of the histograms studied here by a constant factor. Afterpulsing, however, adds clicks in a way that is independent of the count rate. As discussed below, in some types of histograms this also merely scales the histogram by a constant factor, while in other types it dramatically changes extracted count rates; the only solution is to fit only to the portion of the histogram at large delays. We choose 1 ms to begin the fits to these histograms. With these caveats in mind, let us consider different types of second-order histograms in more detail. Fig.~\ref{fig:HistogramsComp} shows several types of second-order histograms, which we discuss in detail in this section.}

\deleted{A}\added{One type of second-order histogram is} first-and-second histogram (Fig.~\ref{fig:Hists_IAT}) is a histogram of waiting times between clicks that are adjacent in time. This type of data has often been called ``interarrival data'' in the literature, and it is the the most common second-order histogram to produce because it has historically been the easiest type of data to \added{gather}\deleted{generate}, using relatively simple click-time processing circuits. These histograms plot the probability that no click occurs for a time $\tau$ after an initial click, and then a click occurs at time $\tau$ after the initial event. This is why the histogram in Fig.~\ref{fig:Hists_IAT} is zero for the first $\approx$ 25 ns: this is the recovery time, where photons that arrive are either lost altogether, or their clicks are delayed until later as twilight counts. The large peak immediately after\added{ the recovery time} is what we call the ``detector effects peak'' (DEP), and it contains both afterpulsing clicks and twilight counts. While twilight counts have a fairly narrow distribution, afterpulsing may extend \deleted{hundreds}\added{tens} of microseconds as shown partially in Fig.~\ref{fig:AP_prof} (which is also a first-and-second histogram). After this time, the detector fully recovers, and the histogram faithfully reproduces the statistics of the input light.

For Poissonian light, once the detector has recovered from previous clicks, the shape of the histogram should fit $e^{-r\tau}$, which is the Poisson probability of no events for a time $\tau$ followed by one event at that time. Since we can fit this exponential using an independent amplitude parameter, the $r$ here is unaffected by dead-time losses, which only scale the histogram\deleted{, as noted above}. However, this type of data suffers from a crucial problem: the afterpulsing feature has an exponential-like shape on a first-and-second histogram, so to avoid this feature affecting the fitted count rate, we \added{must }fit data at large delays. But when the count rate\deleted{ $r$} is large, one has to wait a very long time to gather data at large $\tau$. \added{In our SPADs, afterpulsing is noticeable for tens of $\mu$s, so let us suppose we wait for 100~$\mu$s to avoid afterpulsing. }\deleted{As an}\added{In this} example, at even a modest continuous count rate of \deleted{$5 \times 10^4$ }\added{250~k}counts/s, the probability of no click for at least \deleted{one millisecond}\added{100~$\mu$s} (assuming no afterpulsing) is about \deleted{$1.9 \times 10^{-22}$}\added{$1.4 \times 10^{-11}$ per click}, so that one needs to wait \deleted{about three times the present age of the universe before expecting a single event beyond one millisecond}\added{nearly four days on average before getting a single event beyond 100~$\mu$s}. Thus, building statistics sufficient for a two-parameter exponential fit without the influence of afterpulsing is impractical. (We do, however, use first-and-second histograms in other ways, as discussed in the next section.)


Another choice is to generate full second-order correlation data, which is a histogram of the delays between all pairs of clicks in a dataset. Such a histogram is shown in Fig.~\ref{fig:Hists_FC}. This data still exhibits the sharp afterpulsing peak, \added{and we observe the effects of this peak for tens of $\mu$s, so we fit to data beginning at a delay of 100~$\mu$s.}\deleted{so we  fit to data beyond 1 ms to extract count rates.} But because we look at all pairs of clicks in this data, the extent of delays for which we can easily get data is much larger; in fact the largest delay for which there is a histogrammed event is necessarily equal to the data collection time. For Poissonian light, there is a constant probability of a click, so the probability for a particular delay between clicks (with any number of events in between) is also constant. Because we only collect data for a finite amount of time, experimentally the shape we see (if we view the entire dataset, not shown in Fig.~\ref{fig:Hists_FC}\deleted{ but illustrated in Fig.~\ref{fig:RT1}}) is a downward-sloped line whose $x$-intercept is equal to the data collection time. The downward slope (a deviation from the theoretically horizontal line) is purely an edge effect. The shape of the histogram at large delays is given  by $r^2\Delta t(t_0-t)$, where $t_0$ is the total data collection time and $\Delta t$ is the bin width. A fit to this line is shown as the \deleted{orange}\added{magenta} dashed line in Fig.~\ref{fig:Hists_FC}\deleted{ (and the blue dashed line in Fig.~\ref{fig:RT1})}. We can think of the expression as a product of $r\Delta t$, the probability of a click per bin, and $r(t_0-t)$, which captures the number of clicks ``available'' for correlations at time $t$ in the data collection. Note the nature of the extracted $r$: it is only an amplitude in the mathematical expression. Since these histograms are scaled by dead time losses as discussed above, the extracted count rate is decreased by dead time losses. The typical correction factor for such an effect is 
\begin{equation}
    r' = \frac{r}{1 - rt_D},
\end{equation}
where $t_D$ is the dead time and $r'$ would be the measured count rate if there were no dead time. In this work we do not rely on count rates extracted this way, but full second-order correlation data is used in other ways, as discussed further below.

A final \deleted{choice we can make for count rate extraction is to use}\added{type of second-order histogram is} higher-$n$ first-and-$n$ histograms, which plot the delays between each click and the $(n-1)$th click immediately after it. (First-and-second histograms are a special case of these, with $n=2$.) Fig.~\ref{fig:Hists_FN} shows several of these histograms generated from the same dataset. A first-and-$n$ histogram is a measure of the probability of having exactly $n-2$ clicks in a time $t$. For Poissonian light\added{ on an ideal detector}, these events follow a probability density function that is $p_n^{\text{ideal}}(t) = r(rt)^{n-2}e^{-rt}/(n-2)!$.\deleted{Note that our fitting expression can include an additional scaling factor that can be varied independently of the count rate $r$, so our fitted count rates are unaltered by dead time losses.}\added{ For a detector that has a recovery time $\tau_r$ and a probability $p_a$ of an afterpulse or twilight count occurring immediately at the end of the recovery time, it can be shown~\cite{CountingStats} that the probability density function for times beyond the recovery time is}
\begin{align}
    p_n(t) &= re^{-r[t-(n-1)\tau_r]} \sum_{k=1}^{n-1} {n-1 \choose k} p_a^{n-1-k}(1-p_a)^k \frac{r^{k-1}\left[t-(n-1)\tau_r\right]^{k-1}}{(k-1)!}, \label{eqn:RefE8} \\
    &\hspace{25em} t > (n-1)\tau_r, n \geq 2. \nonumber
\end{align}
\added{(This is Eqn. S24 of Ref.~\cite{CountingStats}, with $\tau_i$ set to equal $\tau_r$ to model the fact that first-and-$n$ histograms assume a click at $t=0$.) The $k=n-1$ term with $\tau_r=0$ and $p_a=0$ corresponds to the ideal detector expression. While the ideal detector expression already produces a good fit to data, Monte Carlo simulations suggest that the fitted count rates vary from the true count rates by amounts on the order of 1-2\%, and there is a deviation from the true count rate depending on the portion of the histogram used for fitting. However, it can be shown using Eqn.~\ref{eqn:RefE8} that, to first-order in $p_a$, the number of events per bin is}
\begin{align}
    P_n(t) &= N\Delta t\, r\left[ 1 - (n-1)p_a + \frac{(n-1)(n-2)p_a}{r(t-(n-1)\tau_r)} \right] \frac{\left[ r(t-(n-1)\tau_r) \right]^{n-2}}{(n-2)!} e^{-r(t-(n-1)\tau_r)}, \label{eqn:Prob_FirstOrder} \\
    &\hspace{27em} t > (n-1)\tau_r, n \geq 2. \nonumber
\end{align}
\added{The factor $N\Delta t$ converts the probability density into a number of events per bin, with $N$ the number of clicks in the dataset and $\Delta t$ the width of the time bin used to discretize the data. Eqn.~\ref{eqn:Prob_FirstOrder} is an expression being fit for $r$, $p_a$, and $\tau_r$, with no independent amplitude parameter. It uses the same assumptions as Eqn.~\ref{eqn:RefE8}, but drops higher-order terms in  $p_a$, which makes the fit  more numerically robust. The term proportional to $p_a/rt$ biases the histogram slightly to early times; this corresponds to the fact that, when an $n$-click sequence involves an afterpulse, the average duration of the sequence will be shorter since afterpulses have a timescale shorter than that of the average time between photon-produced clicks. The term proportional to $p_a$ then scales down the entire histogram to make up for this addition, renormalizing the probability density function.}

\added{In using Eqn.~\ref{eqn:Prob_FirstOrder} we make two notes. First, $p_a$ and $\tau_r$ empirically have a large covariance, so we constrain $\tau_r$ by setting it equal to the time of the first nonzero bin in the histogram. Second, $p_a$ has only some relation to the ``true'' afterpulsing+twilight probability. The above equations model these phenomena as occurring immediately after the end of the recovery time, which results in maximum distortion of the histogram toward early times; to the extent that afterpulsing and twilight counts have any temporal width in a real detector, this distortion is reduced, so we expect the fitted $p_a$ to be lower than the true $p_a$. We see this empirically in using Eqn.~\ref{eqn:Prob_FirstOrder}. As a result, we generally discount the fitted value of $p_a$ and measure the afterpulsing+twilight probability in other ways, as discussed below. We note also that, while the fitted $p_a$ has a somewhat large uncertainty in our fits, the covariance between $p_a$ and $r$ is small, so the fitted count rate is robust to this uncertainty.}

\added{From Monte Carlo simulations, Eqn.~\ref{eqn:Prob_FirstOrder} gives nearly constant fitted count rates regardless of where the fitting is begun, as long as it is at delays of at least two recovery times. Furthermore, these simulations suggest that, for $n \geq 4$, the fitted count rate is below but well within 1\% of the true rate, and that this fitted rate converges as $n$ increases and improves with both smaller $p_a$ and afterpulsing that has greater temporal width. That is, we can fit nearly the entire histogram, without regard to the limit of 100~$\mu$s set for avoiding afterpulsing, and still achieve very good fitted count rates.}\deleted{Furthermore, these first-and-$n$ histograms deal with the disadvantageous effects of afterpulsing. In a first-and-second histogram, the afterpulsing peak is measured faithfully, because we look at each click and the click immediately following it. For a first-and-third histogram, the first nonzero peak at early delays is due to \emph{second} afterpulses, a phenomenon with a probability $p_a^2$, much lower than $p_a$, so the effect of this initial peak is sharply decreased. For higher $n$ the size and effect of this peak decreases, as can be seen in Fig.~\ref{fig:Hists_FN}. Of course, the afterpulsing events are not gone completely. Instead, the afterpulsing ``diffuses'' over the histogram, amounting to a constant scaling. As an example, if a first click is followed by a second click which afterpulses, a first-and-third histogram will include an event connecting the first click to the afterpulse; the effect extends in time for higher $n$. This means that high-$n$ first-and-$n$ histograms are merely scaled by afterpulsing, but the structure is unchanged. Since we do not obtain the measured count rate from the amplitude of the histogram, the count rate extracted from these histograms is not affected by afterpulsing or by recovery time effects. At the same time the large, long features that influence the first-and-second histograms become negligible and diffuse for higher-$n$ histograms, so, for larger $n$, we can begin our fitting much earlier delays than 1 ms without concern that afterpulsing will influence the fitted count rate. First-and-$n$ histograms therefore solve all of our needs and allow us to extract very high count rates.} To our knowledge, this technique is not used elsewhere in the literature (although we are not the first to use first-and-$n$ histograms for some form of detector characterization \cite{Autocorrelations}).

\added{To extract a count} rate unaltered by dead time losses, we use first-and-$n$ histograms with $n=6$, as this was found to be sufficiently high for the count rates to converge. \deleted{We begin the fit at delays equal to the maximum of the histogram divided by 1.5}\added{We choose the fit start time in the following way (though there is nothing special about this choice except that it exceeds twice the recovery time): we find the first nonzero bin, and to this delay we add $n/5$ times the difference between the recovery time and the maximum of the first-and-third histogram}, which \deleted{are}\added{leads to fit start times} as early as 100 ns in Fig.~\ref{fig:Hists_FN}.\added{ We do not use the fitted values of $p_a$ or $\tau_r$ from any of these fits, as discussed above.} The code that generates these histograms, along with some sample data and an example implementation of count rate extraction from a first-and-$n$ histogram, is available in Code 1 \cite{Git}.

\subsection{Detector Characterization Methodologies}

\begin{figure*}[t!]
\centering
\includegraphics[width=\textwidth]{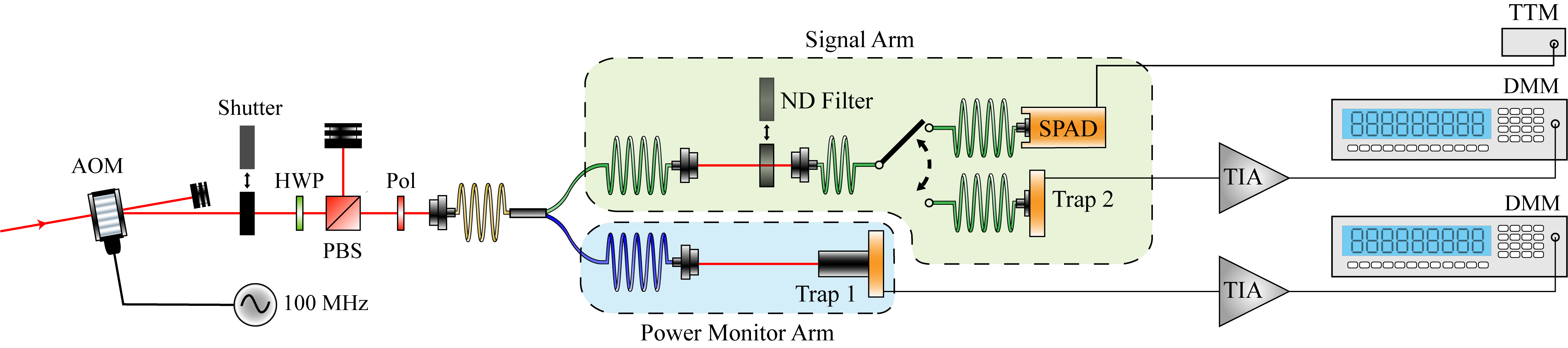}
\caption{\label{fig:ExperimentalSetup} The experimental setup for detector calibration. Light at 780 nm is sent through an acousto-optic modulator (AOM) and a laser shutter, and then a system for power control consisting of a half-waveplate (HWP), a polarizing beam splitter (PBS), and a polarizer (Pol) on a motorized rotation mount. The light is then coupled into a $1 \times 2$ fiber splitter whose outputs are the power monitor arm (blue) and the signal arm (green). The power monitor beam is free-space-coupled to a trap detector, Trap~1. The signal arm light may be sent through a calibrated ND filter on a motorized flip mount, and then it is coupled into another fiber which may be coupled to a second trap detector, Trap~2, or to the SPAD under test. Both trap detectors are attached to calibrated high-gain transimpedance amplifiers (TIAs) whose output voltages are read by 8-1/2-digit digital multimeters (DMMs), while the SPAD signals are sent to a time tagger module (TTM). The DMMs and TTM send data to a computer for processing. For the calibration measurements presented in this work, the AOM is always on, and both arms of the fiber splitter are used; for the \deleted{tomography}\added{reconstruction} measurements, the AOM is used as an optical switch to produce pulses, and the power monitor arm is not used, reducing the setup to the one shown in Fig.~1 of the main text.}
\end{figure*}

To measure the inputs to our reconstruction algorithm, we developed the experimental setup shown in Fig.~\ref{fig:ExperimentalSetup}, which allows us to measure everything we need with minimal reconfiguration. We use 780 nm light generated by a commercial external cavity diode laser. The beam passes through an acousto-optic modulator (AOM) (which is always on for the measurements of this section but is used to shape pulses for the \deleted{tomography}\added{reconstruction} measurements presented in the main text), as well as a laser shutter. The light is passed through a set of polarization optics for power control: a half-waveplate and polarizing beamsplitter followed by a polarizer on a motorized rotation mount. This light is coupled into a $1\times2$ fiber splitter which splits the light into a power monitoring arm and a signal arm. The power monitor arm is free-space-coupled to a Gentec TRAP7-Si-C-BNC \cite{NISTDisc} trap detector \cite{Traps}. The signal arm contains a calibrated neutral density (ND) filter with transmittance $T_{\text{ND}}=0.001412(1)$ mounted on a motorized flip mount. After this filter, the light is coupled into a fiber whose other end may be coupled either to a second trap detector or to a SPAD to be tested. Both traps are connected to high-precision transimpedance amplifiers (Gentec SDX-1226-1) with gains set to $10^8$ V/A for all experiments here, and the resulting voltages are measured by two Keysight 3458A  voltmeters. The SPADs that we calibrate and test in this work, two Excelitas SPCM-780-13-FC modules, are connected to an eight-channel Roithner TTM8000 time tagger, which receives SPAD clicks and sends data about the channel and time of the events to a computer for processing. 

To calibrate our detectors, measuring some of the necessary quantities requires having first measured others, so we gather three types of data and analyze them in this order: \deleted{(i) clicks from the SPAD due to cw input light at various (uncalibrated) input rates, to gather information about afterpulsing and the reset time; (ii) clicks from the SPAD with the detector capped, to extract the background count rate and the afterpulsing profile;}\added{(i) clicks from the SPAD with no signal light, to measure the background count rate, afterpulsing probability, and afterpulsing profile; (ii) clicks from the SPAD due to cw input light at various (uncalibrated) input rates, to gather information about recovery time effects;} and (iii)~clicks from the SPAD due to calibrated cw input light, to be compared with readings from the power monitor trap detector, to determine the detector efficiency. 

\paragraph{Background Count Rate.} \deleted{The next type of data to analyze is}\added{First, we analyze} background count data. \deleted{Since we want to measure the full background count rate, w}\added{W}e take care to set up the environment exactly as it will be for \deleted{tomography}\added{reconstruction}, including coupling the same fiber to the SPAD \deleted{under test }that will be used for \deleted{tomography}\added{reconstruction}; we simply block all \deleted{input}\added{signal} light. Then we collect time-tagged SPAD data for 12 hours to generate adequate statistics. We create full second-order correlation histograms, and\deleted{, as before,} fit to the linear background using the data corresponding to delays greater than \deleted{1 ms}\added{100~$\mu$s}. This line can be written  as $r^2 \Delta t (t_0-t)$, where $\Delta t$ is the bin size and $t_0$ is the data collection time. The extracted $r_{b}$ is the background count rate. (Although this extracted count rate is affected by dead time losses, our background count rates are so low that the error due to this\added{ effect} is negligible.)

An alternative strategy to measure the background count rate would be to simply count the total number of clicks in a dataset like the above (taken over a long time, with no input signal) and divide by the data collection time. This number will need to be corrected for the total afterpulsing probability, i.e. the probability for the detector to afterpulse at all times; however, to find this, we use a methodology (discussed in the next paragraph) that requires knowledge of the background count rate. Our method removes this circularity problem by eliminating the afterpulsing effects altogether while not being much more complicated.

\paragraph{Afterpulsing Profile and Probability, and Recovery Time.} After extracting this count rate, we also use the background counts dataset to generate a full afterpulsing profile. We form a first-and-second histogram from that data, which accurately maps the profile of afterpulsing clicks. In a first-and-second histogram, the background counts contribute an additional exponential factor $e^{-r_{b}t}$. Since the count rate is \added{very} low, the first-and-second data extends to several tens of milliseconds, so we can perform a fit beginning at \deleted{1 ms}\added{100~$\mu$s} delay. We fit to an exponential with the known background rate but amplitude to be fitted, then extrapolate the fit to zero time delay and subtract it from the histogram to obtain the afterpulsing profile. (We manually set the bins corresponding to the recovery time to zero.) The profile obtained this way is the one shown in Fig.~\ref{fig:AP_prof} above. \deleted{Finally, the afterpulsing profile can be scaled appropriately by noting that we have already measured the probability of afterpulsing within two recovery times; we simply scale the entire profile so that the sum of the profile in the first two recovery times matches this measured value. This also allows us to determine the afterpulsing probability over any amount of time by summing the profile over the corresponding interval of time. In particular, we can determine a total afterpulsing probability by summing over the entire profile; these values are given in Table I of the main text.}\added{If the histogram is then normalized by dividing by the total number of clicks in the dataset, then summing the remaining profile will give the probability of afterpulsing. Summing over the full profile gives the values in Table I of the main text.}

\added{We note that, purely due to statistical noise, the afterpulsing profile after the subtraction contains bins with slightly negative counts. However, because the calculation of $p_a$ in Eqn. \ref{eqn:AP_prof} involves cumulative probabilities over the profile, we will never actually have negative numbers in the calculation. In some sense the integration in Eqn. \ref{eqn:AP_prof} performs an average over these long times that mitigates the effects of the negative bins. These bins should not be set to zero; doing so would bias the resulting probabilities.}

\added{From the first-and-second histogram, we also identify the recovery time, defined to be the time of the first nonzero bin of the first-and-second histogram \cite{Autocorrelations}. This choice can be shown to correctly account for missing clicks, as described at the end of the next subsection.}

\paragraph{Dead and Reset Times.} To measure afterpulsing and recovery time effects, we use the DEP, which contains both afterpulsing and twilight counts, the latter being delayed until after the end of the recovery time \cite{TwilightDelays}. Although these two phenomena are overlapped in time, we have a way to tell them apart due to their physical origins: afterpulsing originates with the detector circuitry, while twilight counts originate from real photons whose clicks simply got delayed. Thus by varying the rate of input photons, we  vary the number of twilight counts, while the fraction of afterpulsing stays constant\added{ (for input photon rates that are not too high)}.

Based on this, our procedure is as follows \cite{ResetTimeMeasurement}. For several different input count rates, which do not need to be known independently, we take cw click data with the SPAD under test\deleted{, and generate a full second-order correlation histogram of the data.}\added{. From these data we generate a first-and-sixth histogram to measure the count rate as discussed above. We then generate a first-and-second histogram from the same dataset, and use this to extract the probability of a DEP click. Considering afterpulsing in more detail, let us suppose it has a temporal profile $a(t)$. At low count rates, this is the entire DEP. Then, after the recovery time ends, afterpulsing is competing with the Poissonian process of clicks from input photons, which leads to the second click having a probability density function}
\begin{equation}
    p(t) = (1-\bar{p}_a)re^{-r(t-\tau_r)} + \bar{p}_a\left[ p_{a}(t) e^{-r(t-\tau_r)} + (1-P_{a}(t))re^{-r(t-\tau_r)} \right], \hspace{0.5em} t \geq \tau_r.
    \label{eqn:IATFit}
\end{equation}

\noindent\added{Here as before, we define the recovery time $\tau_r$, and we defined the magnitude $\bar{p}_a = \int_0^\infty a(t)\,dt$, the probability $p_{a}(t) = a(t)/\bar{p}_a$, and the cumulative probability $P_a(t) = \int_0^t p_a(t')\,dt'$. At long times, $p_a(t) \rightarrow 0$ and $1-P_a(t) \rightarrow 0$, so only the first term is non-negligible. We use this as a background function for subtraction, multiplied by $N\Delta t$ to convert it to a number of events per bin. To handle the recovery time region, we find all negative bins preceding the maximum of the subtracted histogram and add back the subtracted function, which sets bins that had zero clicks to zero, and treats all clicks in this region as legitimate clicks to be included in the counting.}

\added{What remains in Eqn.~\ref{eqn:IATFit} after the subtraction consists of two terms. The second of those terms, in which afterpulsing is interrupted by a count from a photon, is negligible for all count rates discussed here. We can see this heuristically as follows. Let us model afterpulsing as an exponential decay, $p_a(t) \approx e^{-t/t_a}/t_a$. Then $P_a(t) = 1-e^{-t/t_a}$. Then the first term is larger than the second by a factor $1/(rt_a)$. Fig.~\ref{fig:AP_prof} suggests our detectors have an effective exponential decay timescale of just a few ns; thus, for the largest count rates in our measurements, $1/(rt_a) \sim 10$. Therefore we can safely neglect the last term in Eqn.~\ref{eqn:IATFit}. Thus, after the subtraction, we are left with the second term of Eqn.~\ref{eqn:IATFit}. Dividing the subtracted profile by $e^{-r(t-\tau_r)}$, we obtain $\bar{p}_a p_a(t)$. }\deleted{Then, beginning at 1~ms to avoid afterpulsing influencing the fit, we fit the data to a line $a(t_0-t)$, where $t_0$ is the total data collection time for the dataset. This linear fitting is shown by the orange dashed line in Fig.~\ref{fig:Hists_FC} for real data, and illustrated in Fig.~\ref{fig:RT1}. We then subtract this ``background'' from the histogram. After subtraction, what remains is only the DEP, which we sum }\added{This is summed} from zero delay to twice the recovery time\added{, which should capture all twilight counting effects while reducing uncertainty from the subtraction process}. \deleted{Summing only to twice the recovery time ensures that we only count adjacent pairs of clicks, and therefore only twilight counts and ``first-order'' afterpulses. }Dividing this sum by the number of clicks in the dataset, we finally obtain the fraction of clicks in the DEP.\deleted{ Fig.~\ref{fig:RT1} illustrates some subtleties of this procedure that are particularly important at high count rates.}

\begin{figure*}[t!]
\centering
\includegraphics[width=0.9\textwidth]{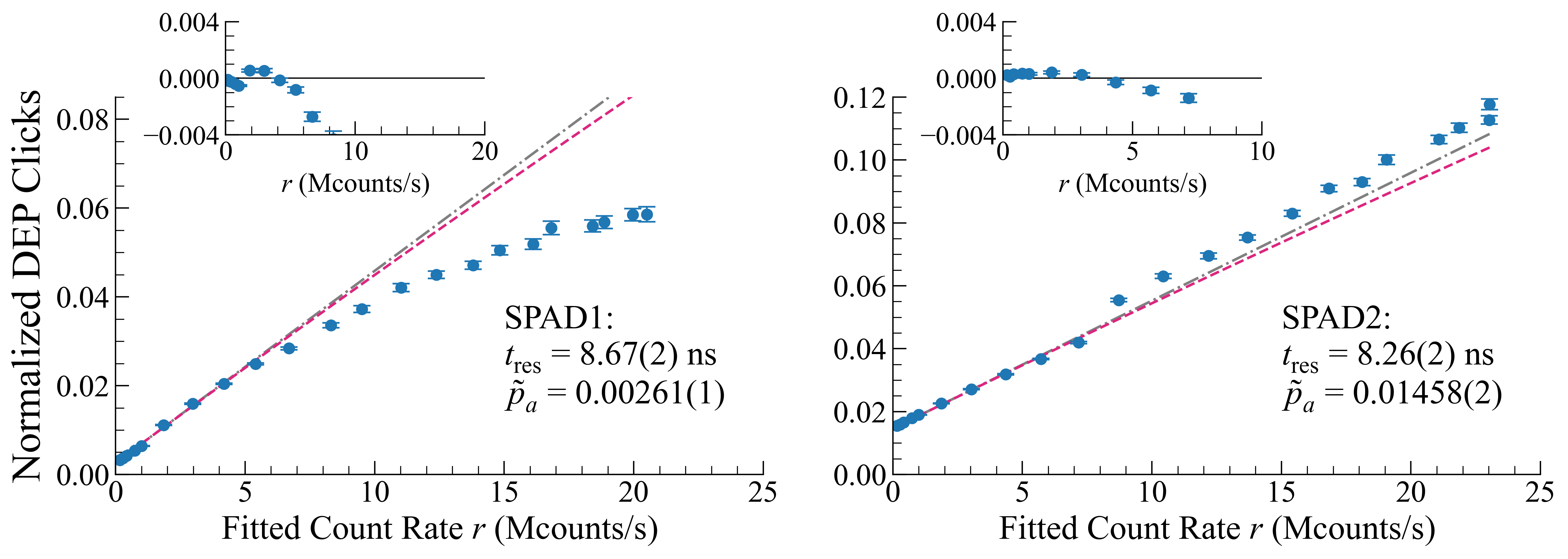}
\caption{\label{fig:ResetTimesFig} \added{Normalized DEP clicks as a function of count rate obtained from first-and-sixth histograms, for SPAD1 (left) and SPAD2 (right). The data (blue dots) are shown with error bars representing two standard deviations. As discussed in the text, the previously measured afterpulsing probability is used as the y-intercept of a linear fit to the data points with count rates below 5~Mcounts/s, which yields a value of the reset time $t_{\text{res}}$. The gray dash-dotted line shows this linear fit extrapolated over the plot. The dashed magenta line is the more accurate Eqn.~\ref{eqn:P_DEP}, using the previously obtained values of $p_a$, $t_{\text{res}}$, $t_{\text{rec}}$. As discussed in the text, as the count rate increases, the DEP clicks deviate from this model, and the two detectors deviate in different ways. Insets show residuals between the data points and the magenta model line, again with two-standard-deviation error bars.} \deleted{Illustration and results of the procedure for obtaining SPAD reset times and afterpulsing probabilities. (a) The illustration shows a full second-order correlation histogram (solid black line), with various features exaggerated for illustration purposes. Region I extends to twice the recovery time, and region III begins at 1 ms. To obtain the fraction of clicks in the DEP, we first fit the histogram in Region III to a linear function $a(t_0-t)$ where $t_0$ is the duration of the data collection (and the x-intercept of the line). This linear fit (dashed blue line) is extrapolated to zero time delay. Then we sum the clicks in the blue areas of Region I. The blue areas are all defined by the extrapolated line. The leftmost blue region represents bins where events are registered, but whose total clicks lie below the extrapolated linear background; we count all of these as real clicks and add them to the sum. We also add the clicks in the DEP above the line (the large light-blue-shaded peak in Region I). After the first bin above the linear background, we subtract any bins below the line, like the dark blue area of Region I. The sum is carried out  up to twice the recovery time, where the recovery time is defined as the first nonzero bin (not illustrated in this depiction). We divide the resulting sum by the total number of events to obtain the fraction of clicks in the DEP. (b) The fraction of clicks in the DEP plotted against detected count rate extracted from first-and-fourth histograms, for two different SPADs under test. Solid lines are linear fits to the data (blue triangles for SPAD1 and red squares for SPAD2), with fit function and fitting parameters indicated in the figure. With our model for the recovery time, the slope of the line is a measurement of half of the reset time $t_{\text{res}}$, while the y-intercept of the line is the probability of afterpulsing within twice the detector's recovery time $\tilde{p}_a$. The different offsets of the lines here thus indicate significantly different afterpulsing probabilities. The insets on the right show the residuals of the data with the linear fits.}}
\end{figure*}

\added{Using this procedure for several input count rates from about $2 \times 10^5$ to over $2 \times 10^7$ counts/s, we obtain the data shown in Fig.~\ref{fig:ResetTimesFig}. At low count rates (below about 5~Mcounts/s), we observe linear behavior, but at high count rates, a deviation from linearity becomes apparent.}\deleted{We obtain the fraction of clicks in the DEP for several different input count rates, and fit the results to a linear function of the count rate; these fits are shown in Fig.~\ref{fig:ResetTimesFig} for two SPADs. We measured the DEP click probability for nearly two dozen different count rates from about $2 \times 10^5$ to over $2 \times 10^7$ counts/s, and found linear behavior across the entire range. Here, the count rates are extracted from first-and-fourth histograms.}\added{  We model the DEP as occurring immediately after the end of the reset time and being composed of two types of clicks: (i) afterpulsing with probability $p_a$, and (ii) twilight counts, with probability related to the product of a photon arriving and the detector sensing it. The twilight count dynamics may be modeled as an inhomogeneous Poisson point process with mean rate}
\begin{equation}
    \langle n_{tw} \rangle = \int_0^{t_{\text{rec}}} r (1-D(t))\,dt = \frac{1}{2}r \left( t_{\text{rec}} - t_d \right).
    \label{eqn:n_tw}
\end{equation}
\added{The factor of $1/2$ is due to our choice of model and the definition of the reset time that we use; $\langle n_{tw} \rangle$ can be thought of as an effective area over which clicks will be detected as twilight counts. Eqn.~\ref{eqn:n_tw} defines a Poisson distribution of photon events which will generate twilight counts; taking the probability of at least one such event occurring, we obtain the twilight probability}
\begin{equation}
    p_t = 1 - e^{-\langle n_{tw} \rangle} = 1 - e^{-(t_{\text{rec}} - t_d)/2}.
\end{equation}
\added{So we can write the total DEP probability as}
\begin{equation}
    p_{\text{DEP}} = p_a + (1-p_a)\,p_t,
    \label{eqn:P_DEP}
\end{equation}
\added{When $rt_{\text{rec}}, rt_{\text{d}} \ll 1$, this reduces to a linear function, $\tilde{p}_{\text{DEP}} = p_a + (1/2)(1-p_a) r (t_{\text{rec}} - t_d)$, in which the offset is the afterpulsing probability and the slope is related to the reset time. As the count rate increases, Eqn.~\ref{eqn:P_DEP} describes saturation due to multi-photon events occurring during the reset time; this is shown by the magenta lines in Fig.~\ref{fig:ResetTimesFig}.}

\added{We perform a linear fit using the data points with count rates below 5~Mcounts/s, with the fit offset constrained using the afterpulsing probability found previously, so only the linear coefficient is a fit parameter. (This constraint produces no inconsistency since the afterpulsing probability was also found with a first-and-second histogram as described above.) This fit yields the reset time.} The final values for the afterpulse probability (within two recovery times) and the reset time for both SPADs under test are given in Fig.~\ref{fig:ResetTimesFig}. Note also that, from the reset time and the recovery time, we can determine the dead time as $t_{\text{dead}} = t_{\text{rec}} - t_{\text{reset}}$.

\added{The deviations from Eqn.~\ref{eqn:P_DEP} in the data are notable and different for the two detectors. SPAD1 exhibits a faster saturation than expected from Eqn.~\ref{eqn:P_DEP}. This fast saturation has been seen in other detectors as well \cite{CountingStats, Autocorrelations}, and it has been suggested that thermal properties of the detector that vary at high count rates might cause the observed deviations \cite{CountingStats}. Similarly, the deviations of SPAD2, which do not exhibit this fast saturation but nonetheless differ from the model, may be attributed to a number of possible mechanisms, such as recovery timings changing differently than for SPAD1, or non-Markovianity which changes the effective afterpulsing probability (both of which were discussed in subsection A.V. above). These deviations are not modeled by our algorithm, but in any case, they do not appear to be the primary limitation of the algorithm at high count rates.}

\added{We note here that there is an alternative method of measuring the dead time using an analysis similar to the above method for producing the reset time. If we compare the actual number of clicks in any of the datasets used to produce Fig.~\ref{fig:ResetTimesFig} to an expected number of clicks, there should be a deficit that yields information about the time for which the detector is dead. At each count rate $r$, we can calculate an expected number of clicks in our dataset as simply $rT$, where $T$ is the time over which click data at the rate $r$ was taken. We can compare this to the actual number of clicks in the dataset $N(1-p_a)$, where $N$ is the number of clicks in the dataset and $p_a$ is the afterpulsing probability of the detector; we remove the afterpulsing clicks with the factor $1-p_a$. The difference, properly normalized, is the expected fraction of clicks lost to the dead time effects, which we can write}
\begin{equation}
    p_{\text{lost}} = \frac{rT - N(1-p_a)}{rT}.
\end{equation}
\added{By comparison, by modeling the process as an inhomogeneous Poisson process similar to above [an integral identical to that of $\langle n_{tw} \rangle$ except for replacing $1-D(t)$ with $D(t)$], we obtain an expected fraction of lost clicks as $1-e^{-(t_{\text{rec}} + t_d)/2}$. Performing this analysis for the data in Fig.~\ref{fig:ResetTimesFig} and fitting to the model, we find recovery times consistent within uncertainty to the recovery times obtained by using the first nonzero bin of the dark-count first-and-second histogram. This validates this choice for measuring the recovery time.}

\added{We note that the method described in the last paragraph cannot be used to determine the recovery time, because the count rates $r$ extracted from the first-and-$n$ histograms rely on an input recovery time according to Eqn.~\ref{eqn:Prob_FirstOrder}. However, this does not invalidate this method as a check on the recovery time, because the fitted count rate is negligibly sensitive to small changes (a few ns) to the input recovery time. Different methods of determining count rates insensitive to dead times may make it possible to use the method of the last paragraph to determine a dead time; in addition, we note the possibility of performing a simultaneous nonlinear optimization on both the count rates and the data $p_{\text{lost}}$ to find a best fit for the recovery time.}

\paragraph{Quantum Efficiency.} Finally, it remains to measure the detector efficiency. To measure the DE, we need to compare a known input photon rate to a count rate measured by the SPAD, and the measured count rate needs to be the count rate the SPAD would measure if there were no recovery time effects, so that the measured detector efficiency is that of a fully armed detector, as required by our model. Typically the independent measurement is done with a conventional photodiode, and there is a gap of several orders of magnitude between what the photodiode can measure reliably and what the SPAD can accept. We decrease this gap from both sides. On one end, our photodiodes are trap detectors \cite{Traps} with high-gain and high-precision readout electronics that allow us to sense 100 pW reliably (the system's noise-equivalent power over a measurement of 300 s is roughly 1 pW). On the other end, as previously shown, extracting count rates using first-and-$n$ histograms allows us to reliably extract count rates up to tens of millions of counts per second, which is higher than conventional methods can achieve without correction for dead time effects. These two techniques reduce the gap between the needed powers to just 2 to 3 orders of magnitude, which can be bridged by commercial ND filters.

The input power is measured in the power monitor arm of the setup in Fig.~\ref{fig:ExperimentalSetup}. The optical power is free-space-coupled to a trap detector with efficiency $\eta_\text{pm}$, which converts the power to a current with a responsivity $R$; this current is converted to a voltage by a transimpedance amplifier (TIA) with gain $G$, and this voltage $V_\text{meas}$ is read by an 8-1/2-digit voltmeter. Thus the power in the power monitor arm is calculated as $P_\text{pm} = V_\text{meas}/(\eta_\text{pm}RG)$. Then, by coupling the signal arm into a fiber coupled to a second trap detector, we independently measure both the transmittance of the ND filter $T_{\text{ND}}$ and the power ratio $R_{\text{AB}}$ between the power monitor and signal branches. The ND filter is then placed in the signal path for the measurements of the detector efficiency. As a result, the power into the SPAD under test is $P_{in} = P_\text{pm}R_{\text{AB}}T_{\text{ND}}$.

Finally, we compare this input power to a count rate extracted from a first-and-\deleted{fourth}\added{sixth} histogram as \deleted{discussed}\added{described above}. The measured count rate must be corrected for background counts, so the rate to compare to is just $\mathcal{R} = \mathcal{R}_\text{meas} - \mathcal{R}_\text{dc}$, where $\mathcal{R}_\text{dc}$ is the independently measured background count rate.

Putting everything together, the detector efficiency is 
\begin{equation}
    \eta = \frac{\mathcal{R} \frac{hc}{\lambda}}{P_{in}} = \frac{\mathcal{R}_\text{meas} - \mathcal{R}_\text{dc}}{V_\text{meas}R_{\text{AB}}T_{\text{ND}}/\eta_\text{pm}RG} \frac{hc}{\lambda}.
    \label{eqn:eta}
\end{equation}
For all of the measurements here, $\lambda = 780$ nm with a linewidth of at most 1 MHz and a drift of not more than about 50 MHz, limited by a laser lock to a commercial wavemeter. In addition, $R = 0.6293$ A/W is the responsivity calculated using a simple one-photon-equals-one-electron rule, and $G=10^8$ V/A for all the measurements here.

\begin{figure}[t!]
\centering
\includegraphics[width=0.8\textwidth]{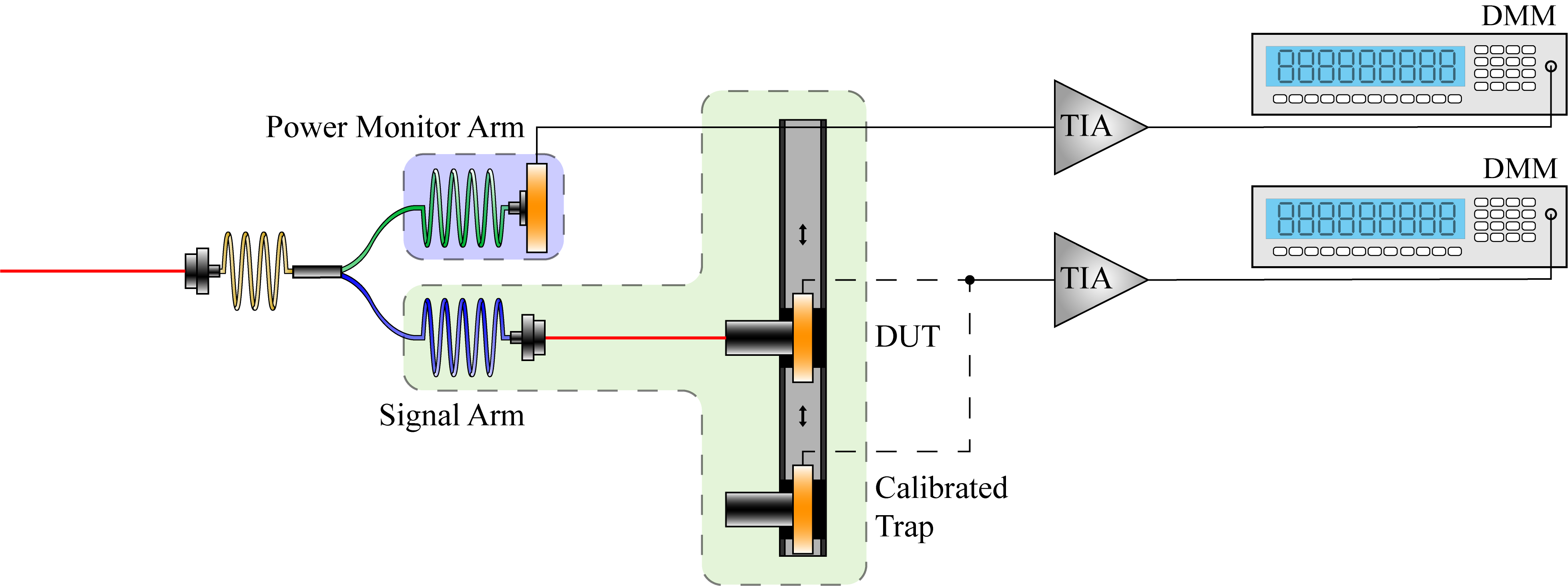}
\caption{\label{fig:TrapCalSetup} The setup used to transfer calibration to the trap detectors used in our measurements. Input light is sent through a fiber splitter into two arms, a power monitor arm and a signal arm. The former is coupled directly by fiber into a trap detector whose efficiency is not necessary to the calculations. The signal arm light is coupled into a fixed free-space path. The calibrated trap and the device under test (DUT) are transferred into and out of this path via a 1D optical rail. All traps are connected to high-gain TIAs which are connected to high-precision voltmeters.}
\end{figure}

To measure our trap detector's efficiency $\eta_{\text{pm}}$, we use a trap detector previously calibrated at the Visible to Near-Infrared Spectral Comparator Facility of the National Institute of Standards and Technology (NIST) \cite{NIST_cal_facility}. We transfer the calibration of this trap to our traps using the relatively simple setup shown in Fig.~\ref{fig:TrapCalSetup}. Light is split by a fiber splitter into two paths, a power monitor path and a signal path. The power monitor path is directly fiber-coupled into a trap detector whose efficiency is not necessary to know. The signal path light is coupled to a fixed free-space path; the NIST-calibrated detector and the device under test are alternately slid into this path using optical rails. First, we take 3 hours of continuous measurements of the voltages and compute an Allan deviation. From this, we set the duration of an individual measurement as 200 seconds. Then we perform 30 measurement cycles, over two days. Each measurement cycle consists of a period of 200 s of collecting voltage data from the NIST-calibrated detector and the power monitor trap, followed by sliding the device under test into the signal beam path and taking another period of 200 s of readings from the device under test and the power monitor traps. Thus each cycle gives a total of four sets of readings: the signal path readings with the NIST-calibrated trap $\{V_\text{i,sg,cal}\}$, the power monitor readings taken at the same time $\{V_\text{i,pm,cal}\}$, the signal path readings with the device under test $\{V_\text{i,sg,DUT}\}$, and the power monitor readings taken at the same time $\{V_\text{i,pm,DUT}\}$, with $i \in [1,N]$ indexing the $N$ readings in each set. In this way, for one measurement cycle, we calculate
\begin{equation}
    \frac{\eta_{\text{pm}}}{\eta_\text{cal}} = \frac{1}{N} \sum_{i=1}^N \frac{\tilde{V}_\text{i,sg,DUT}/\tilde{V}_\text{i,pm,DUT}}{\tilde{V}_\text{i,sg,cal}/\tilde{V}_\text{i,pm,cal}}
    \label{eqn:eta_trap}
\end{equation}
where $\tilde{V}$ denotes the background-subtracted versions of the sets $V$, and $\eta_\text{cal}$ is the efficiency of the NIST-calibrated trap detector.

\begin{figure*}[t!]
\centering
\subfloat{\includegraphics[width=0.33\columnwidth]{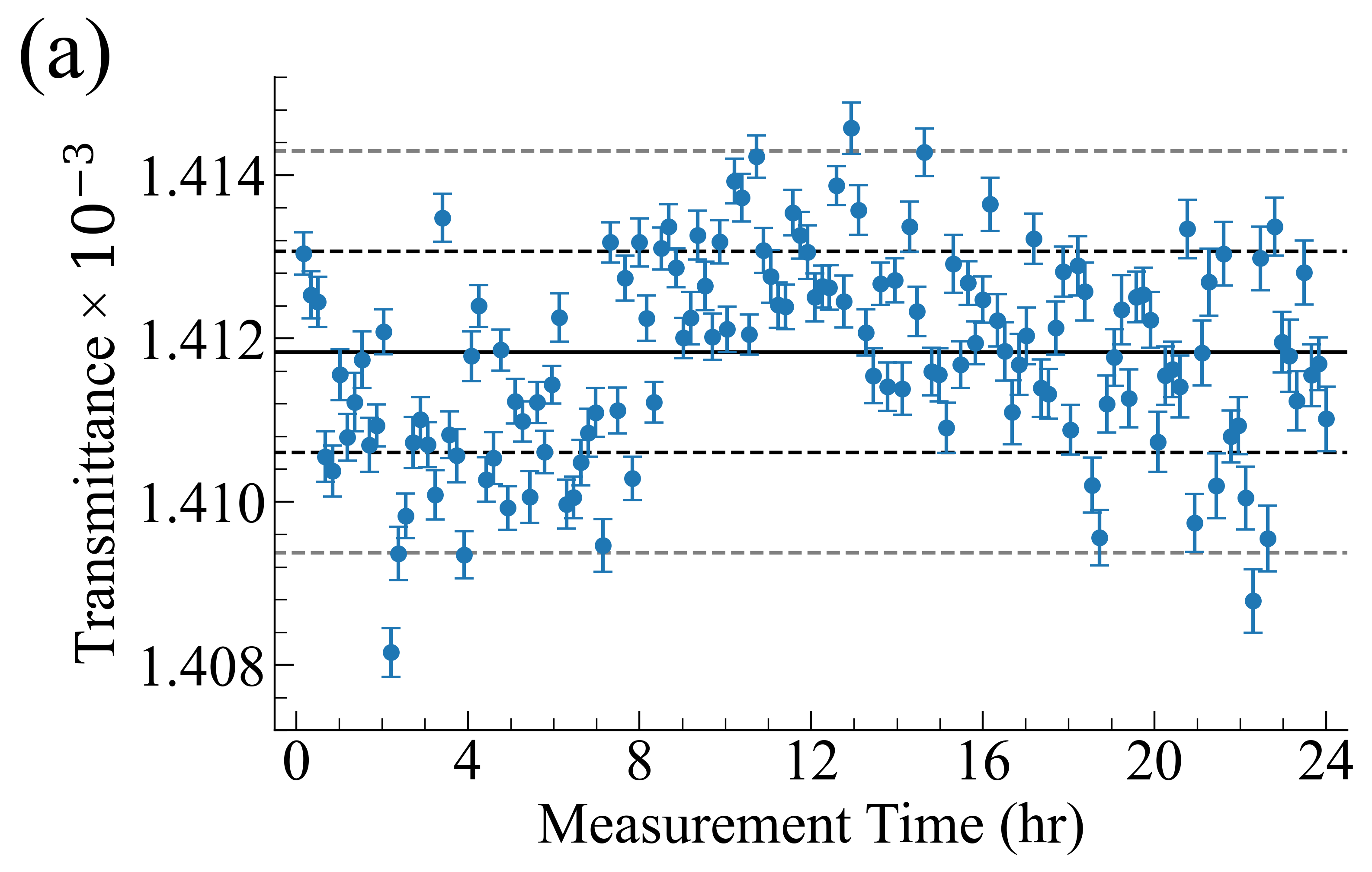} \label{fig:NDTransmissions}}
\subfloat{\includegraphics[width=0.33\columnwidth]{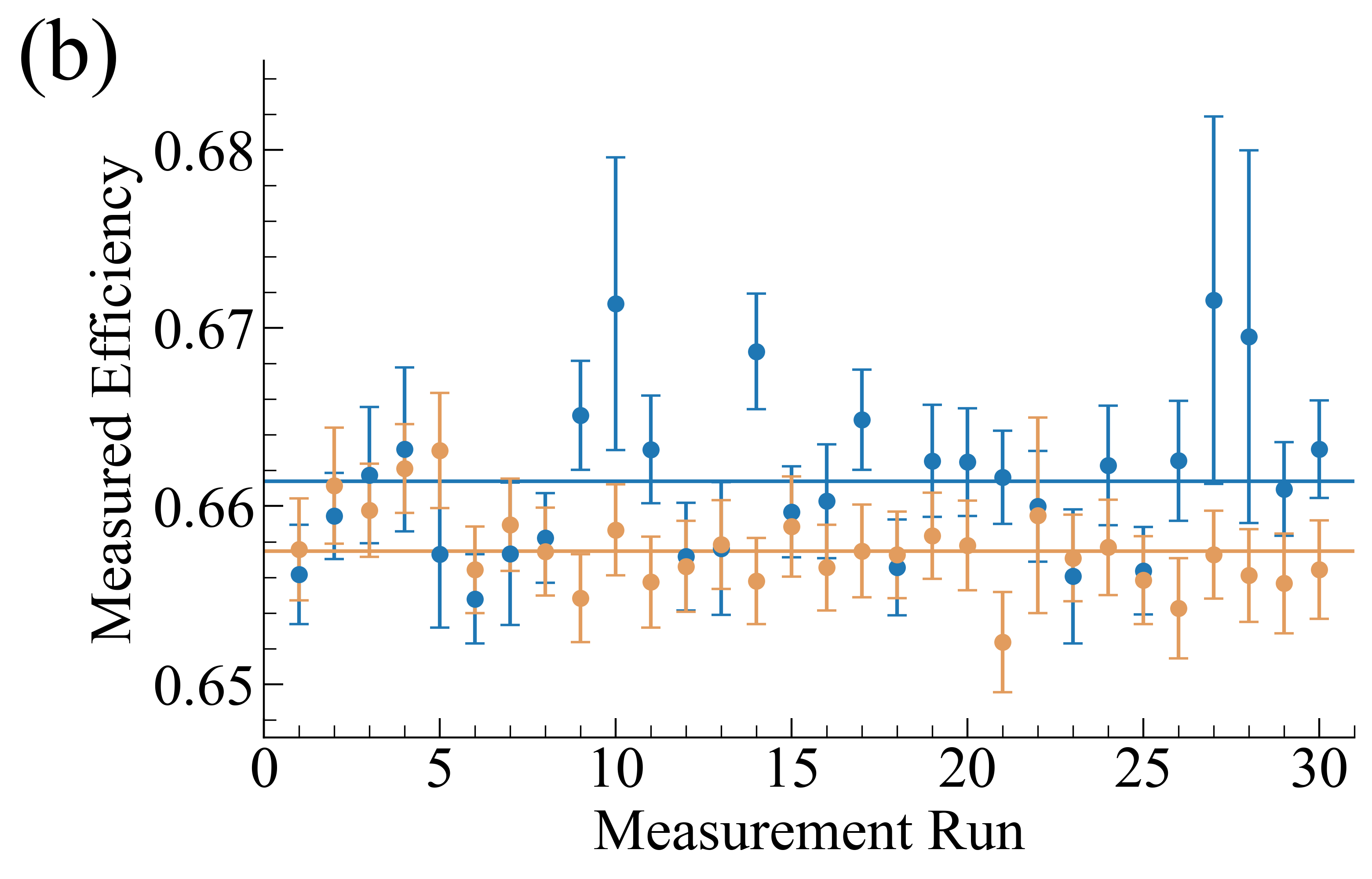} \label{fig:QE1}}
\subfloat{\includegraphics[width=0.33\columnwidth]{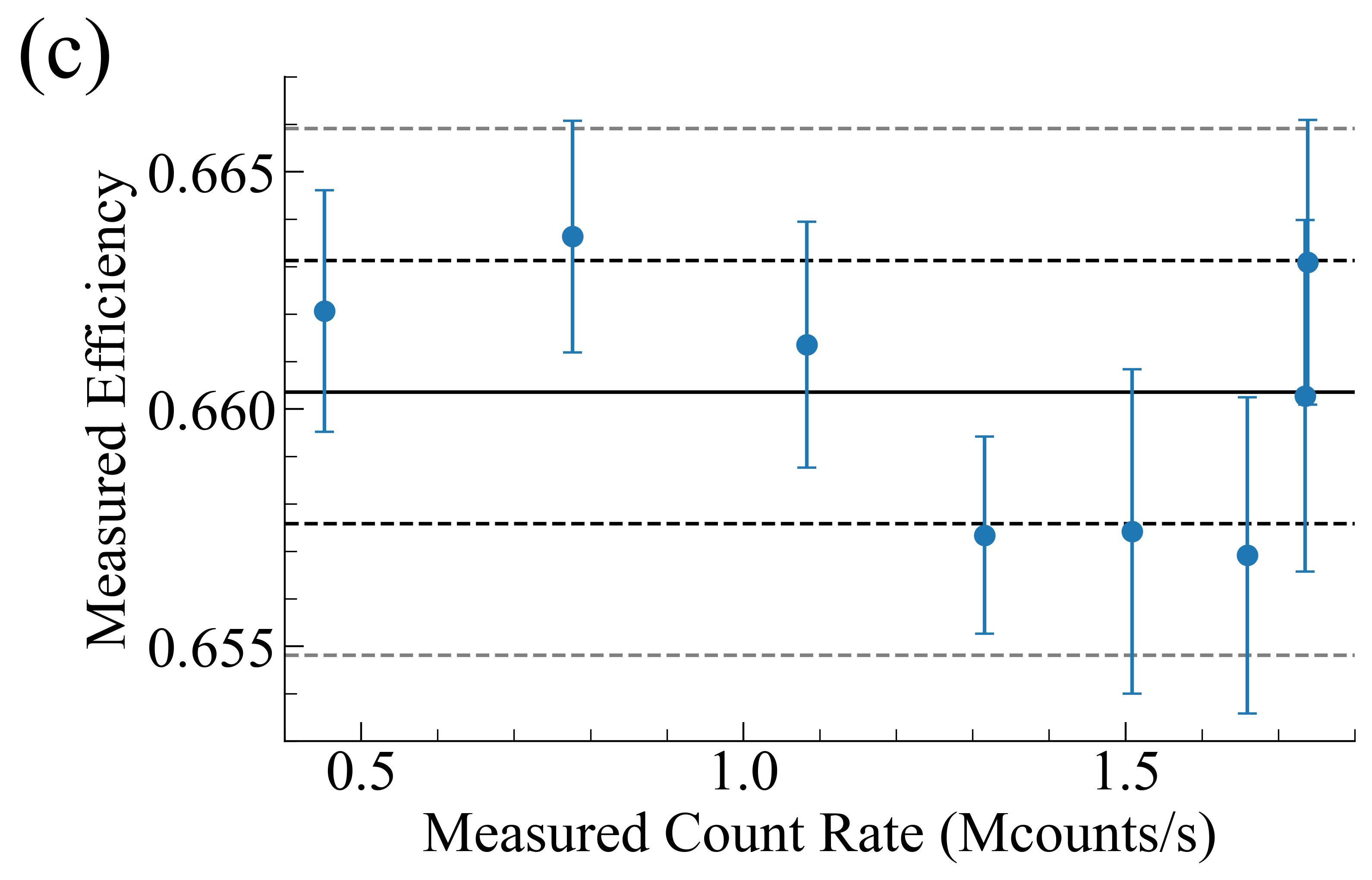} \label{fig:QE2}}

\caption{\label{fig:QEData} Detector efficiency measurement data. (a) Measured transmittance of the ND filter from 140 runs over 24 hours, with result $T_{\text{ND}} = 0.001412(1)$. Error bars represent one standard deviation uncertainty. The solid black line indicates the mean, and the dashed lines indicate one (black) and two (gray) standard deviations of the data. We attribute the variation to environmental changes over the course of the 24 hr measurement. (b) Two sets of 30 measurements of the detector efficiency of SPAD2, taken at two different input powers. Error bars represent one standard deviation uncertainty. The somewhat large variation is due primarily to a small number of voltage readings accompanying each measurement (as few as eight). Horizontal lines show the corresponding means of each data set. (c) Average detector efficiency measurements at several input count rates for SPAD2. Error bars represent one standard deviation uncertainty. The count rates on the horizontal axis are extracted using first-and-fourth histograms of the data. The solid black line indicates the mean, and the dashed lines indicate one (black) and two (gray) standard deviations of the data. We attribute most of the variation to environmental changes over the course of the 24 hr measurement.}
\end{figure*}

The quantities $T_{\text{ND}}$ and $R_{\text{AB}}$ also require separate measurement, but the measurement principles are similar to the measurement of $\eta_{\text{trap}}$. To calibrate the ND filter, we follow a technique similar to Ref.~\cite{NDFilters} (which also inspired the measurement of $\eta_{\text{pm}}$ just described). Using the setup in Fig.~\ref{fig:ExperimentalSetup}, the signal arm's fiber is attached to a trap detector, and readings from the signal and monitor arms' voltmeters are taken and compared. We compare 300 sec of readings with the filter in to 300 sec of readings with the filter out, so one measurement cycle gives four sets of readings: the signal path readings with the filter in $\{V_\text{i,sg,in}\}$, the power monitor readings taken at the same time $\{V_\text{i,pm,in}\}$, the signal path readings with the filter out $\{V_\text{i,sg,out}\}$, and the power monitor readings taken at the same time $\{V_\text{i,pm,out}\}$, with $i \in [1,N]$ indexing the $N$ readings in each set. The final calculation of the ND filter transmittance is then calculated as the mean
\begin{equation}
    T_{\text{ND}} = \frac{1}{N} \sum_{i=1}^N \frac{\tilde{V}_{i,\text{sg,in}}/\tilde{V}_{i,\text{pm,in}}}{\tilde{V}_{i,\text{sg,out}}/\tilde{V}_{i,\text{pm,out}}},
    \label{eqn:T_ND}
\end{equation}
where $\tilde{V}$ denotes the background-subtracted versions of the sets $V$. The results of 24 hr of measurements are shown in Fig.~\ref{fig:NDTransmissions}. We obtained the final value $T_{\text{ND}} = 0.001412(1)$. A discussion of uncertainties is reserved for the next section.

The measurement of the ratio of power in the two branches $R_{\text{AB}}$ is performed somewhat similarly, except that there are not two independent measurements needed: we simply take voltage readings for 300 seconds, and then divide the background-subtracted readings element-wise to obtain the ratio. Mathematically,
\begin{equation}
    R_{\text{AB}} = \frac{1}{N} \sum_{i=1}^N \frac{V_{i,\text{sg}}}{V_{i,\text{pm}}}
    \label{eqn:R_AB}
\end{equation}
where the notation is similar to the last paragraphs. Note that, in the signal arm, the trap is fiber-coupled, so that $R_{\text{AB}}$ includes the fiber coupling in addition to other factors such as the imbalance of the fiber splitter. Through a 24-hr measurement we obtained $R_{\text{AB}} = 61.5(1)~\%$.

The dark current of the traps was measured using a similar methodology to the above. We blocked all of the light and then took 12 hr of readings from the traps and performed an Allan deviation. This determines the time over which to measure and average the dark current, which was also 300 seconds. This measurement must be done for each setting of the TIA gain used. For a gain of $10^8$ V/A, the dark currents in both traps were measured to be $\approx 10^{-4}$ V.

Having performed measurements of all of the necessary quantities, we then measured the detector efficiency. We  chose eight input powers ranging from detected count rates of $2\times 10^5$ counts/s to $2 \times 10^6$ counts/s, and for each rate we took 30 measurements of $\eta$. A measurement cycle consists of opening a shutter to expose both the SPAD and the power monitor trap to the laser power for varied amounts of time between 4 s and 40 s, with the highest count rates using the shortest times and vice versa. We then match up the power monitor readings to the SPAD time tags from each cycle to obtain values of $\eta$ for each run using Eqn.~\ref{eqn:eta}. Two sets of 30 measurement runs at two different input powers are shown in Fig.~\ref{fig:QE1}. All 30 measurements were then averaged to obtain a measurement of $\eta$ at each count rate. These average values are shown plotted against count rate in Fig.~\ref{fig:QE2}. Since these did not show significant variation over the count rates used here, we averaged all of these measurements to obtain a final value for $\eta$ for each detector under test. The results are given in Table I of the main text.

\subsection{Calibration of Input Coherent States}

To validate our reconstruction technique, we need to compare the reconstruction to an independently measured distribution, so we separately calibrated the coherent states we used for reconstruction. Fig.~\ref{fig:ExperimentalSetup} shows the experimental setup we used for this purpose; before coupling the signal fiber into the SPAD for \deleted{tomography}\added{reconstruction} measurements, we coupled it into Trap 2, with efficiency $\eta_{\text{trap}}$ measured similarly to $\eta_{\text{pm}}$ above. We took voltmeter readings from Trap 2 with no ND filter in the path. The power onto Trap 2 is calculated as $P_{\text{in}} = V_{\text{meas}}/(\eta_{\text{trap}}RG)$, with notation similar to previous sections. For the \deleted{tomography}\added{reconstruction} measurement, the signal arm fiber is coupled into the SPAD and the calibrated ND filter is then inserted to deliver power $T_{\text{ND}}P_{\text{in}}$ to the SPAD. The power monitor arm was not used for this measurement.

A final factor to consider in this calibration is that the measurement of $P_{\text{in}}$ is done with continuous-wave (cw) light, so we need to multiply the power by a factor that tells us the total average energy in a pulse. The shape of the pulse as well as the peak power delivered during data collection must be taken into account by this factor. (The latter can be different from the cw power, for instance due to AOM thermalization effects.) We measure this factor directly using the SPADs. We take $2\times 10^6$ experimental cycles of each of two situations: (i) the AOM fully on and thermalized, and (ii) the AOM set to modulate its drive as appropriate to provide the pulse we wish to analyze. For these data we attenuate the beam using the polarizer after the AOM but before the detector, so that the count rate into the SPAD is low enough to be unaffected by recovery time effects while still probing the AOM behavior accurately. We subtract a background from both the pulsed and cw datasets, then choose a window of time of duration $T$ (equal to the duration used for the number state \deleted{tomography}\added{reconstruction} analysis) and divide the total number of clicks in the pulsed dataset in this window by the total number in the cw dataset in the same window. We denote this ``shape factor'' $f_s$.

Putting all of this together, we calculate the expected number of photons in any given coherent pulse as (Eqn. 5 of the main text)
\begin{equation}
    \bar{n}_{\text{exp}} = \frac{P_{\text{in}}f_s}{hc/\lambda} = \frac{V_{\text{meas}}T_{\text{ND}}f_s}{\eta_{\text{trap}}RG} \frac{1}{hc/\lambda}.
    \label{eqn:n_exp}
\end{equation}

\subsection{Effects of Detection System Timing}

\added{ The SPADs that we calibrate and test in this work, two Excelitas SPCM-780-13-FC modules, are connected to an eight-channel Roithner TTM8000 time tagger module (TTM).}

\added{The SPADs have timing jitter of 350 ps. However, our algorithm does not depend on the timing of any particular click, except in estimating the photon profile $\gamma(t)$. For this, a large number of experimental cycles allows us to average over detector jitter, giving a good estimation of $\gamma(t)$.}

\added{The TTM has an effective resolution of about 164 ps. Because we use bins of about 1 ns width (6 TTM bins), we believe the TTM resolution does not limit our results in any way.}

\section{\label{sec:unc_app}Discussions of Uncertainties}

In this section we discuss the calculation of all uncertainties presented in the paper.

\subsection{Uncertainties in Detector Characterization}

\added{The first quantity we measure in detector characterization is the background count rate. This is estimated using the data from a full second-order correlation histogram after 100~$\mu$s. To estimate the uncertainty, we bootstrap off of the fitted number of clicks in each bin, re-fitting the count rate each time; the standard deviation of these fitted count rates gives an estimate of the uncertainty in the background count rate.}

\added{Next, the afterpulsing probability is determined by subtracting a background from a first-and-second histogram of dark count data and summing what remains. Recall that the exponential fit, which begins at 100~$\mu$s delay, is constrained by the previously measured dark count rate, so that only the amplitude is a free parameter. To estimate the uncertainty in the resulting afterpulsing probability, we use the fit to bootstrap random backgrounds; we then fit to the bootstrapped data beyond 100~$\mu$s using a random dark count rate sampled from a normal distribution whose mean and width are the previously determined dark count rate and uncertainty in that rate. Then, as before, for each bootstrapped dataset, the fitted background is extrapolated to zero and subtracted, and what remains is summed to 100~$\mu$s, giving a bootstrapped estimate of the afterpulsing probability. The standard deviation of these probabilities gives our estimate of the uncertainty in the total afterpulsing probability $u_{a,0}$. To estimate the uncertainty on the probability of afterpulsing over any time window, we fractionally scale $u_{a,0}$. That is, if the probability of afterpulsing in a time window from $t=0$ to $t=t_0 < 100$~$\mu$s is $p_a(t_0)$, we calculate the uncertainty as $u_{a}(t_0) = u_{a,0} p_a(t_0)/p_{a,0}$.}

\deleted{We begin with a discussion of the reset time, recovery time, and afterpulsing uncertainties, as these are the first calculated quantities in the calibration analysis. The recovery time is simplest: it}\added{The recovery time} is estimated by multiplying the bin size by the number of bins with zero events at the start of a second-order histogram. A simple way to estimate the uncertainty is to thus use half a bin size. In our case $\Delta t = 164.6$ ps, the limit of our time tagger resolution according to the device specifications. Because we have many of these histograms at many different count rates, we can take an average over the measured recovery times, and divide the half-bin-width uncertainty by the square root of the number of histograms analyzed to reduce the estimated uncertainty.

The reset time \deleted{and afterpulsing probability are}\added{is} obtained by a linear fit to the fraction of clicks in the DEPs of \deleted{full correlation histograms}\added{first-and-second histograms} at different count rates, as shown in Fig.~\ref{fig:ResetTimesFig}. Each of the points in this fitting procedure has an uncertainty, which we estimate by combining in quadrature two uncertainty sources: (1) the total number of clicks in the peak (whose uncertainty we take to be the square root), and (2) the uncertainty in the slope of the linear background subtracted from the histogram to obtain the DEP. To estimate the latter, we perform a simple bootstrapping procedure on the data\deleted{ (we bootstrap so that we can model the errors as Poissonian rather than Gaussian)}. Then we perform the subtraction-and-summation process \deleted{shown in Fig.~\ref{fig:RT1}}\added{discussed above} in Monte Carlo fashion, randomly sampling \deleted{slopes}\added{count rates} from a Gaussian whose mean is the fitted \deleted{slope}\added{count rate} and whose uncertainty is the bootstrapped uncertainty. The standard deviation of the resulting set of counts in the DEP gives an estimate of the uncertainty in that number due to the uncertainty in the slope of the subtracted background. We sum (1) and (2) in quadrature to obtain the uncertainty in the fraction of clicks in the DEP. After fitting the data to a line using these estimated uncertainties as weights, we constrain the reduced $\chi^2$ to unity and use the resulting factor to scale the fitted parameter uncertainties. These are the final results we present in Table I of the main text and in Fig.~\ref{fig:ResetTimesFig}. We note also that we compute the dead time simply as $t_{\text{dead}} = t_{\text{rec}} - t_{\text{reset}}$, so the uncertainty in the dead time is propagated from the uncertainties in the recovery and reset times.

\deleted{One of the uncertainties we obtain from this linear fit is the uncertainty $u_{a,0}$ on the probability $p_{a,0}$ of afterpulsing within two recovery times. To estimate the uncertainty on the probability of afterpulsing over any time window, we fractionally scale $u_{a,0}$. That is, if the probability of afterpulsing in a time window from $t=0$ to $t=t_0$ is $p_a(t_0)$, we calculate the uncertainty as $u_{a}(t_0) = u_{a,0} p_a(t_0)/p_{a,0}$.}

\begin{table*}
\caption{\label{tab:QEUncertainties}%
Summary of sources of uncertainty for the detector efficiency as calculated in Eqn.~\ref{eqn:QE_AppE}. From left to right, columns are: quantity that contributes to the uncertainty in $\eta$, and its value; form of that quantity's contribution to $d\eta/\eta$, and the value of that contribution as a percentage of $\eta$; and a brief explanation of that source of uncertainty. Some of the numerical values given are typical values from a single measurement run, as appropriate. In the last row we give the calculated value and uncertainty for the DE using the values of this particular measurement run. These values come from a run measuring the DE of SPAD2, whose final measured DE is $\eta=0.660(3)$, with a fractional uncertainty (one standard deviation) of 0.45~\%. (Note that the value in the final row of the table is the value of $\eta$ obtained for this particular run, so it differs slightly from the final averaged value just given.) In the last column, (W)SD stands for (weighted) standard deviation.}
\begin{tabular}{lcccl}
\hline\hline
\multicolumn{2}{c}{Measurand} & \multicolumn{2}{c}{Contribution to $\delta\eta/\eta$ (\%)} & Source of Uncertainty \\
\hline
$\mathcal{R}_\text{dc}$ & 203(1) counts/s & $\delta \mathcal{R}_\text{dc}/(\mathcal{R}_{\text{meas}} - \mathcal{R}_\text{dc})$ & $7\times 10^{-5}$ & Fit to second-order histogram \\
$\mathcal{R}_{\text{meas}}$ & 1.512(2) Mcounts/s\footnotemark[1] & $\delta \mathcal{R}_{\text{meas}}/(\mathcal{R}_{\text{meas}} - \mathcal{R}_\text{dc})$ & 0.13\footnotemark[1] & Fit to first-and-fourth histogram \\
$V_{\text{meas}}$ & 41.96(9) mV\footnotemark[1] & $\delta V_{\text{meas}}/V_{\text{meas}}$ & 0.22\footnotemark[1] & SD of the mean of points \\
$R_{\text{AB}}$ & 0.615(1) & $\delta R_{\text{AB}}/R_{\text{AB}}$ & 0.16 & WSD of 140 points over 24 hr \\
$T_{\text{ND}}$ & 0.001412(1) & $\delta T_{\text{ND}}/T_{\text{ND}}$ & 0.09 & WSD of 140 points over 24 hr \\
$\eta_{\text{trap}}$ & 0.994(3) & $\delta \eta_{\text{trap}}/\eta_{\text{trap}}$ & 0.29 & WSD of 30 points over two days \\
\hline
$\eta$ & 0.661(3) & --- & 0.43 & --- \\
\hline\hline
\end{tabular}
\small{${}^a$ Typical value for one measurement cycle}
\end{table*}

\deleted{The uncertainty in the background count rates is estimated in a similar way to the above. We find the background count rate by fitting to the linear part of a full correlation histogram of dark count data, and the uncertainty of this slope is bootstrapped. Here there is typically very little data in each bin, so the bootstrapping process is important to model the errors as Poissonian rather than Gaussian.}

Finally we discuss the uncertainty in the detector efficiency. The sources of uncertainty are summarized in Table~\ref{tab:QEUncertainties} with values for one measurement of the DE of SPAD2. Recall that we compute the detector efficiency as
\begin{equation}
    \eta = \frac{\mathcal{R} \frac{hc}{\lambda}}{P_{in}} = \frac{\mathcal{R}_\text{meas} - \mathcal{R}_\text{dc}}{V_\text{meas}R_{\text{AB}}T_{\text{ND}}/\eta_\text{trap}RG} \frac{hc}{\lambda}.
    \label{eqn:QE_AppE}
\end{equation}
The final value for $\eta$ and its uncertainty for each detector is a weighted average and standard deviation, respectively, of the values for $\eta$ at several powers, as shown in Fig.~\ref{fig:QE2}. Each of these and its uncertainty are in turn the weighted average and standard deviation of 30 individual measurements of $\eta$ at a given input power (two sets of 30 at two input powers are shown in Fig.~\ref{fig:QE1}). For each individual run, we calculate the uncertainty in $\eta$ using standard error propagation techniques:
\begin{equation}
    \left( \frac{\delta\eta}{\eta} \right)^2 = \frac{\delta\mathcal{R}_{\text{meas}}^2 + \delta\mathcal{R}_{\text{dc}}^2}{\left( \mathcal{R}_{\text{meas}} - \mathcal{R}_{\text{dc}} \right)^2} + \left( \frac{\delta V_{\text{meas}}}{V_{\text{meas}}} \right)^2 + \left( \frac{\delta R_{\text{AB}}}{R_{\text{AB}}} \right)^2 + \left( \frac{\delta T_{\text{ND}}}{T_{\text{ND}}} \right)^2 + \left( \frac{\delta \eta_{\text{trap}}}{\eta_{\text{trap}}} \right)^2.
\end{equation}
We discuss the nature of the uncertainties in each quantity below, with typical values and contributions to the final uncertainty given in Table~\ref{tab:QEUncertainties}.

First, the uncertainty in the dark count rate $\delta\mathcal{R}_\text{dc}$ is found by bootstrapping uncertainties on the fit to a full second-order correlation histogram of dark count data, as previously discussed. Similarly, the uncertainty in the measured click rate $\mathcal{R}_\text{meas}$ is found by bootstrapping uncertainties on the fit to the first-and-\deleted{fourth}\added{sixth} histograms that yield $\mathcal{R}_\text{meas}$. 

Next, $V_\text{meas}$ is the average of several background-subtracted readings of the voltmeter attached to the power monitor trap detector during the time the SPAD is collecting count rate data. We take the uncertainty in $V_\text{meas}$ to be the standard deviation of the individual readings, divided by the square root of the number of readings (which is supported by Allan variance curves indicating the Gaussian nature of the noise), summed in quadrature with the uncertainty in the independently-measured dark current. Because we only sample the voltmeter about 3 times per second (this is software-limited; the voltmeter's integration time is 50 ms), and the shutter is sometimes open for as little as a few seconds, this can be the largest source of uncertainty for some of the measurement runs.

The uncertainties on $R_{\text{AB}}$, $T_{\text{ND}}$, and $\eta_{\text{trap}}$ are all estimated using similar methods. For $R_{\text{AB}}$ and $T_{\text{ND}}$, the values and uncertainties are taken to be the weighted average and standard deviation of 140 measurements each taken continuously over 24 hr, while for $\eta_{\text{trap}}$ there are 30 total measurements taken at various times over two days. We use the full standard deviation (as opposed to a standard deviation of the mean) because the measurement samples environmental variation that we wish to include in the uncertainty. Each individual measurement is in turn a simple standard deviation of the element-wise ratios used to compute the value of the quantity, the quantities in the sums of Eqs.~\ref{eqn:R_AB}, \ref{eqn:T_ND}, and \ref{eqn:eta_trap} respectively, divided by the square root of the number of readings in the cycle.

We note that it is possible to characterize the variations and uncertainties in an ND filter much more carefully than we did \cite{NDFilters}. For this work we adopted a simplifying approach by only measuring the transmittance of the filter \emph{in situ}. The filter was mounted on a motorized flip mount that allowed for precise replacement of the filter in the beam path, so that we always sampled the same location on the filter (as an alternative to measuring different locations on the filter and including the results in an uncertainty analysis). By measuring the filter over the course of one full day we believe we adequately sample the variation of the transmittance caused by temperature and humidity variations in the lab. We also believe the effects of detector nonlinearity and interference within the filter, which is absorptive, to be negligible. This means that not only do we neglect etaloning effects on the measured powers, but we also neglect any additional wavelength dependence of the transmittance due to these effects. Furthermore, the laser is sufficiently narrow, and its drifts are sufficiently small, that we can neglect all variations in transmittance as a function of wavelength.

For $\eta_{\text{trap}}$, the quantity we really measure is $\eta_r \equiv \eta_{\text{trap}}/\eta_\text{cal}$. The calibration itself has some uncertainty, so that we compute the overall uncertainty as
\begin{equation}
    \left( \frac{\delta \eta_{\text{trap}}}{\eta_{\text{trap}}} \right)^2 = \left( \frac{\delta \eta_r}{\eta_r} \right)^2 + \left( \frac{\delta \eta_\text{cal}}{\eta_\text{cal}} \right)^2.
\end{equation}
For us, $\eta_\text{cal} = 0.989(3)$.

We suppose that the electronics---the TIA gain and the voltmeter---add negligible statistical uncertainty, but we mention the small systematic uncertainties here. The gain of the transimpedance amplifier has a quoted 0.025~\% accuracy. We estimate the voltmeter to contribute a systematic uncertainty of no more than 0.01~\%. We also suppose the wavelength of the light to have negligible uncertainty, and therefore no contribution to the uncertainty of the detector efficiency, either directly or through variations in $T_{\text{ND}}$ or $\eta_{\text{trap}}$ with wavelength.

\subsection{Uncertainties in Input Coherent State Calibration}

Finally, we briefly discuss the uncertainty in the calibrated input coherent state average photon numbers  $\bar{n}_{\text{exp}}$ of Eqn.~\ref{eqn:n_exp} (or Eqn. 5 of the main text). The uncertainty is calculated as
\begin{equation}
    \left( \frac{\delta\bar{n}_{\text{exp}}}{\bar{n}_{\text{exp}}} \right)^2 = \left( \frac{\delta V_{\text{meas}}}{V_{\text{meas}}} \right)^2 + \left( \frac{\delta T_{\text{ND}}}{T_{\text{ND}}} \right)^2 + \left( \frac{\delta \eta_{\text{trap}}}{\eta_{\text{trap}}} \right)^2 + \left( \frac{\delta f_s}{f_s} \right)^2.
    \label{eqn:delta_n_bar}
\end{equation}
We have discussed the nature of $\delta T_{\text{ND}}$ and $\delta\eta_{\text{trap}}$ already. Further, because $V_{\text{meas}}$ is simply the average of a set of voltage readings, we take $\delta V_{\text{meas}}$ to be the standard deviation of those readings, divided by the square root of the number of readings.

The shape factor $f_s$ is calculated as a ratio $\tilde{N}_\text{pd}/\tilde{N}_\text{cw}$ where $\tilde{N}_\text{pd}$ is the total number of clicks in the pulsed dataset and $\tilde{N}_\text{cw}$ that in the cw dataset, and the tildes denote background-subtracted quantities. The number of background clicks over the time window of interest is $N_\text{bg}$. We suppose the uncertainties in $N_\text{pd}$, $N_\text{cw}$, and $N_\text{bg}$ to be the square roots of the respective quantities, so standard uncertainty propagation gives

\begin{equation}
    \left(\frac{\delta f_s}{f_s}\right)^2 = \frac{N_\text{pd}}{\tilde{N}_\text{pd}^2} + \frac{N_\text{cw}}{\tilde{N}_\text{cw}^2} + \left( \frac{N_\text{pd} - N_\text{cw}}{\tilde{N}_\text{pd}\tilde{N}_\text{cw}} \right)^2 N_\text{bg}.
\end{equation}

\subsection{Uncertainties on Reconstructed Distributions}

We obtain uncertainties on the reconstructed distributions by Monte Carlo sampling. There are two types of uncertainties involved. (1) Sampling uncertainties, in which the detected click distribution varies from $\mathbb{D} \mathbf{P}$ (using the notation from the beginning of Section A) due to the experiment being finite and probabilistic. This produces statistical uncertainties on the number components of the distribution. We estimate this uncertainty by assuming each number component of the measured distribution follows a Poissonian distribution, which, while not really correct, should produce a reasonable estimate of the magnitude of the effects. (2) Detector parameter uncertainties, those described in subsection C.I. Changes in the values of the detector parameters produce highly correlated changes in the number components of the reconstructed distribution; for instance, a lower detection efficiency will cause larger population in higher number components \emph{and} smaller population in lower number components. 

\begin{figure*}[t!]
\centering
\includegraphics[width=\textwidth]{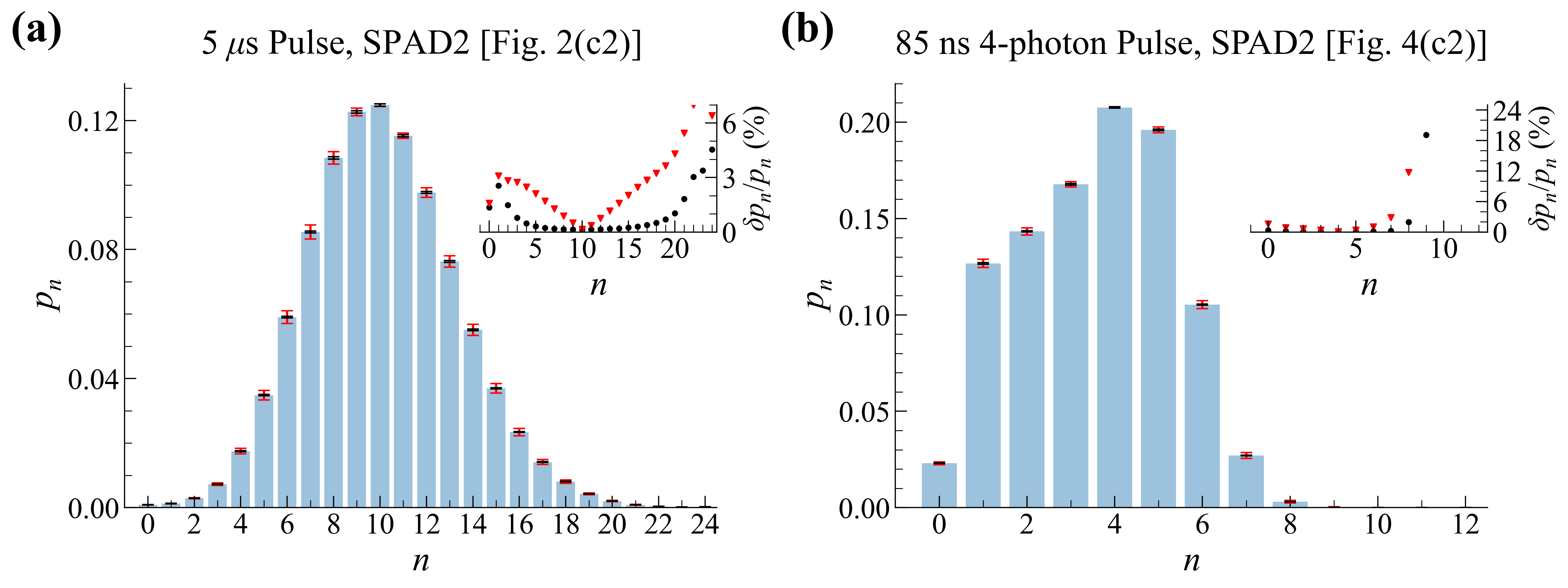}
\caption{\label{fig:ReconUncs1} Reconstructions of two of the datasets from the main paper, showing uncertainties in each number component. The left figure is Fig. 2(c2) of the main paper, the reconstruction of a 5 $\mu$s coherent pulse from SPAD2; the right is Fig. 4(c2) of the main paper, the reconstruction of the 85 ns 4-photon coherent pulse from SPAD2. To increase visual clarity, the main figures' error bars show two standard deviations. As discussed in the text, the black error bars represent statistical uncertainty due only to sampling error, while the expanded red bars show effects due to the detector parameter uncertainties, which are highly correlated across number components. The insets show (one standard deviation) percent uncertainties $\delta p_n/p_n \times 100$, where $\delta p_n$ is the uncertainty in number component $p_n$. Black dots indicate sampling error and red triangles indicate additional uncertainty due to uncertainty in detector parameters.}
\end{figure*}

We show these two types of uncertainties in Fig.~\ref{fig:ReconUncs1} for two of the pulses reconstructed in the main text. We use two error bars on each number component, the smaller black one indicating the purely statistical uncertainty due to sampling error, and the larger red one indicating the additional uncertainty due to detector parameter uncertainties. The magnitude of the error bars is one standard deviation of the results of 1,000 Monte Carlo simulations. The interpretation is that, accounting for all uncertainties, the value of each number component lies (within one standard deviation) within the larger red error bars, but that variation of any one number component outside the smaller black bars is correlated with variations of other components elsewhere. Thus the larger red error bars give a sense of the magnitude of the variation due to the uncertainty in the detector parameters, but are not representative of uncorrelated statistical uncertainties.

\begin{table*}[h!]
\centering
\caption{\label{tab:ReconUncs}%
Breakdown of contributions to the uncertainties of a few number components of the reconstruction of the 5 $\mu$s SPAD2 coherent pulse. All uncertainties are one standard deviation. As discussed in the main text, the uncertainties from all but sampling error are highly correlated with one another. The quadrature sum of the numbers given matches the total uncertainty $\delta p_n$ found by a Monte Carlo simulation with all uncertainties accounted for. We also give the total fractional uncertainty $\delta p_n/p_n$. The uncertainty contributions from the recovery time effects are zero due to the way we discretize $D(\tau)$, as discussed in the main text, so we omit $\delta\mathbb{R}$ from the table.}
\sisetup{
table-alignment-mode = format,
table-number-alignment = center,
}
\begin{tabular}{@{}
lrrrr
@{}}
\hline\hline
Contribution ($\times 10^{-8}$) & {$n=2$} & {$n=6$} & {$n=10$} & {$n=14$} \\
\hline
Sampling error & 4200 & 14000 & 17000 & 11000 \\
$\delta\mathbb{L}$ & 7200 & 102000 & 9900 & 88000 \\
$\delta\mathbb{B}$ & 1 & 21 & 2 & 19 \\
$\delta\mathbb{A}$ & 21 & 2000 & 2800 & 3100 \\
\hline
Total, $\delta p_n$ & 8300 & 103000 & 20000 & 89000 \\
$\delta p_n/p_n$ & {2.7~\%} & {1.7~\%} & {0.16~\%} & {1.6~\%} \\
\hline\hline
\end{tabular}
\end{table*}

The insets of Fig.~\ref{fig:ReconUncs1} show the percent uncertainties in each component, with the black and red dots corresponding to the black and red error bars of the main figures. \added{For the data in the left panel of Fig.~\ref{fig:ReconUncs1}, whose reconstruction closely matched the expected Poissonian,}\deleted{In the two pulses analyzed,} the total percent uncertainties, including the correlated uncertainties, stay below 5~\% for all components with at least 0.1~\% of the reconstructed population.\added{ For the right panel of Fig.~\ref{fig:ReconUncs1}, in which the algorithm failed to reconstruct a Poissonian, the uncertainties on the large-$n$ components are large, perhaps a consequence of the algorithm's failure to accurately reconstruct.} To further quantify the magnitudes of the uncertainties and gain insight in their sources, in Table~\ref{tab:ReconUncs} we give a breakdown of uncertainty by source for several number components of the reconstructed 5 $\mu$s SPAD2 pulse shown in Fig.~\ref{fig:ReconUncs1}(a). To generate these values, we ran sets of 1,000 Monte Carlo simulations with all uncertainties set to zero except the one under test.

The \added{general form of the }sampling error as a function of $n$ is determined by the shape of the measured click distributions. For both of the pulses in Fig.~\ref{fig:ReconUncs1}, the click distributions are peaked, with a relatively small number of zero-click events, followed by an increase to some maximum number, and then a decrease to zero events for a large number of clicks. Thus the percent uncertainty due to sampling error is large at low and high $n$, but is small at intermediate $n$. 

Though the sampling error is a non-negligible contribution for many of the components, the uncertainties due to detector parameters dominates for most number components, as shown in Table~\ref{tab:ReconUncs} and inferred from the insets of Fig.~\ref{fig:ReconUncs1}. Among these, the uncertainty from the detector efficiency is by far the largest contribution. Though all of our detector parameters' estimated relative uncertainties are roughly the same (0.1~\% to 0.5~\%), these uncertainties contribute at different levels because the parameters themselves contribute to the reconstruction at different levels. Since the total effect of afterpulsing is at most $\approx$ 2~\% and the total effect of background counts is even smaller in all of our data, the effects of the detector efficiency are the largest corrections we make. We consider the DE an order unity correction, the afterpulsing an order 0.01 correction, and the background counts even smaller. The uncertainty contributions track this hierarchy fairly closely, though afterpulsing becomes more important at larger $n$.

The dead time and recovery time uncertainties contribute nothing to the reconstruction uncertainties, which is due to the way we discretize $D(\tau)$. In constructing this function, we round both $t_{\text{dead}}$ and $t_{\text{rec}}$ to the nearest bin, and then interpolate between these; since the uncertainty in these quantities is much less than the typical bin width of 1 ns, every Monte Carlo simulation chooses the same bins after rounding, resulting in no difference in the matrix $\mathbb{R}$ across simulations.

The uncertainties being dominated by the uncertainty in the detector efficiency largely explains the patterns of fractional uncertainty seen in the insets of Fig.~\ref{fig:ReconUncs1}. A change in the DE would shift the peak of the reconstructed distribution, so the components on the ``sides'' of the peak are highly sensitive to changes in the DE, while components near the peak are first-order insensitive. A second effect also plays a part: at larger $n$, generally more factors of the efficiency are involved in the reconstruction, so the large-$n$ components are also increasingly sensitive to changes in the DE. Both of these patterns are visible in the insets of Fig.~\ref{fig:ReconUncs1}.

For the anti-bunched light reconstructed in this work (Fig. 5 of the main text), we used Monte Carlo simulation in a similar manner as above to investigate the uncertainty on the reconstructed value of $g^{(2)}(0)$. The balance of uncertainties is somewhat different than that of the number components of coherent states above. The primary contribution to the uncertainties on the reconstructed $g^{(2)}$ presented in the main paper is sampling error. This can of course be reduced by taking more data. The next largest contributor, a factor of $2$-$3$ lower, is the uncertainty in the afterpulsing probability. In this work, the detector efficiency uncertainty contributes negligible uncertainty to the reconstructed value of $g^{(2)}$. This is sensible since, theoretically, losses should not affect the value of $g^{(2)}$ \cite{MigdallSPDs}. The balance of other effects may be due to our particular setup's very low efficiency of photon production from our Rydberg ensemble.

\subsection{A Note to the User}

\begin{figure*}[t!]
\centering
\includegraphics[width=\textwidth]{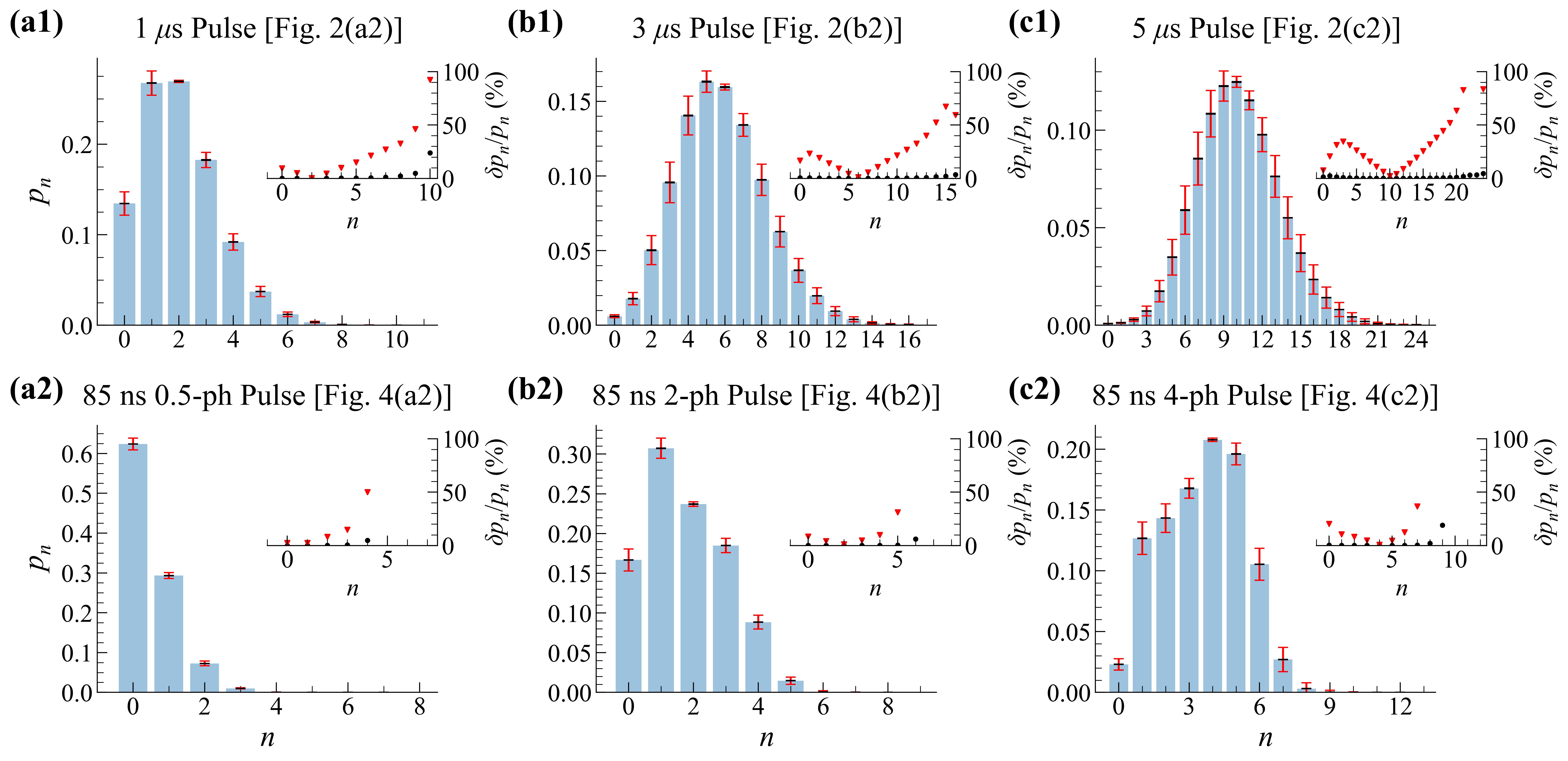}
\caption{\label{fig:ReconUncs2} Uncertainties (one standard deviation) on reconstructions when the DE has a relative uncertainty of 5~\%. All of these data are from SPAD2, corresponding to Figs. 2 and 4 of the main text, as indicated in the square brackets in each plot's subtitle: the top row are long coherent pulses (1 $\mu$s, 3 $\mu$s, and 5 $\mu$s from left to right), and the bottom row are short coherent pulses (85 ns, with roughly 0.5 photons, 2 photons, and 4 photons per pulse from left to right). The error bars are one standard deviation; black bars indicate only sampling error, while the expanded red bars indicate the correlated errors due to uncertainties in the detector parameters. The insets show (one standard deviation) percent uncertainties $\delta p_n/p_n \times 100$, with black dots indicating sampling error and red triangles indicating additional uncertainty due to uncertainty in detector parameters. The notable trend is that the magnitude of the uncertainties gets larger with larger $n$, for both long and short pulses.}
\end{figure*}

In our detector characterization efforts detailed in Section B, by far the most involved measurement was that of the detector efficiency, in terms of both time and equipment. Therefore it is worthwhile to consider the consequences of a less-involved efficiency measurement procedure on the reconstructions. We found that with inexpensive commercial power meters we were able to measure the efficiencies of our detectors to $\approx$ 5~\% relative uncertainty. We used this relative uncertainty in new Monte Carlo simulations to estimate the resulting uncertainties on the reconstructions. We performed this procedure for several of the pulses of the main text, with results shown in Fig.~\ref{fig:ReconUncs2}.

Besides the uncertainties on the components all being noticeably larger, the uncertainties also generally increase with larger $n$. This is the same effect as discussed in the last section: heuristically speaking, a number state component $n$ can involve up to $n$ powers of the DE in the reconstruction, so larger $n$ components are more sensitive to variations in the DE. This can give the impression of a larger ``average'' uncertainty across the reconstructed distribution when the average $n$ is larger.

From this, it is clear that using this tomographic method precisely requires a good detector efficiency measurement.

We also ran the same Monte Carlo simulations on our anti-bunched light data (Fig. 5 of the main text) to investigate the effects on the reconstructed value of $g^{(2)}(0)$. These simulations (not shown) reveal that uncertainty on the detector efficiency, even at the level of 5~\%, has negligible effect on the reconstructed $g^{(2)}$. This implies that, if one is seeking to characterize anti-bunched light and is only interested in a value of $g^{(2)}$, one does not need to carefully characterize the detector efficiency. As previously mentioned, our setup produces photons only rarely, and this fact may be responsible for the balance of uncertainties here.

\clearpage
\bibliography{refs}